\documentclass[sigconf,9pt]{acmart}
\makeatletter
\def\@ACM@checkaffil{% Only warnings
	\if@ACM@instpresent\else
	\ClassWarningNoLine{\@classname}{No institution present for an affiliation}%
	\fi
	\if@ACM@citypresent\else
	\ClassWarningNoLine{\@classname}{No city present for an affiliation}%
	\fi
	\if@ACM@countrypresent\else
	\ClassWarningNoLine{\@classname}{No country present for an affiliation}%
	\fi
}
\usepackage{graphicx}
\usepackage{algorithmic}
\usepackage[linesnumbered,ruled,lined]{algorithm2e}
\usepackage{xcolor,colortbl}
\usepackage{tcolorbox}
\usepackage[nolist,nohyperlinks]{acronym}
\usepackage[utf8]{inputenc}
\usepackage{bm}
\usepackage{lipsum}
\usepackage[inline]{enumitem}
\usepackage{url}
\usepackage{natbib}
\usepackage{multirow}
\usepackage{amsmath}
\usepackage[export]{adjustbox}
\usepackage{balance}
\usepackage{float}
\usepackage{graphicx}
\usepackage[subrefformat=parens,labelformat=parens]{subfig}
\usepackage{svg}

\setcitestyle{sort&compress}

\graphicspath{ {./} }
\include{misc_macros}

\newcommand{\ourmethod}{\textit{MARS}\;}

\begin{document}

\title{Continuous Multi-user Activity Tracking via Room-Scale mmWave Sensing}

\author{Argha Sen*, Anirban Das*, Swadhin Pradhan**, Sandip Chakraborty*}
\affiliation{%
	\institution{*IIT Kharagpur, India, **Cisco Systems, USA}
 % \country{India*, USA**}
}
\email{{arghasen10, anirbanfuture, swadhinjeet88, sandipchkraborty}@gmail.com}

% \author{Argha Sen}
% \affiliation{%
% 	\institution{Indian Institute of Technology Kharagpur}
% 	% \country{Kharagpur, India}
% 	}
% \email{arghasen10@gmail.com}

% \author{Anirban Das}
% \affiliation{%
% 	\institution{Indian Institute of Technology Kharagpur}
% 	% \country{Kharagpur, India}
% }
% \email{anirbanfuture@gmail.com}

% \author{Swadhin Pradhan}
% \affiliation{%
% 	\institution{Cisco Meraki, USA}
% 	% \country{}
% }
% \email{swadhinjeet88@gmail.com}

% \author{Sandip Chakraborty}
% \affiliation{%
% 	\institution{Indian Institute of Technology Kharagpur}
% 	% \country{Kharagpur, India}
% }
% \email{sandipc@cse.iitkgp.ac.in}

\renewcommand{\shortauthors}{Argha Sen, et al.}

\begin{abstract}
Continuous detection of human activities and presence is essential for developing a pervasive interactive smart space. Existing literature lacks robust wireless sensing mechanisms capable of continuously monitoring multiple users' activities without prior knowledge of the environment. Developing such a mechanism requires simultaneous localization and tracking of multiple subjects. In addition, it requires identifying their activities at various scales, some being macro-scale activities like walking, squats, etc., while others are micro-scale activities like typing or sitting, etc. In this paper, we develop a holistic system called \ourmethod{} using a \textit{single} \ac{COTS} \ac{mmWave} radar, which employs an intelligent model to sense both macro and micro activities. In addition, it uses a dynamic spatial time-sharing approach to sense different subjects simultaneously. A thorough evaluation of \ourmethod{} shows that it can infer activities continuously with a weighted F1-Score of $>94\%$ and an average response time of $\approx 2$ sec, with $5$ subjects and $19$ different activities.        
\end{abstract}
\maketitle


\section{Introduction}
\label{sec:introduction}
Imagine living in an intuitively interactive space without the need to understand its grammar. One doesn't need to interact in a specific way or use \textit{select} voice commands~\cite{amazon_echo} or always wear something~\cite{HexiWear}. Sharing this intelligent space with others does not also degrade the individual user experience. Interestingly, this vision of seamless smart spaces is not novel and quite dated~\cite{lam2018refining, clark2011seamless, bonanni2005counterintelligence}. However, we are yet to occupy this kind of space regularly. For this vision to become an everyday reality, we argue that there is a need for multi-user continuous room-scale activity tracking through passive sensing mechanisms. Hence, this paper attempts to create an activity-sensing system that can be used to make indoor living spaces truly intelligent.

\begin{figure}[!t]
    %Figure font: Times New Roman
    \centering
    \includegraphics[width=0.33\textwidth]{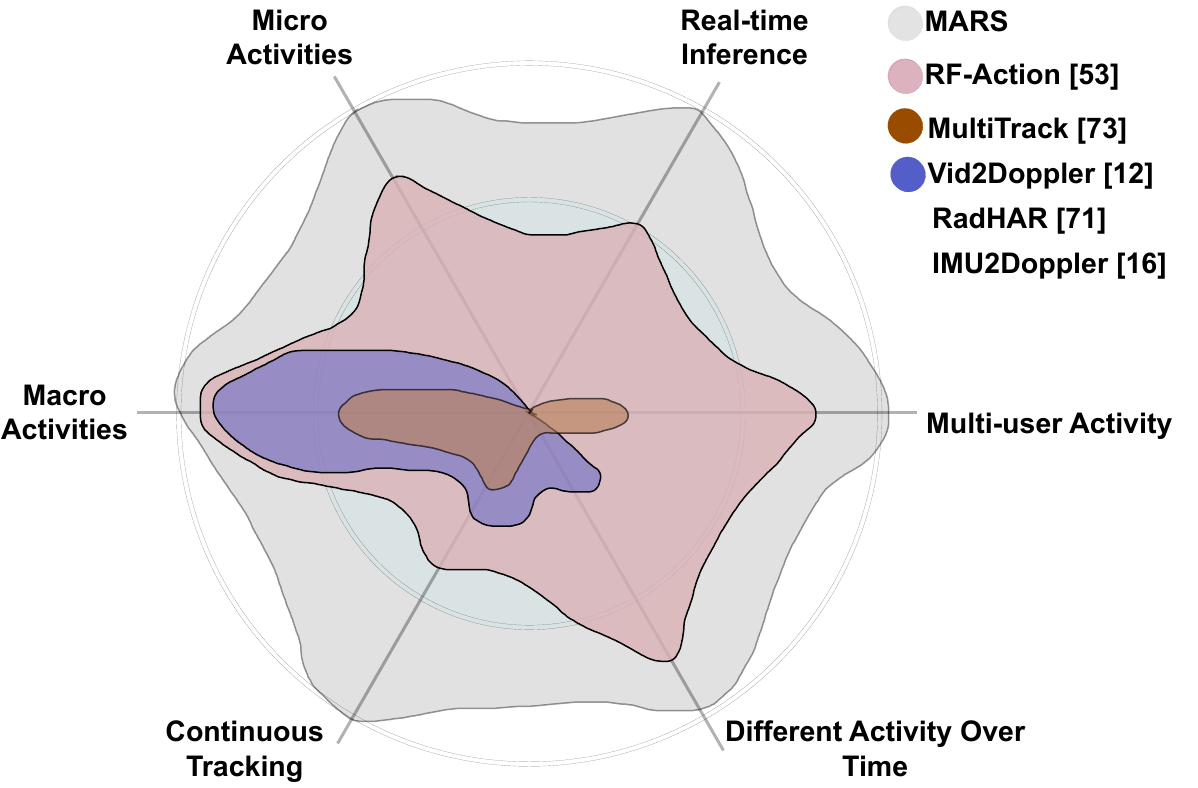}
    %The removed work, mmgaitnet uses two radars
    \caption{Our vision compared to other works}
    \label{fig:existingworks}
\end{figure}

However, it is crucial to establish what features make such a passive activity-sensing system desirable for wide-scale deployment. Learning from the decades of research in wireless sensing~\cite{adib2013see, adib2015smart, fan2020home, li2019making, zhao2018through}, we argue that: 1) continuous subject tracking, 2) monitoring multiple subjects, 3) monitoring different activity over time (for single subject), 4) real-time inference of activities, and 5) multi-activity support (both macro and micro-scale activities) are critical for such a system to be successful and widely adopted. \figurename~\ref{fig:existingworks} shows that the works~\cite{singh2019radhar, ahuja2021vid2doppler, bhalla2021imu2doppler, li2019making, tan2019multitrack, ding2020rf,palipana2021pantomime} closest to our vision primarily focus on a subset of the above objectives. For example,~\cite{ahuja2021vid2doppler, bhalla2021imu2doppler, singh2019radhar, ding2020rf,palipana2021pantomime} supports single-user macro activity tracking. \cite{li2019making} can track multiple users; however, it considers short-duration actions, such as hand-shaking, falling, etc., in addition to continuous activity. On the other hand,~\cite{tan2019multitrack} provides long-range localization and a few macro-activity tracking for multi-users but is unsuitable for continuous multi-activity monitoring.
% For example,~\cite{wang2021m} provides continuous monitoring but fails to give multi-user support; whereas ~\cite{singh2019radhar} can detect multiple macro-activities but is dependent on user orientation.
%\swadhincomment{[SP:Fig. 1 explanation with ref here]}
%This system definitely needs to be robust and adaptable to changes in the environment and users. Furthermore, this has to be embedded within the environment without compromising our privacy and security~\cite{liao2020security, aqeel2016security}. 

We examined different modalities for our envisioned system to achieve continuous passive wireless sensing, including Wi-Fi, UWB, acoustic, and mmWave. As our goal is to make the system self-contained, easy to maintain, and readily deployable, we intentionally ignore the multi-modal approaches~\cite{tesla_news,haciane2018comma,fan2020home} due to their added complexity and modalities where users need to carry or wear something~\cite{schall2018barriers,jacobs2019employee,liu2019beyond, wang2018modeling, fan2018tagfree,bocanegra2020rfgo}. Acoustic sensing is compelling because of its cheap hardware but suffers from range, resolution, and the impact it has on users~\cite{li2022experience}. Wi-Fi sensing has also been deployed extensively in the literature~\cite{adib2013see,adib2015smart,guo2017wifi,wang2014we, ali2015keystroke, cominelli2023exposing}. However, as discussed in~\cite{chen2021rf}, Wi-Fi has limited resolution when simultaneously separating activities at different scales (macro and micro) due to its narrow bandwidth. UWB and mmWave are the most compelling technologies that provide high bandwidth and hence, better resolution for sensing a wide range of activities~\cite{noori2021ultra,zhang2022mobi2sense,venon2022millimeter,singh2019radhar,jiang2018towards,palipana2021pantomime}. However, UWB overlaps with the Wi-Fi in the supported frequency range and thus may experience high interference from Wi-Fi deployments; therefore, we consider mmWave to realize the above vision. Recent exploration in the direction of identification~\cite{zhao2019mid}, position tracking~\cite{wei2015mtrack}, action recognition~\cite{singh2019radhar, jiang2018towards, palipana2021pantomime}, vital-signs~\cite{yang2016monitoring, chen2021movi, zhang2023pi}, speech sensing~\cite{liu2021wavoice, li2020vocalprint, wang2022mmeve}, etc. justifies the practicality of mmWave sensing for daily activity monitoring. However, we observe that current mmWave sensing literature does not address continuous human activity monitoring over a longer time or space in an indoor environment, thereby restricting it from being a pervasive practical solution. We summarize these gaps below.

\textbf{Gaps}: The majority of previous studies~\cite{singh2019radhar, gong2021mmpoint, yu2022noninvasive, ahuja2021vid2doppler, bhalla2021imu2doppler, palipana2021pantomime} have reported a high level of sensing accuracy as the subject is kept within the main lobe ($-15$\textdegree < radar lobe angle < $15$\textdegree) of the radar's field-of-view (FoV). In practice, however, the indoor movement of a subject can be completely random, and thus, activities cannot be detected when the subject is outside the radar's FoV. Incorporating multiple mmWave radars to track multiple users within the same room will significantly increase the complexity due to the complex interference patterns from multi-path signals over the same or overlapping frequency bands. Furthermore, there are countless activities a user can engage in, ranging from macro activities involving major body movements (like cleaning the room) to micro activities involving lesser body movements (like typing on the phone). Most of the previous works~\cite{singh2019radhar, palipana2021pantomime, ahuja2021vid2doppler} primarily consider macro activities for a single user, which are easy to detect due to the rich doppler patterns in the reflected mmWave signals. Nevertheless, in reality, a subject can perform both macro and micro activities over time, whereas different subjects can work on different things simultaneously. Also, tracking activities from multiple subjects is challenging as household objects and motion artifacts across subjects can cause noise from static and dynamic multi-path reflections.

% Furthermore, household objects cause multi-path reflections, increasing the noise level. In addition, dynamic multi-path is not well studied in mmWave due to motion artifacts across multiple users. Moreover, most previous works have only considered a small number of activities~\cite{singh2019radhar, wang2021m}. Conversely, in an indoor environment, there are countless activities a user can engage in, ranging from macro activities involving major body movements to micro activities involving lesser body movements.

\begin{figure}[t]
    \centering
    \includegraphics[width=0.90\columnwidth]{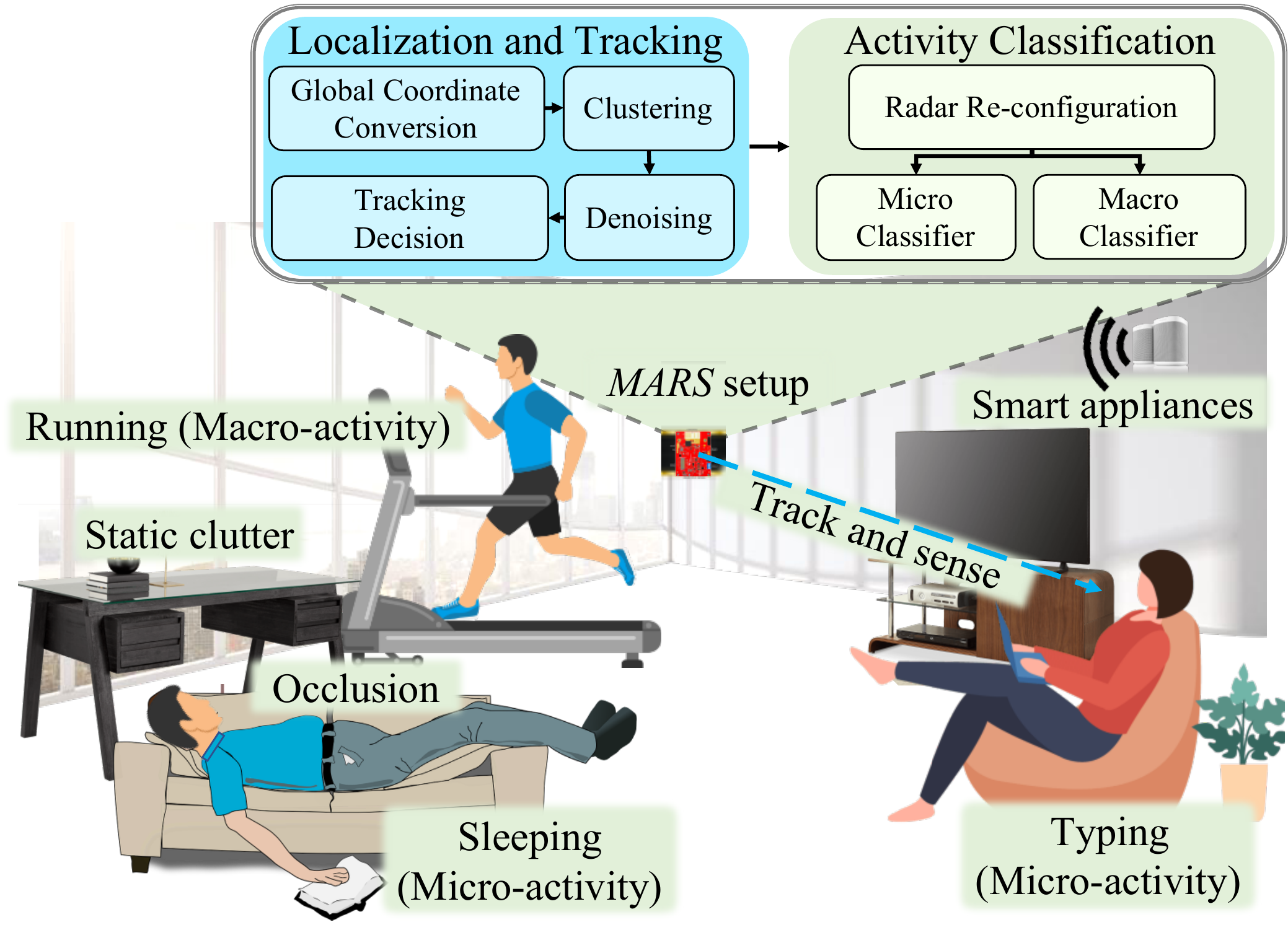}
    \caption{Overview of \ourmethod{}}
    %\vspace{-0.9cm}
    \label{fig:radar_overview}
\end{figure}

Motivated by these gaps and empowered by our vision, in this work, we first divide the activity grammar into two subsets -- (i) \textit{Macro activities} that involve significant body movements (like \textit{changing clothes}) and (ii) \textit{Micro activities} that need minor movements of body parts (like \textit{typing}). Next, to track users' activities seamlessly, we divide the problem statement into two parts (\figurename~\ref{fig:radar_overview}): (i) \textit{localization and tracking} in such a way that multiple users' positions can be tracked in every scenario with a single mmWave radar and (ii) continuous opportunistic \textit{activity monitoring} to distinguish both macro and micro activities. The primary challenges involved in the multi-user localization with a single mmWave radar are: (i) Scenarios when the subject is present inside the room but not within the FoV of the mmWave sensor, (ii) Creation of \textit{zombie} subjects due to multi-path reflections, (iii) Associating subjects based on their \ac{RF} reflections in scenarios when the users cross each other, and (iv) Blind-spots during multi-user tracking due to occlusions by other subjects. Additionally, for continuous activity monitoring across multiple subjects, detecting both macro and micro-scale activities simultaneously with the same mmWave radar configuration is not feasible. For example, the radar with a high-doppler resolution can capture better micro movements but adds more noise in capturing macro activities. In contrast, low-doppler resolution can capture macro movements but fails to detect micro activities.

%Then we explore the underlying issues in localizing multiple users and sensing their activity in an indoor environment, as shown in \figurename~\ref{fig:radar_overview}. 

%\vspace{-0.8cm}

%Users' activities can have macro-scale and micro-scale body movements. Detecting these activities under the same mmWave configuration is not feasible. For example, higher velocity resolution can capture better micro movements but adds more noise in capturing macro activities; similarly, low-velocity resolution can capture macro movements but fails to capture micro activities. (b) Scenarios when the user is not oriented towards the radar, (c) Continuous users' activity monitoring and capturing the activity switching over time is challenging as it needs capabilities to sense both macro and micro-movements.}

\textbf{Contributions}: To mitigate these challenges, in this work, we propose \textbf{\ourmethod{}}, a mmWave-based sensing system:~\textbf{M}ulti-user \textbf{A}ctivity tracking via \textbf{R}oom-scale \textbf{S}ensing.
%To mitigate the above-mentioned challenges, in this work, we propose \textbf{\ourmethod{}}, an \textbf{mm}Wave based sensing system to \textbf{L}ocalize and \textbf{T}rack human activity in an \textbf{E}nvironment \textbf{I}ndependent \textbf{O}pportunistic manner. 
In summary, we contribute in the following ways: 
\begin{enumerate}[leftmargin=*]
    \item We build an end-to-end prototype for continuous multi-user activity monitoring using a single mmWave \ac{FMCW} radar using a novel technique that rotates and scans full 360{\textdegree} opportunistically. The approach develops methods for dealing with zombies, static clutters, and blind spots utilizing a single rotating radar, thereby avoiding the complex interference patterns that arise from multiple radars.
    \item \ourmethod{} employs a novel method of differentiated stacking of the captured range-doppler frames as well as opportunistic switching of radar configurations in order to detect macro and micro activities simultaneously. By doing so, to the best of our knowledge, we design a system that can monitor the \textit{highest} number of human activities in the mmWave domain ($1.6\times$ nearest baseline Vid2Doppler~\cite{ahuja2021vid2doppler}). In contrast to the existing works, \ourmethod{} can run on an edge device for real-time monitoring of activities performed by multiple subjects within a room.
    \item We performed a thorough evaluation of \ourmethod{} at diverse setups and have shown its superiority compared to several other baselines. In classifying the macro and micro activities, we can achieve a weighted F1-Score of $98\%$ and $94\%$, respectively, with an average response time of $\approx 2$s. We open-source our implementation and sample dataset to reproduce our results: \url{https://anonymous.4open.science/r/MARS/}.
\end{enumerate}

\begin{figure*}[t]
    \def\wide{0.19}
    \subfloat[Clapping]{
    \includegraphics[trim={22mm 0 0 7mm}, width=\wide\columnwidth]{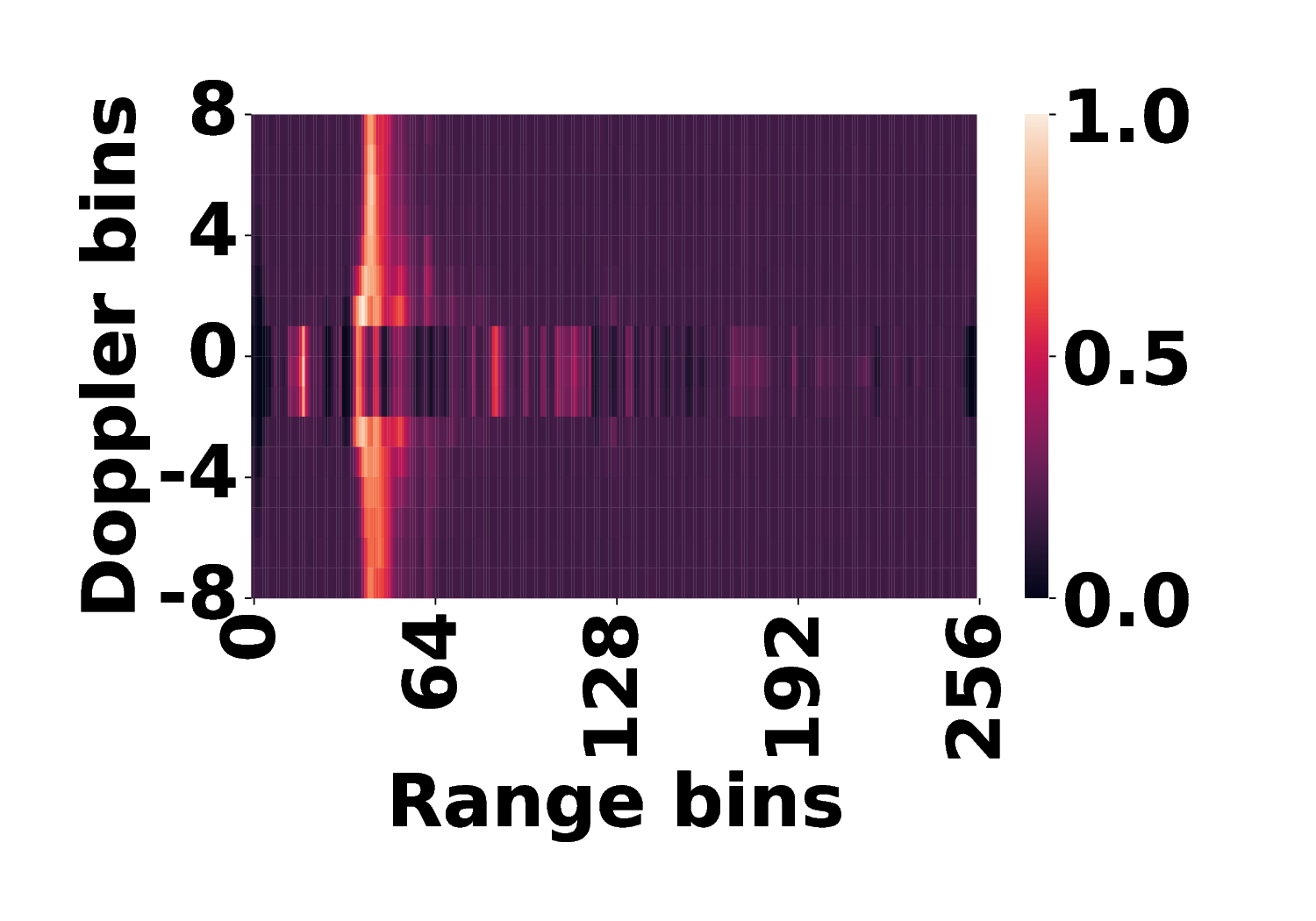}
    }
        \subfloat[Jumping]{
    \includegraphics[trim={22mm 0 0 7mm},width=\wide\columnwidth]{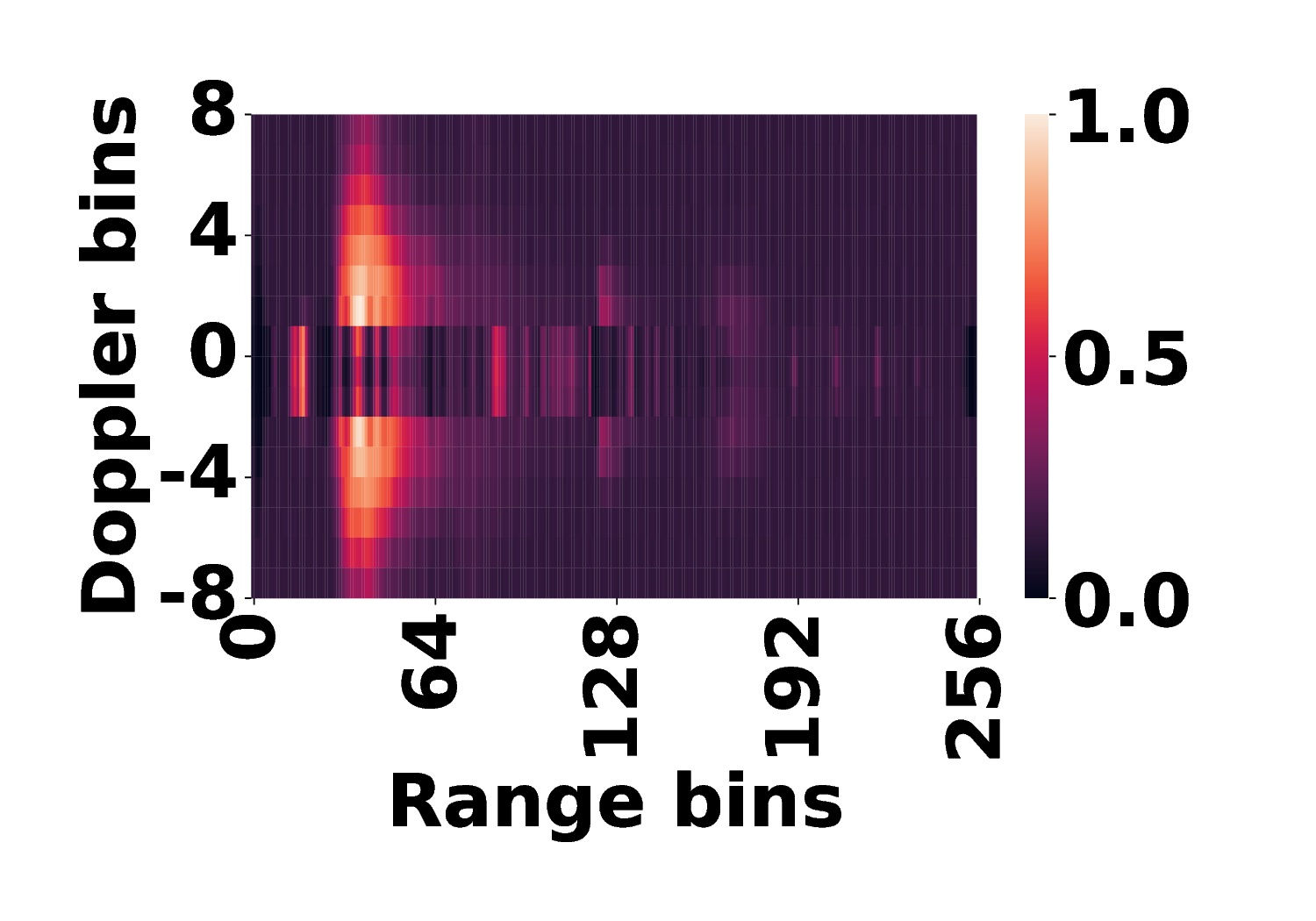}
    }
        \subfloat[Lunges]{
    \includegraphics[trim={22mm 0 0 7mm},width=\wide\columnwidth]{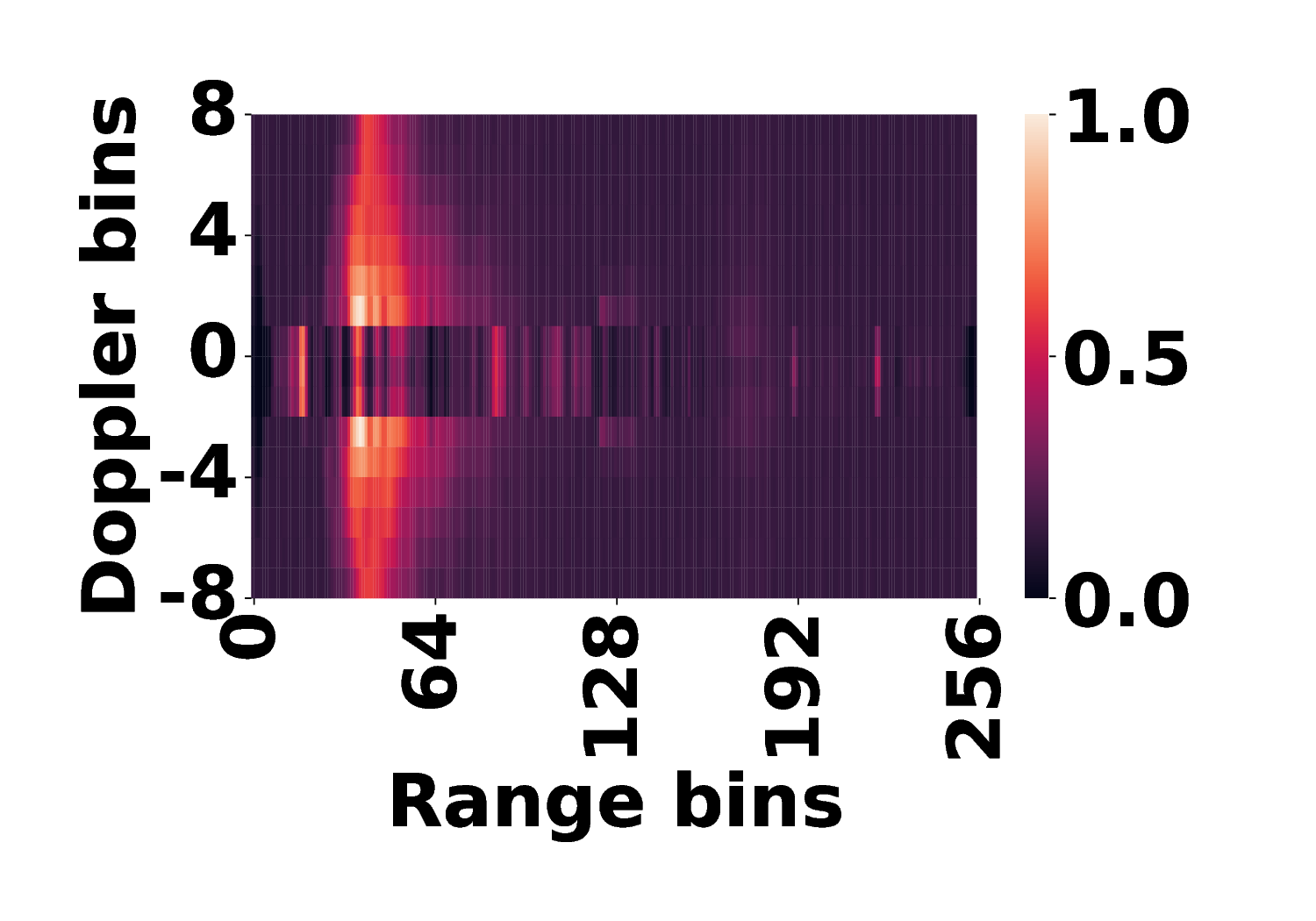}
    }
        \subfloat[Walking]{
    \includegraphics[trim={22mm 0 0 7mm},width=\wide\columnwidth]{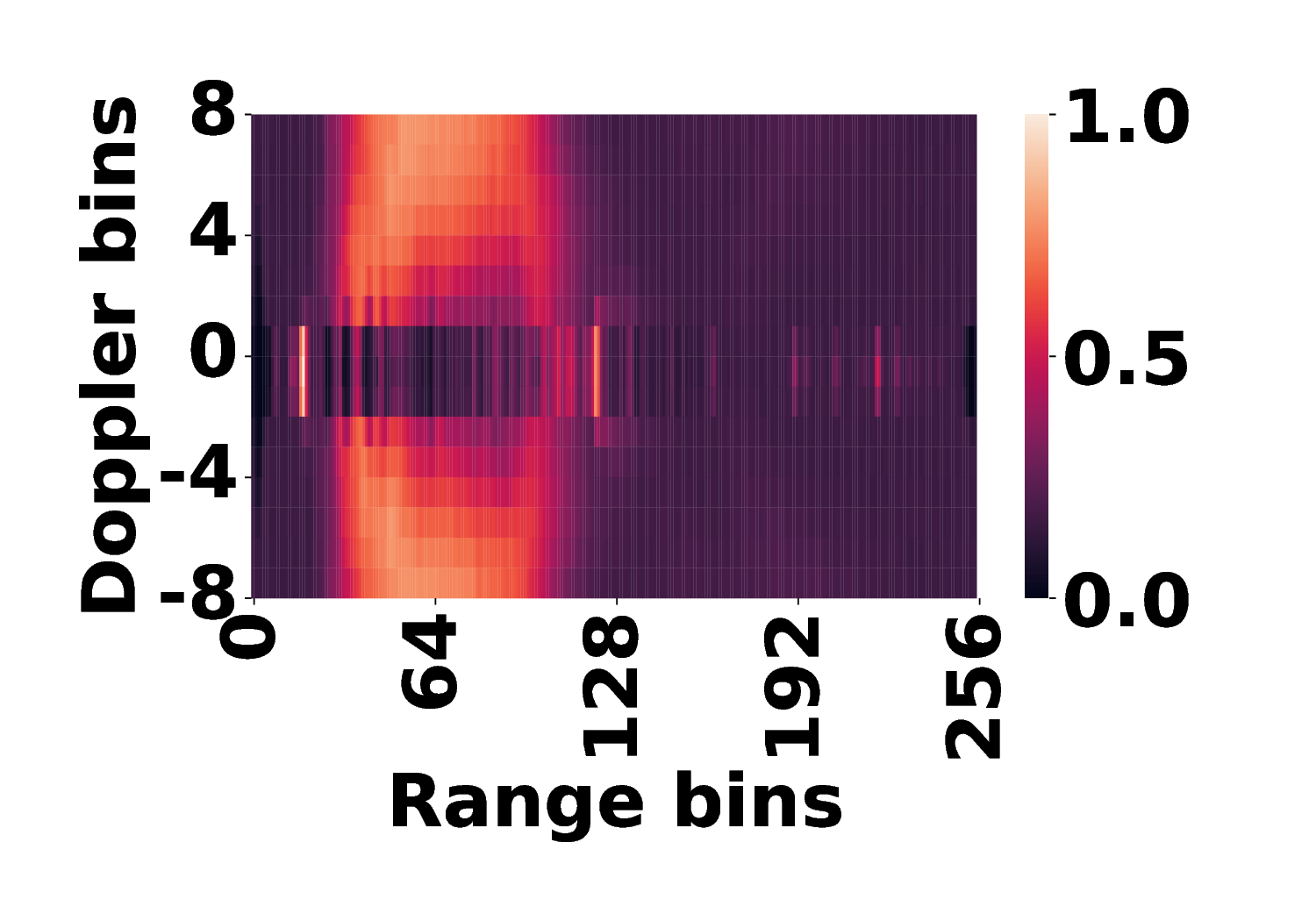}
    }
        \subfloat[Squats]{
    \includegraphics[trim={22mm 0 0 7mm},width=\wide\columnwidth]{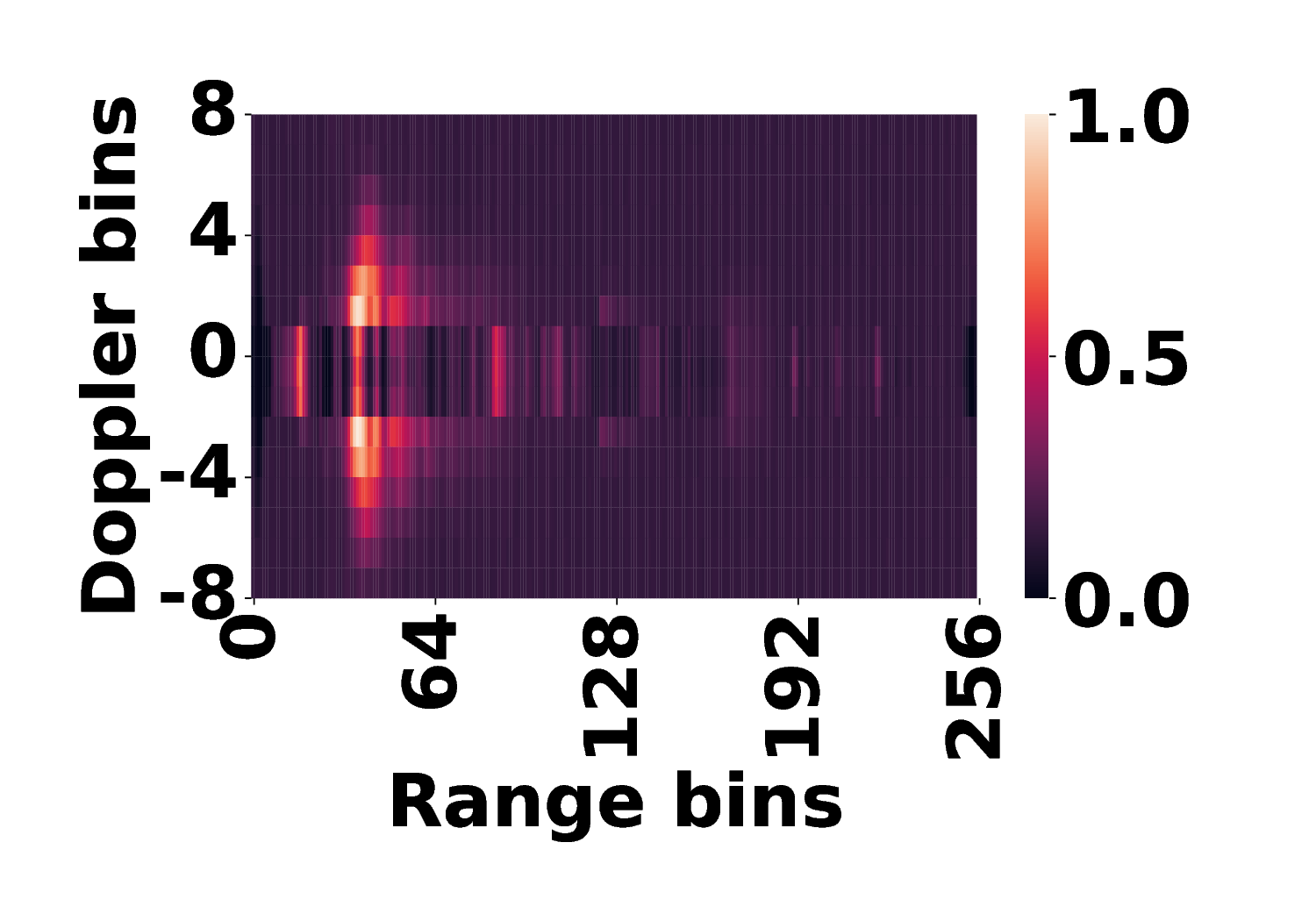}
    }
        \subfloat[Waving]{
    \includegraphics[trim={22mm 0 0 7mm},width=\wide\columnwidth]{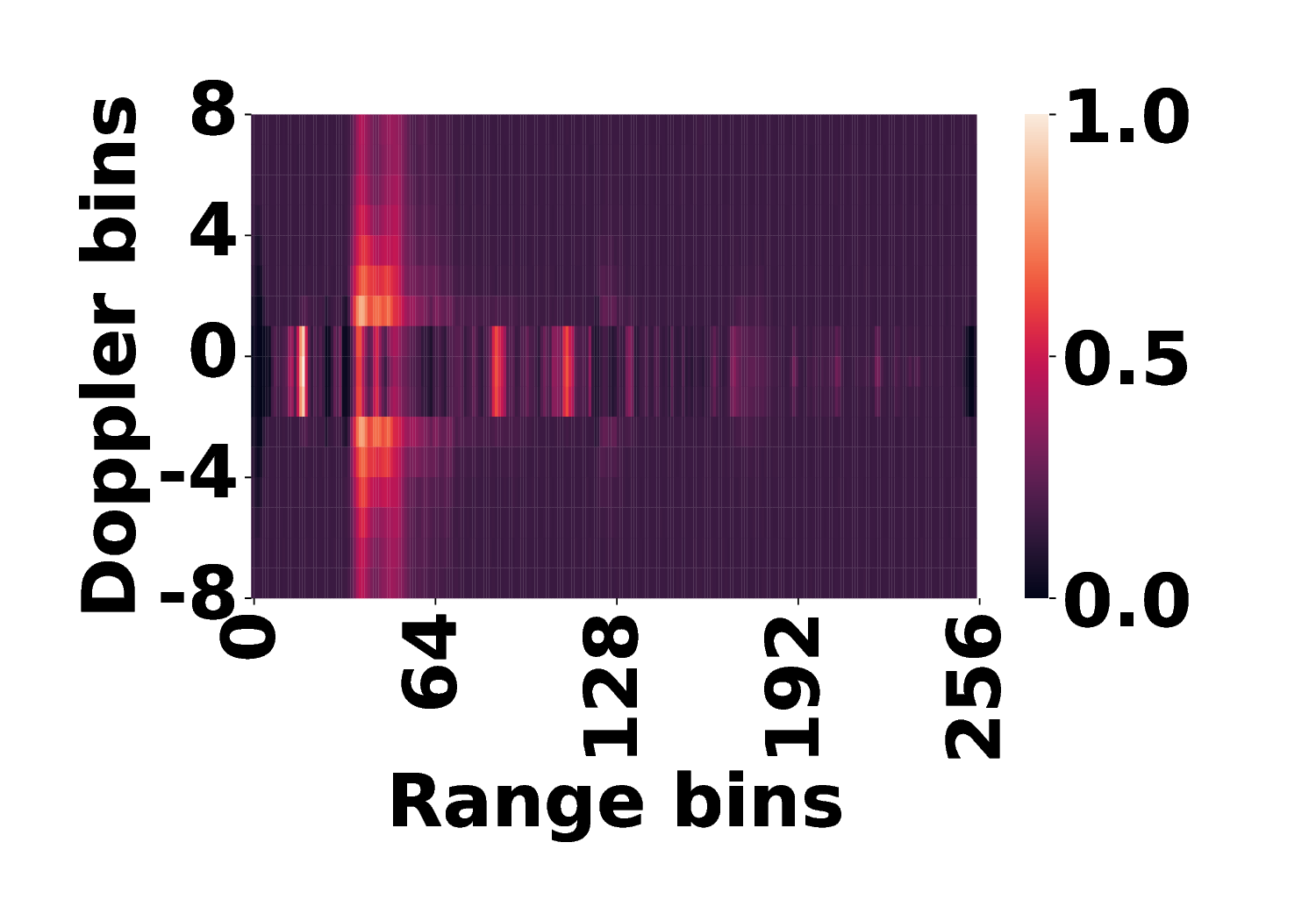}
    }
        \subfloat[Fold \\Cloth]{
    \includegraphics[trim = {22mm 0 0 0mm}, width=\wide\columnwidth]{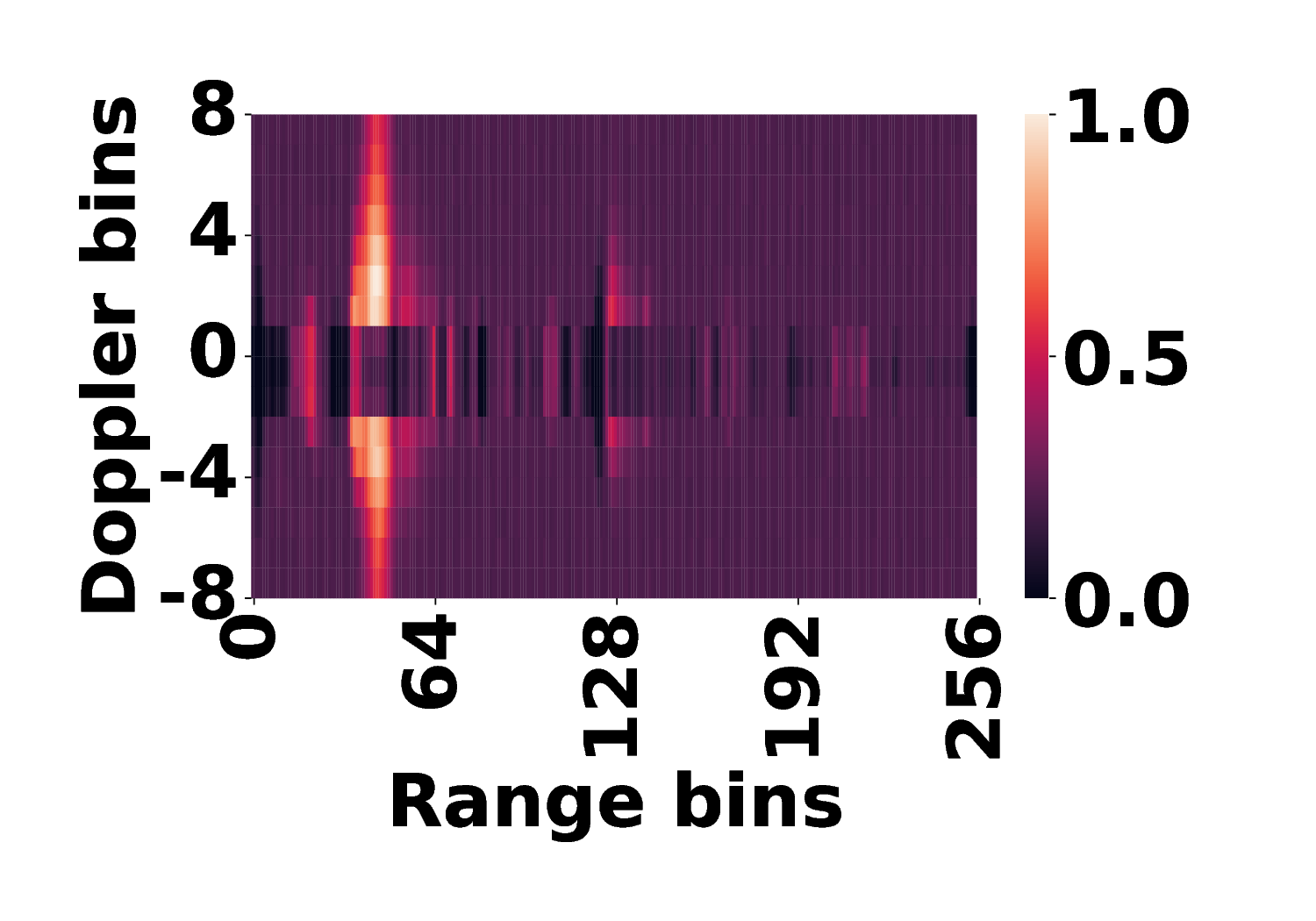}
    }
    \subfloat[Change \\Cloth]{
    \includegraphics[trim = {22mm 0 0 7mm}, width=\wide\columnwidth]{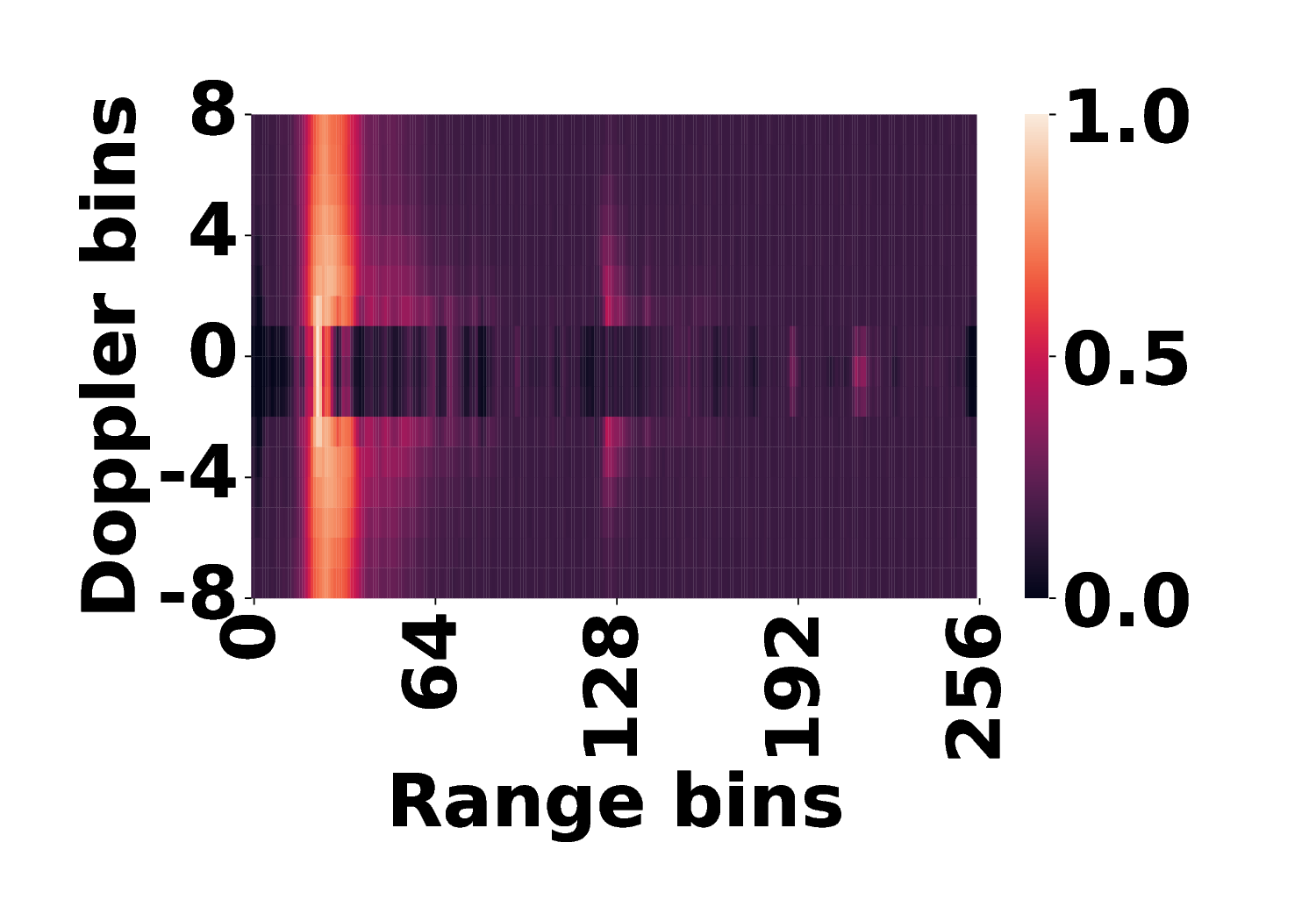}
    }
    \subfloat[Vacuum \\Clean]{
    \includegraphics[trim = {22mm 0 0 7mm}, width=\wide\columnwidth]{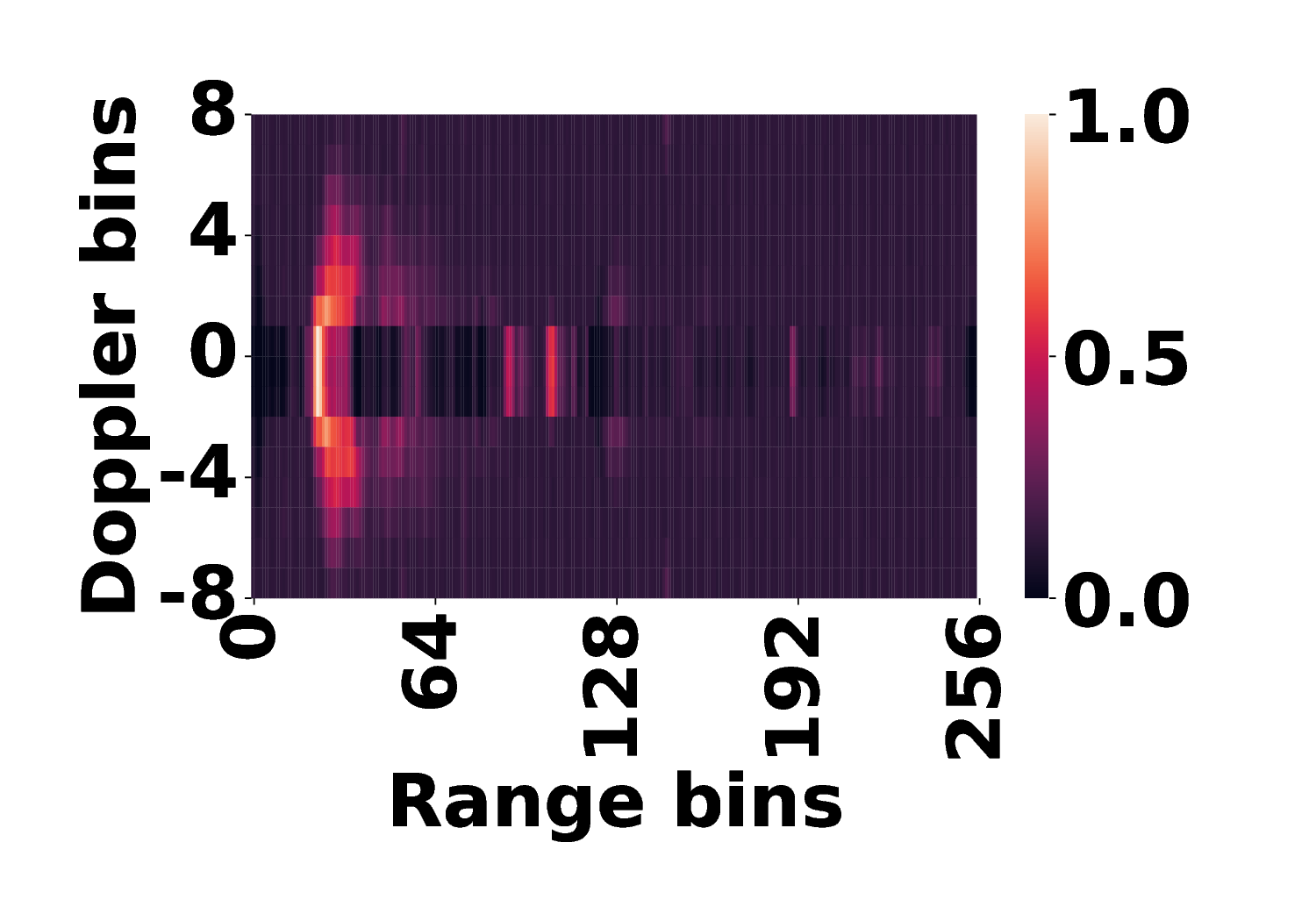}}
    \subfloat[Running]{
    \includegraphics[trim = {22mm 0 0mm 7mm}, clip, width=\wide\columnwidth]{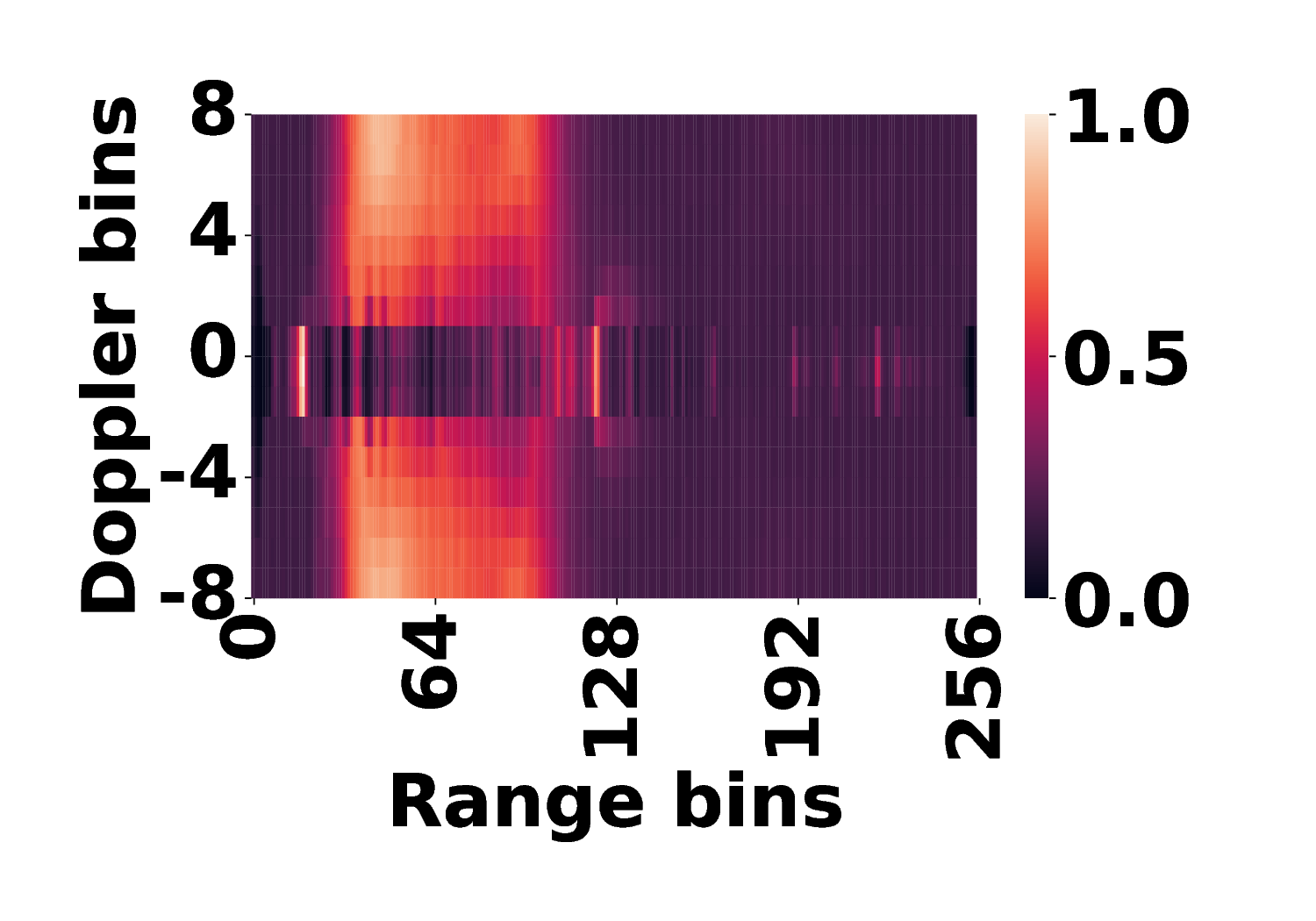}
    }\\ \vspace{-3mm}
        \subfloat[Phone \\Type]{
    \includegraphics[trim={22mm 0 0 7mm},width=\wide\columnwidth]{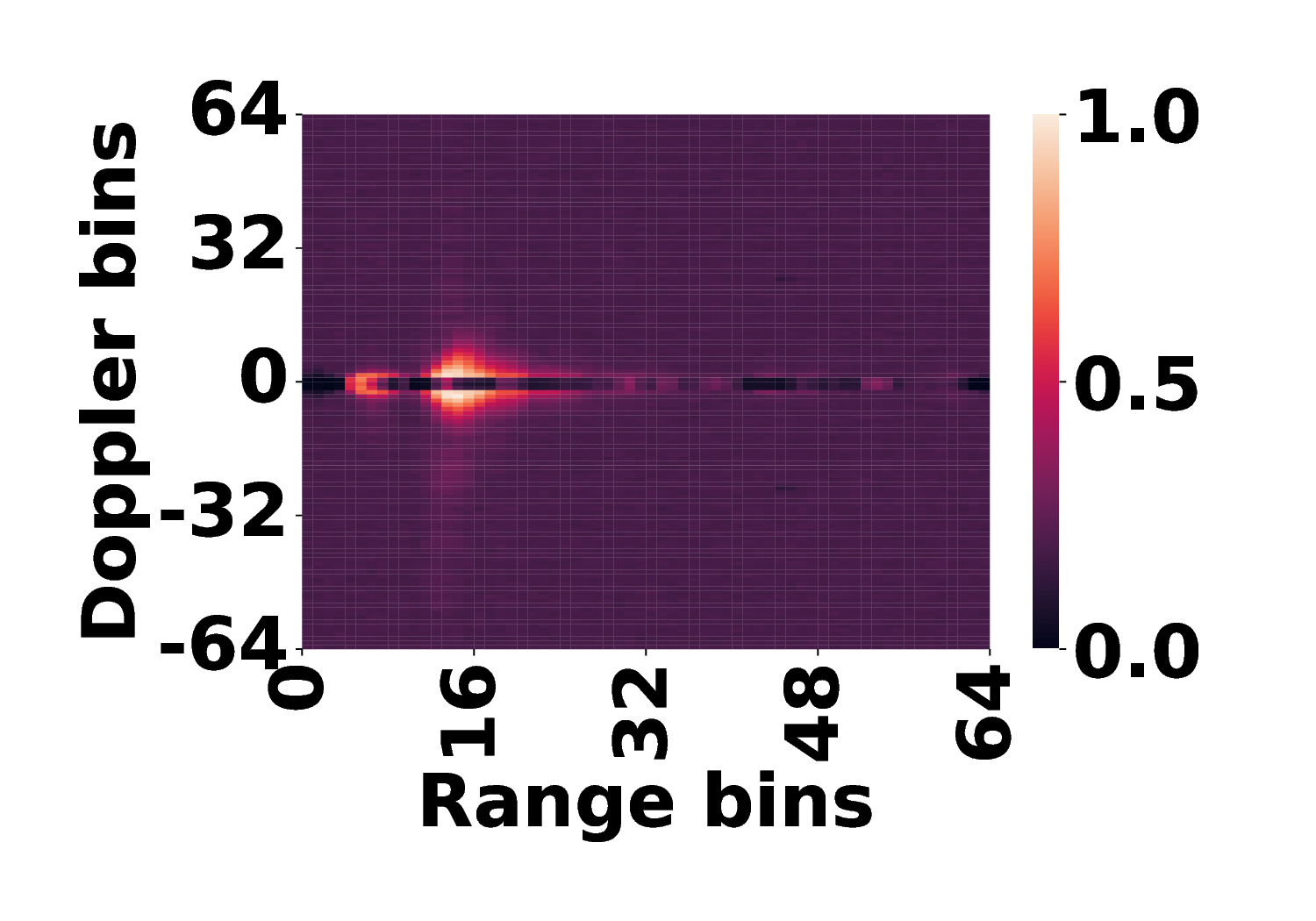}
    }
        \subfloat[Laptop \\Type]{
    \includegraphics[trim={22mm 0 0 7mm},width=\wide\columnwidth]{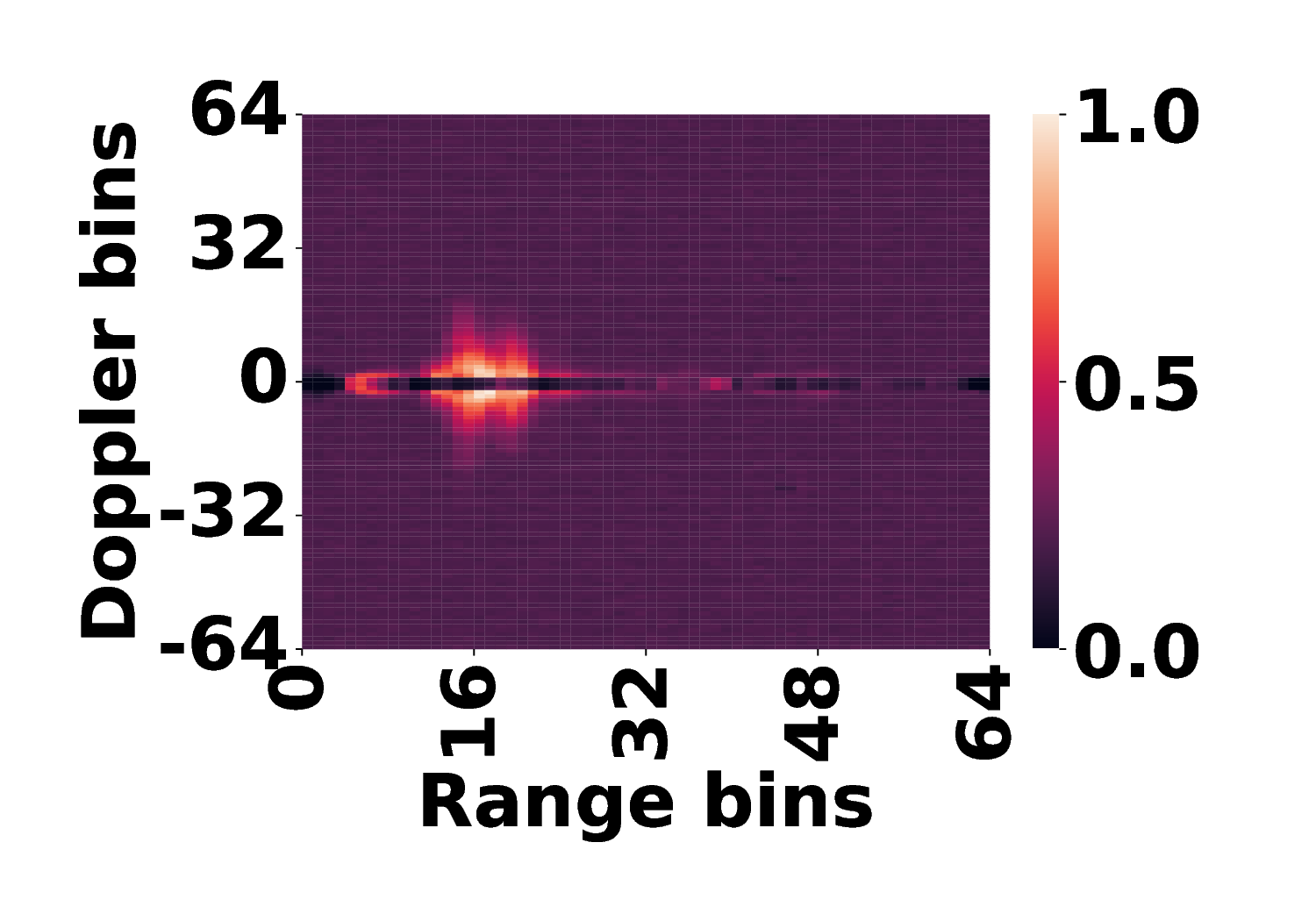}
    }
        \subfloat[Sitting]{
    \includegraphics[trim={22mm 0 0 7mm},width=\wide\columnwidth]{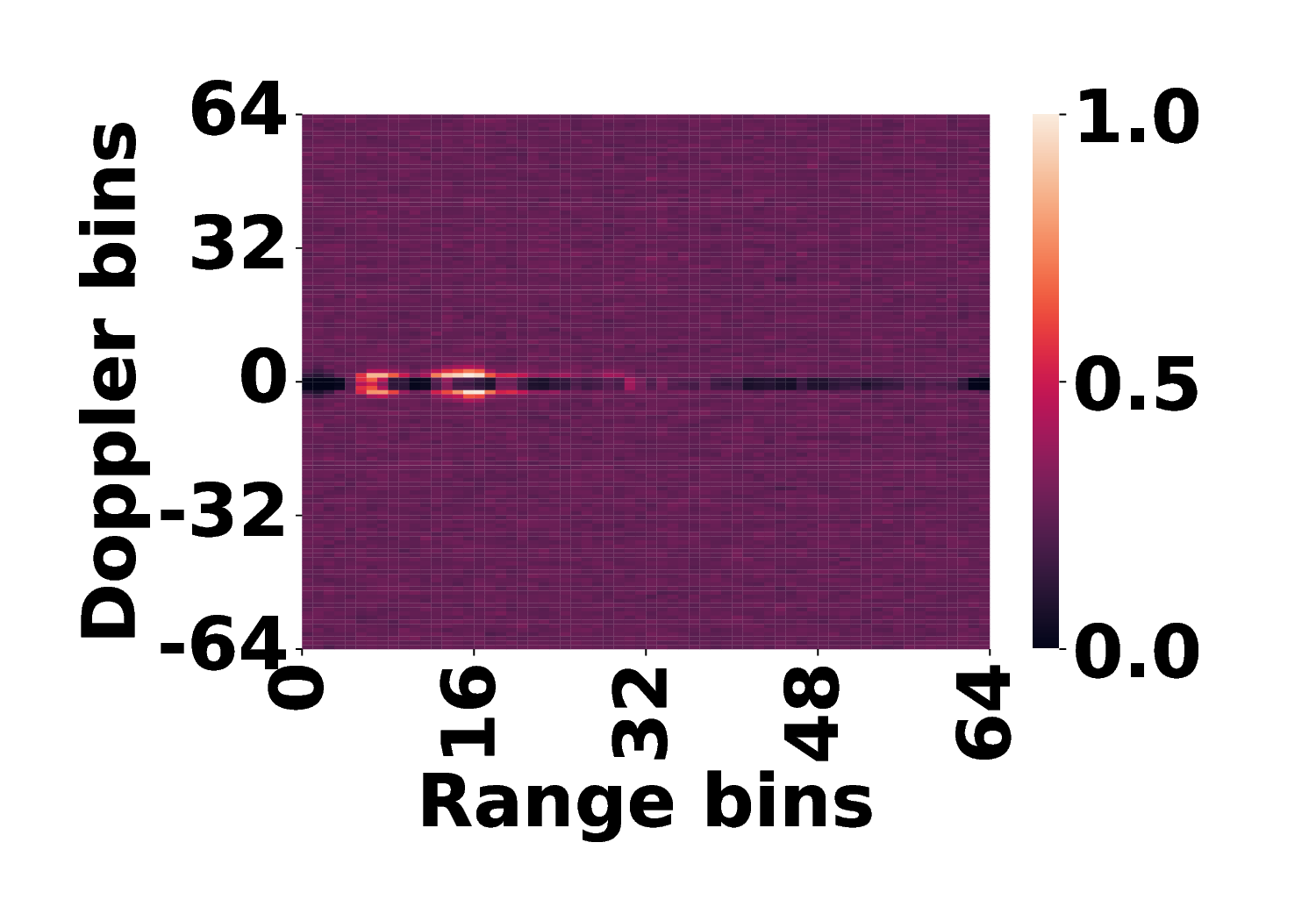}
    }
        \subfloat[Eating]{
    \includegraphics[trim={22mm 0 0 7mm}, clip, width=\wide\columnwidth]{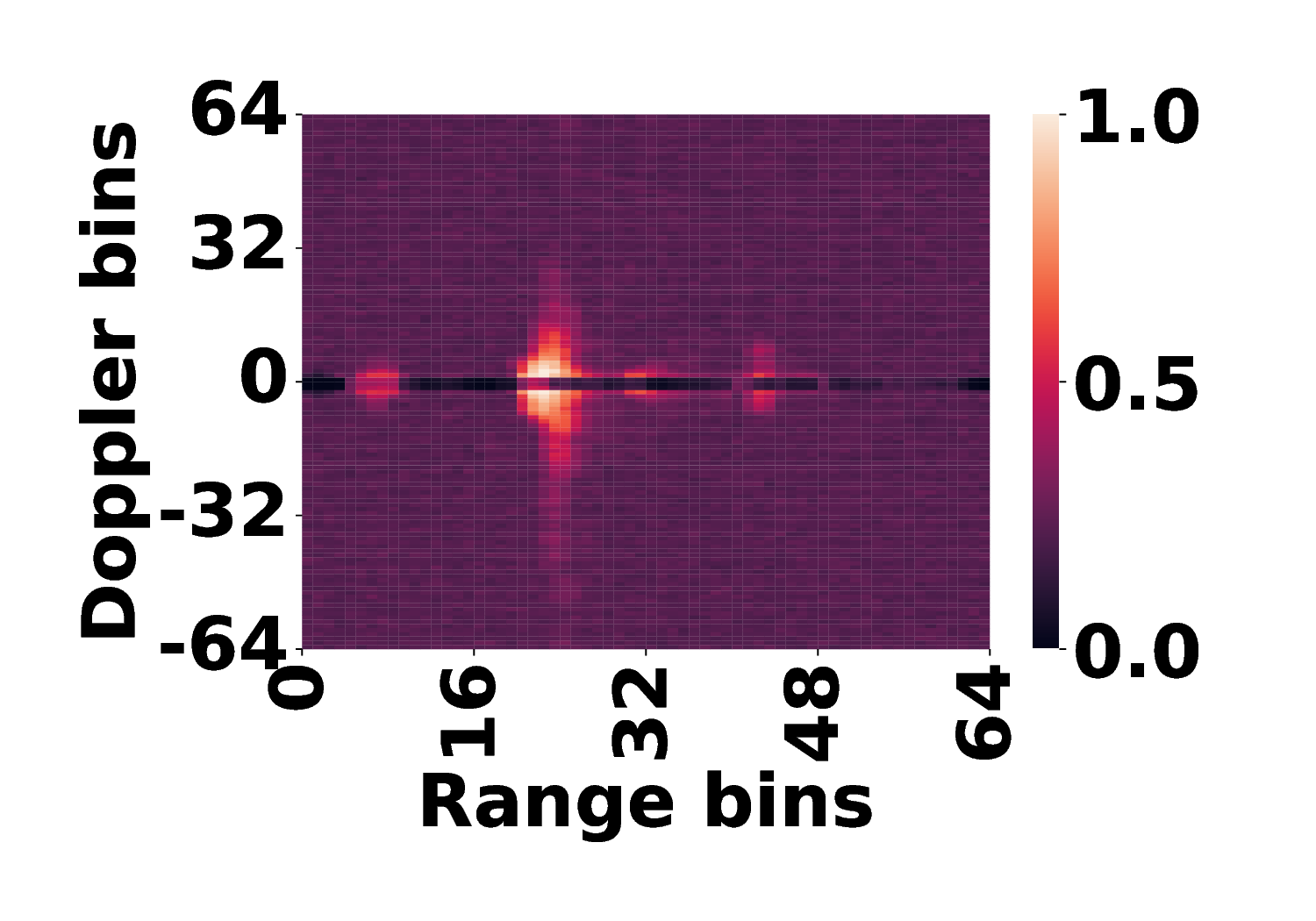}
    }
    \subfloat[Phone \\Talk]{
    \includegraphics[trim={22mm 0 0 7mm},width=\wide\columnwidth]{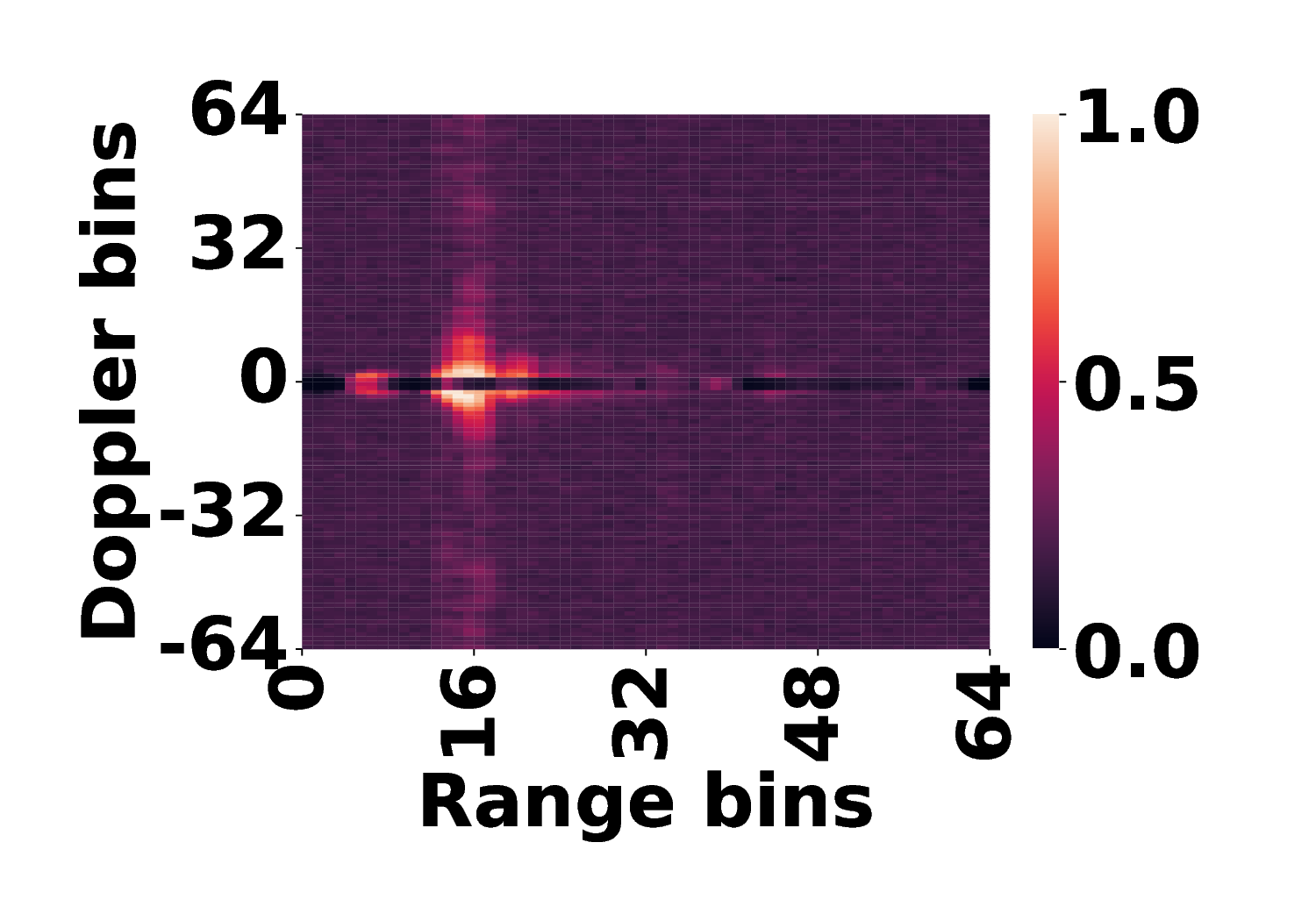}
    }   
    \subfloat[Play \\Guitar]{
    \includegraphics[trim = {22mm 0 0 7mm}, width=\wide\columnwidth]{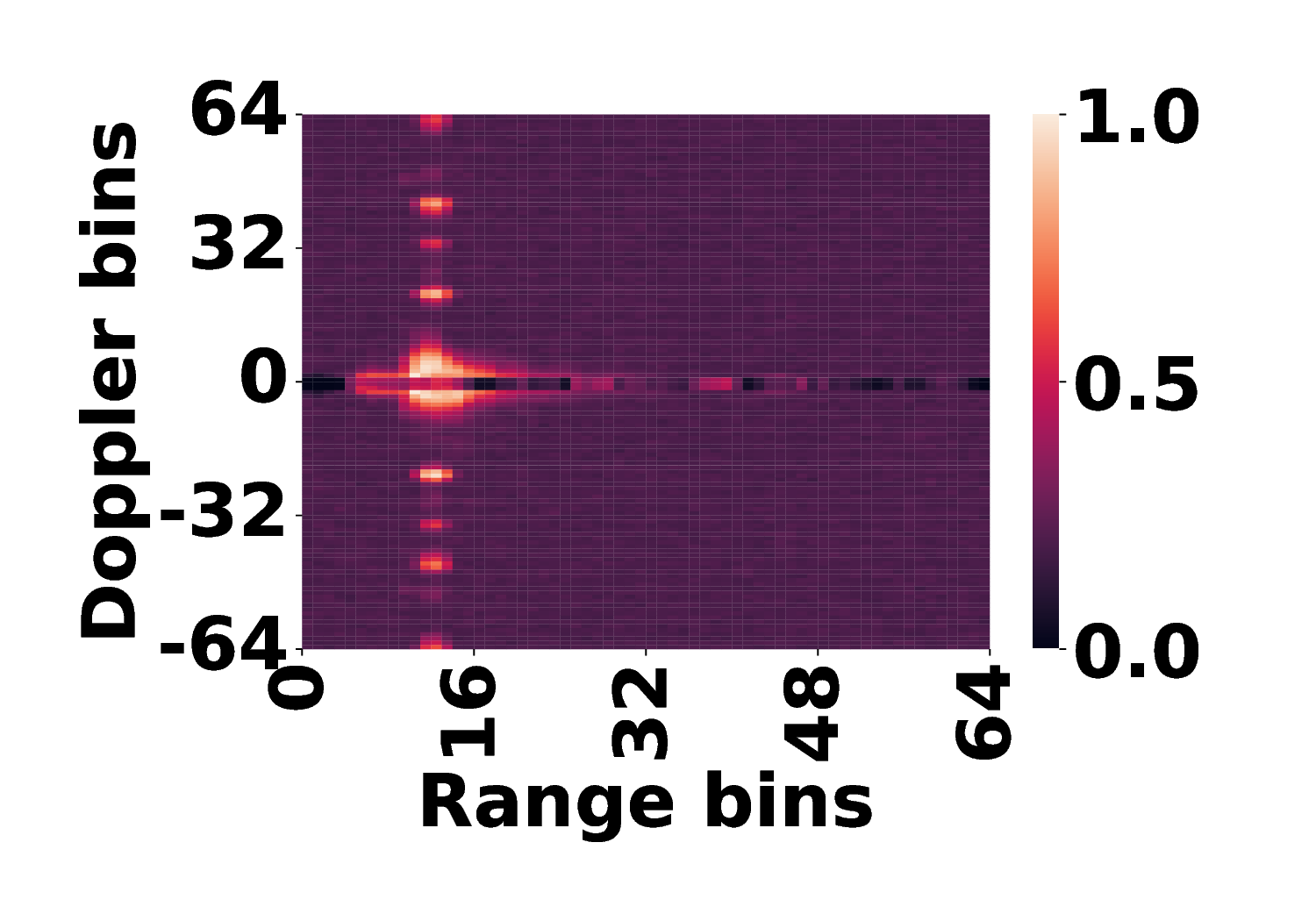}
    }
    \subfloat[Brushing]{
    \includegraphics[trim = {22mm 0 0 7mm}, clip, width=\wide\columnwidth]{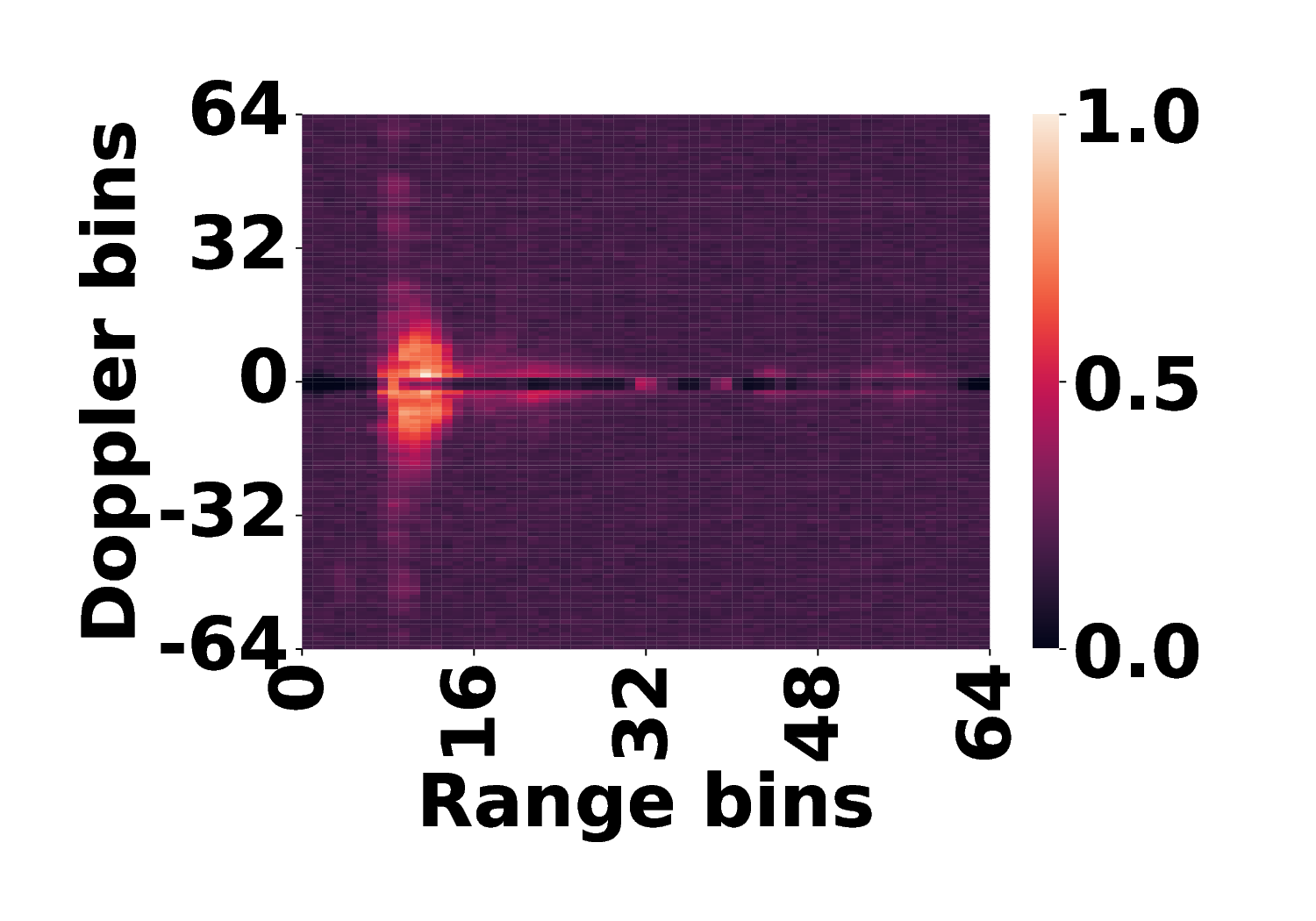}
    }
    \subfloat[Combing]{
    \includegraphics[trim = {22mm 0 0 7mm}, clip, width=\wide\columnwidth]{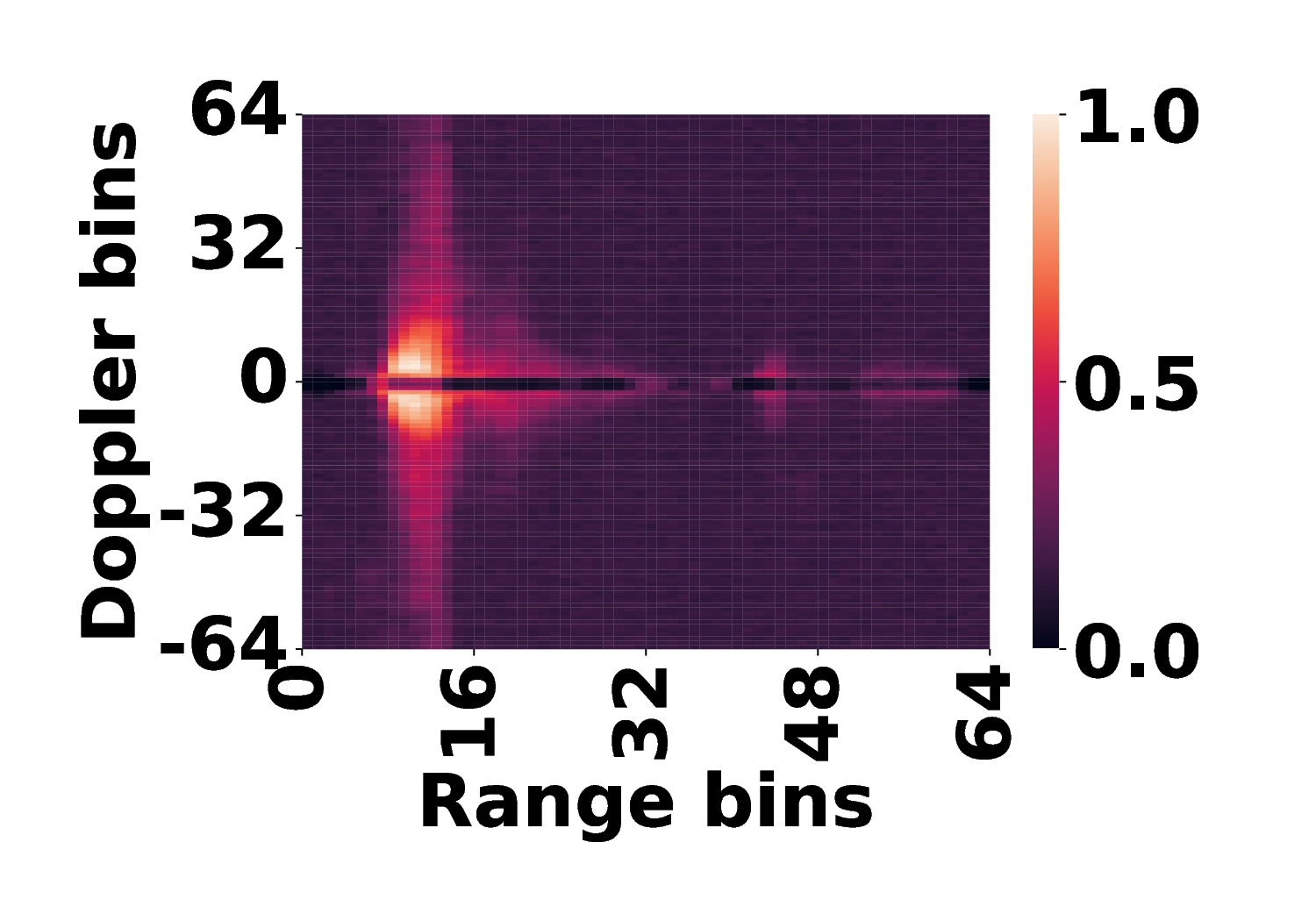}
    }
    \subfloat[Drinking]{
    \includegraphics[trim = {22mm 0 0 7mm}, width=\wide\columnwidth]{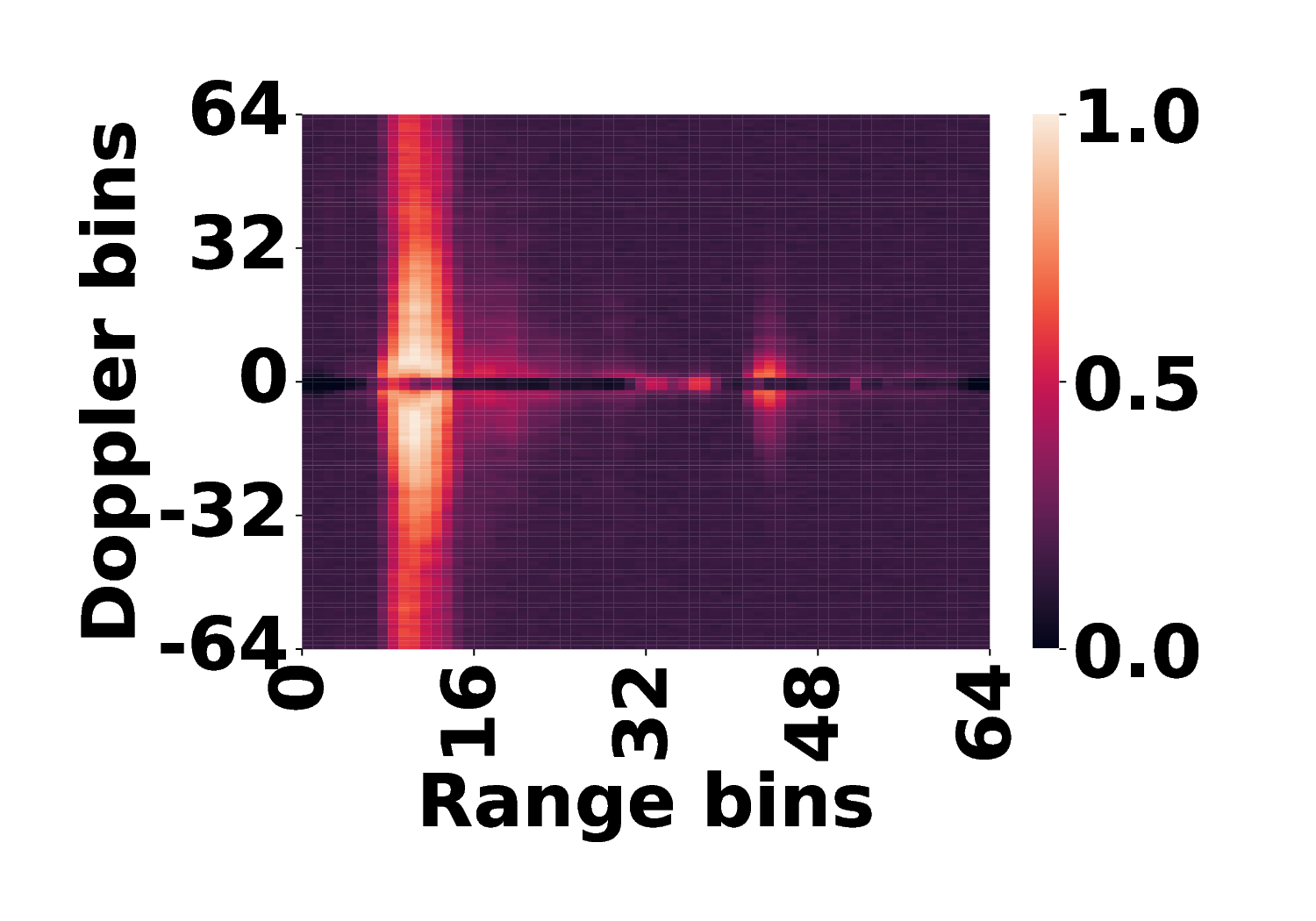}
    }
 \caption{Standard deviation (std) in the range-doppler heatmaps captured during the entire activity duration. (a)-(j): Macro activities with low doppler resolution, (k)-(s): Micro activities with high doppler resolution. Activities having similar body movements have similar patterns, but the difference can be captured in the temporal domain.}
\label{fig:dopplerpatterns}
 \end{figure*}
%\input{tex/Sec_2_System_overview.tex}
%\color{blue}
\section{Preliminaries and Pilot Study}
%In this section, we empirically illustrate the key foundational underpinnings of \ourmethod through pilot studies.
\subsection{Preliminaries} \label{sec:prelims}
The primary working principle of \ac{COTS} mmWave radars is centered on \ac{FMCW}~\cite{rao2020introduction} that transmits continuous frequency chirps and performs \textit{dechirp} operation by combining the transmitted signal (TX) with the signal reflected (RX) from objects to create an \textit{Intermediate Frequency} (IF) signal. From this IF signal, we extract (1) \textit{Pointcloud}, a discrete set of points representing the detected objects~\cite{schumann2018semantic} and (2) \textit{Range}, the distance of the detected objects from the radar. 
%We now introduce the signal processing involved in extracting the above information.
\subsubsection{Range estimation}\label{subsec:range}
 The distance information between the object and the radar can be obtained by measuring the frequency difference between the reflected and transmitted signals~\cite{iovescu2020fundamentals}. This frequency gap, also known as \textit{beat frequency }($f_b$), arises after a Round Trip Time (RTT) of, say $\tau$. If $T_C$ is the transmit time of the mmWave chirp across a bandwidth of $B$, then the slope of the FMCW chirp can be given as $S = \frac{B}{T_C} = \frac{f_b}{\tau}$.
% \begin{figure}
%     \centering
%     \includegraphics[width=0.25\textwidth]{figures/fmcw_chirps.pdf}
%     \caption{ Working principle of an FMCW Radar} 
%     \label{fig:mmwavefunction}
% \end{figure}
%\vspace{-0.1in}
The RTT delay, $\tau$, can be specified as $\tau = \frac{2d}{c}$ where $d$ is the \textit{distance of the detected object} and $c$ is the \textit{speed of light}. Thus, the detected object's distance can be given as, $d = \frac{c}{2} . \frac{T_C}{B} . f_b$. To calculate $f_b$, a Fast Fourier Transform (FFT), called \textit{range-FFT}, is performed on the IF signal, which produces frequency peaks at locations where the reflecting object is present. Locating these frequency peaks in turn estimates the range.
% \begin{figure}
%     \centering
%     \includegraphics[width=0.5\textwidth]{figures/pointcloud_estimation.drawio.pdf}
%     \caption{Point cloud Estimation}
%     \label{fig:pointcloud}
% \end{figure}
\subsubsection{Velocity estimation}\label{sec:velocity}
To measure the velocity of a moving target, the radar transmits $N$ number of chirps separated by a transmission time of $T_C$.  If a subject moves with a speed of $v$, the phase difference between two successive RX chirps corresponding to the motion, $vT_C$, can be given as, $\Delta \phi = \frac{4\pi v T_C}{\lambda}$. A second FFT, called \textit{doppler-FFT}, is performed on these phasors to determine the movement or velocity of the object. This information is captured in a 2D matrix called \textit{range-doppler} $\mathbb{D}_{D \times R}$ where $D$ and $R$ correspond to the numbers of \textit{doppler bins} and \textit{range bins}, respectively. 
% Equation \eqref{eq:range_doppler} shows the range-doppler for $R$ number of \textit{range bins} and $D$ number of \textit{doppler bins}.  On applying range-FFT, the objects present at different locations are captured.
% \begin{equation}
%     \label{eq:range_doppler}
%      \mathbb{D} =  \begin{bmatrix}
%  \mathbf{P}_{1,1}& \mathbf{P}_{1,2} & \cdots & \mathbf{P}_{1,R}\\ 
%  \mathbf{P}_{2,1}& \mathbf{P}_{2,2}& \cdots & \mathbf{P}_{2,R} \\ 
%  \vdots & \vdots & \ddots & \vdots\\ 
%  \mathbf{P}_{D,1} & \mathbf{P}_{D,2} & \cdots & \mathbf{P}_{D,R} 
% \end{bmatrix}
% \end{equation}
%For each object, an FFT of the angle is performed across multiple doppler-FFTs. 
\subsubsection{Pointcloud estimation}\label{sec:pointcloud}
The pointcloud is estimated through the standard CFAR algorithm~\cite{nitzberg1972constant} that detects peaks of the range-doppler matrix corresponding to the detected objects. The pointcloud consists of the coordinates $(x_i, y_i, z_i)$, doppler variation $(d_i)$, and the received power $(p_i)$ of the detected objects. The pointcloud set $(S)$ for $N$ number of detected objects can be given as $S = \bigcup_{i=1}^N \{\left ( x_i, y_i, z_i, d_i, p_i \right )\}$.

\subsection{Pilot Study}\label{sec:pilot_study}
We consider $19$ different activity classes from \textit{Activities of Daily Living} (ADLs), \textit{Instrumental Activities of Daily Living} (IADLs)~\cite{adlsiadls}, and \textit{daily indoor exercises} -- (i) \textit{macro activities} like walking, running, jumping, clapping, lunges, squats, waving, vacuum cleaning, folding clothes, changing clothes, and (ii) \textit{micro activities} like laptop-typing, phone-talking, phone-typing, sitting, playing guitar, eating food, combing hair, brushing teeth, and drinking water. In contrast to the existing literature that primarily uses voxelized pointcloud~\cite{singh2019radhar, wang2021m, palipana2021pantomime, cai2023millipcd} or 1D doppler~\cite{ahuja2021vid2doppler, bhalla2021imu2doppler}, in this paper, we explore range-doppler 2D heatmaps for activity classification; the primary motive is to find a parameter that can detect both macro and micro activities simultaneously from different users. For this purpose, we conduct a set of \textit{pilot experiments} to explore to what extent range-doppler information can be used in capturing human activity signatures and how the indoor setting impact such sensing capability.  

%? 2) What is the impact of static clutters on captured signatures? 3) How relevant is the Non-Line of Sight (NLoS) reflection of the transmitted radar waves? And finally, 4) How do the radar configuration parameters affect the sensing capability? 

 %************ Fig: Merged ************
\begin{figure*}[!t]
    \begin{minipage}{0.32\textwidth}
        \centering
  \setlength\tabcolsep{4pt}
  \def\arraystretch{0.5}
  \def\wide{0.45}
 \begin{tabular}{ll}
\raisebox{-\totalheight}{\includegraphics[width=\wide\textwidth, trim={2.5cm 0 1.5cm 0}, clip]{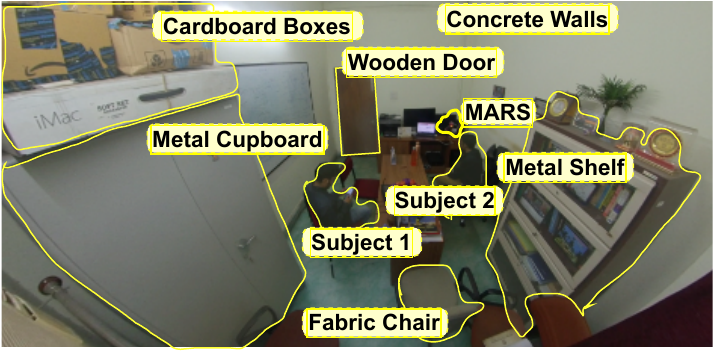}}\label{fig:static_scenario}

& 
\raisebox{-\totalheight}{\includegraphics[width=\wide\textwidth, trim={3cm 0 0.5cm 0}, clip]{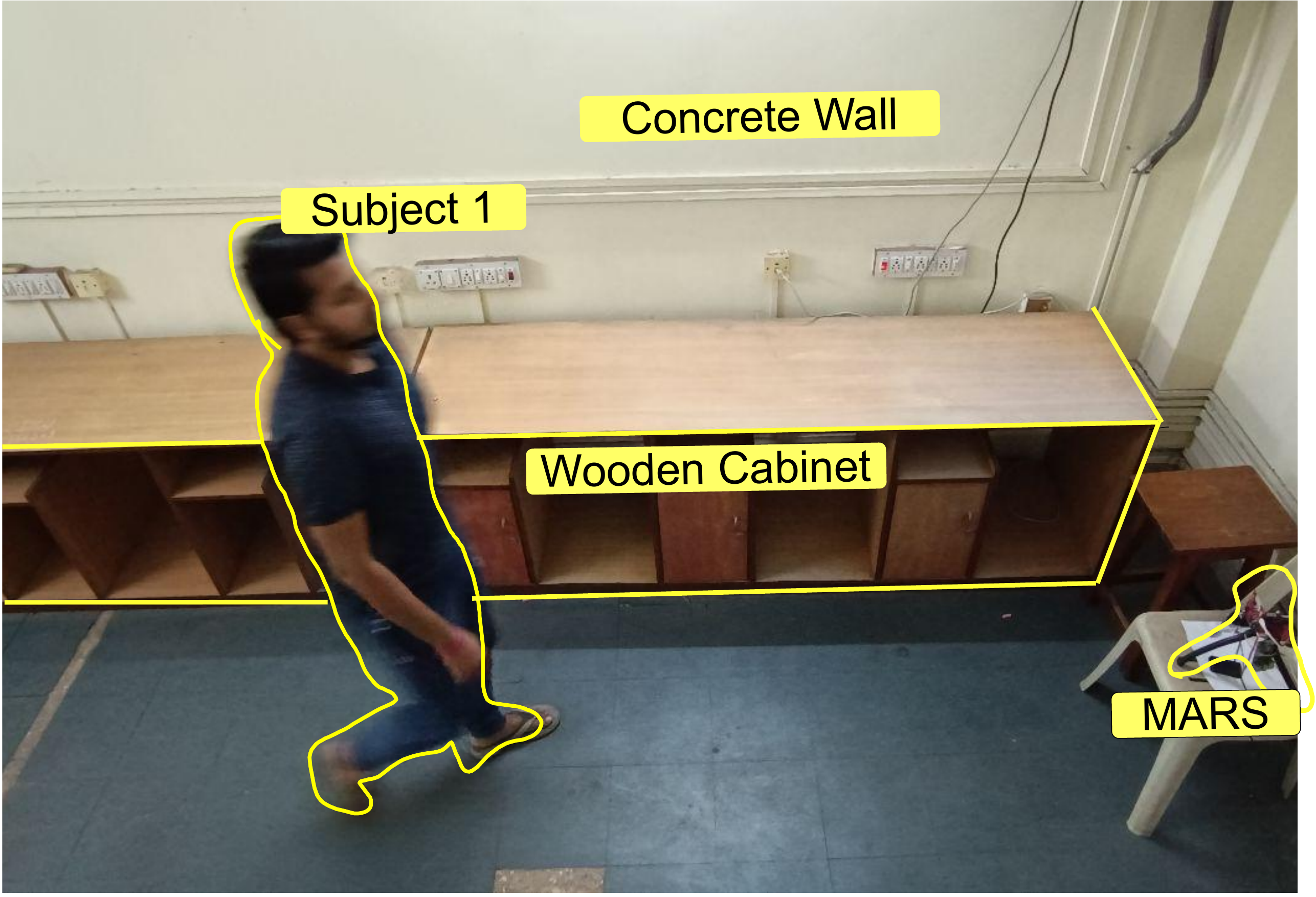}}\label{fig:zombie_data}
\\[-8pt]
 \subfloat[Static clutter]{\raisebox{-\totalheight}{\includegraphics[width=\wide\textwidth]{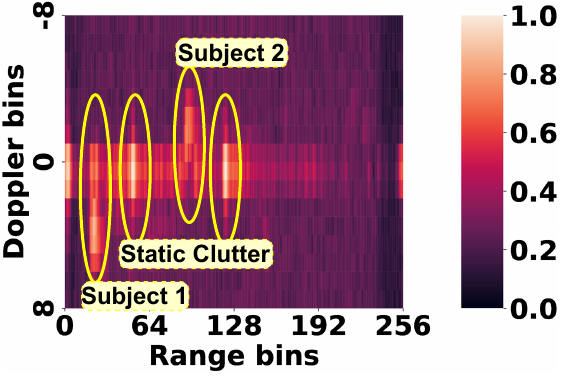}\label{fig:static_scenario_data}}}
& 
\subfloat[NLoS movements]{\raisebox{-\totalheight}{\includegraphics[width=\wide\textwidth]{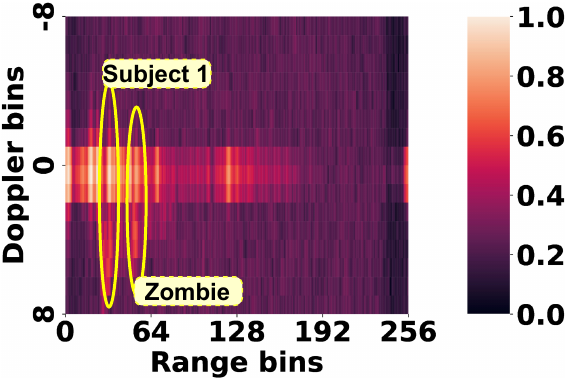}}\label{fig:zombie_scenario}}
 \end{tabular}
 \caption{Range-doppler signatures for two scenarios}
\label{fig:pilot1}
    \end{minipage}
    \hspace{0.5em}
    \begin{minipage}{0.3\textwidth}
                \centering
  \setlength\tabcolsep{0.7pt}
  \def\arraystretch{0.5}
  \def\wide{0.5}
  % \centering
    \includegraphics[width=\textwidth, trim={5cm 0 6.9cm 0cm}, clip]{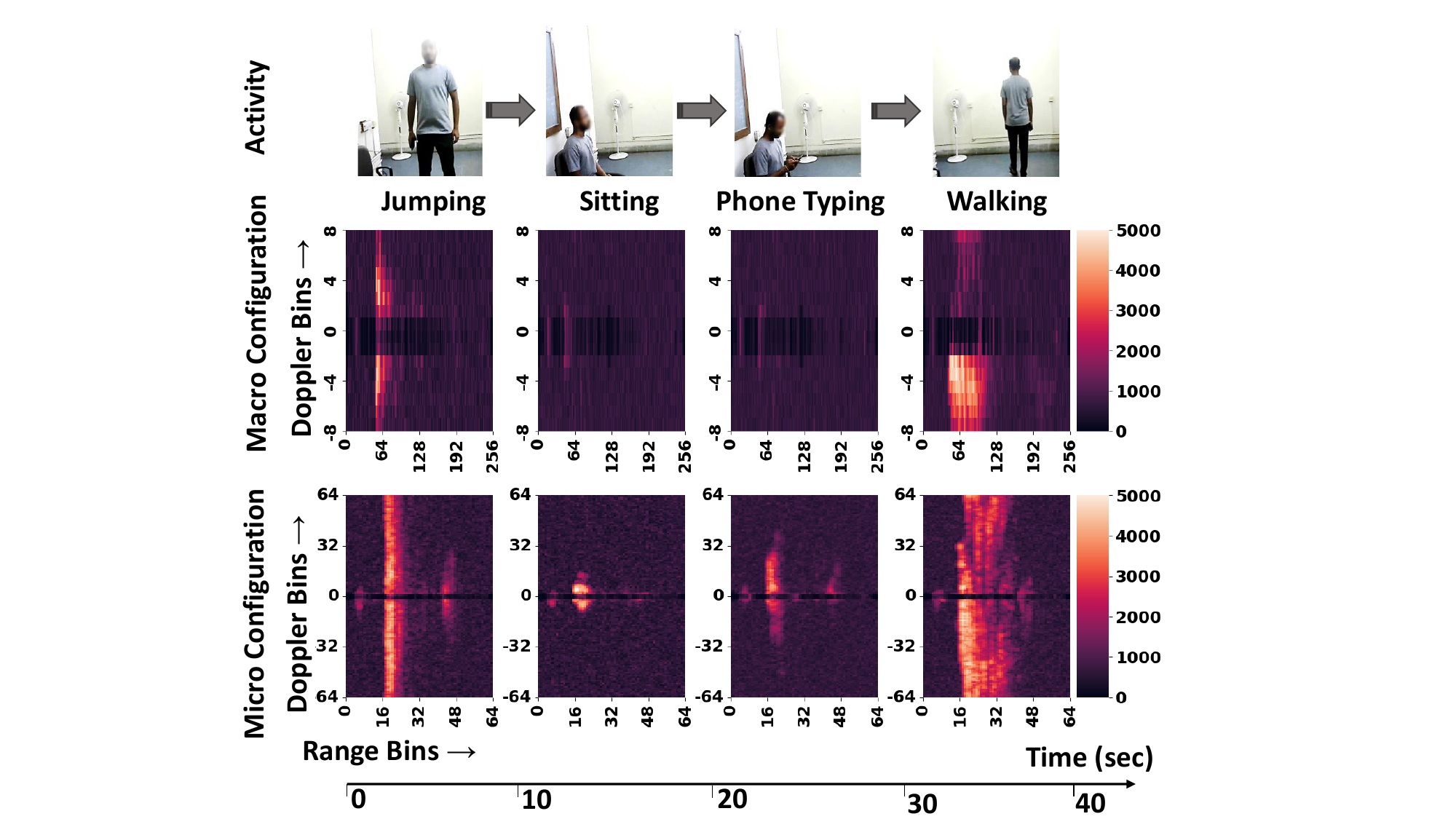}
    \caption{Range-doppler (standard deviation for individual activity windows) over time}
    \label{fig:conf_motivation}
    \end{minipage}
    \hspace{0.5em}
    \begin{minipage}{0.32\textwidth}
        \centering
        \includegraphics[width=0.5\columnwidth]{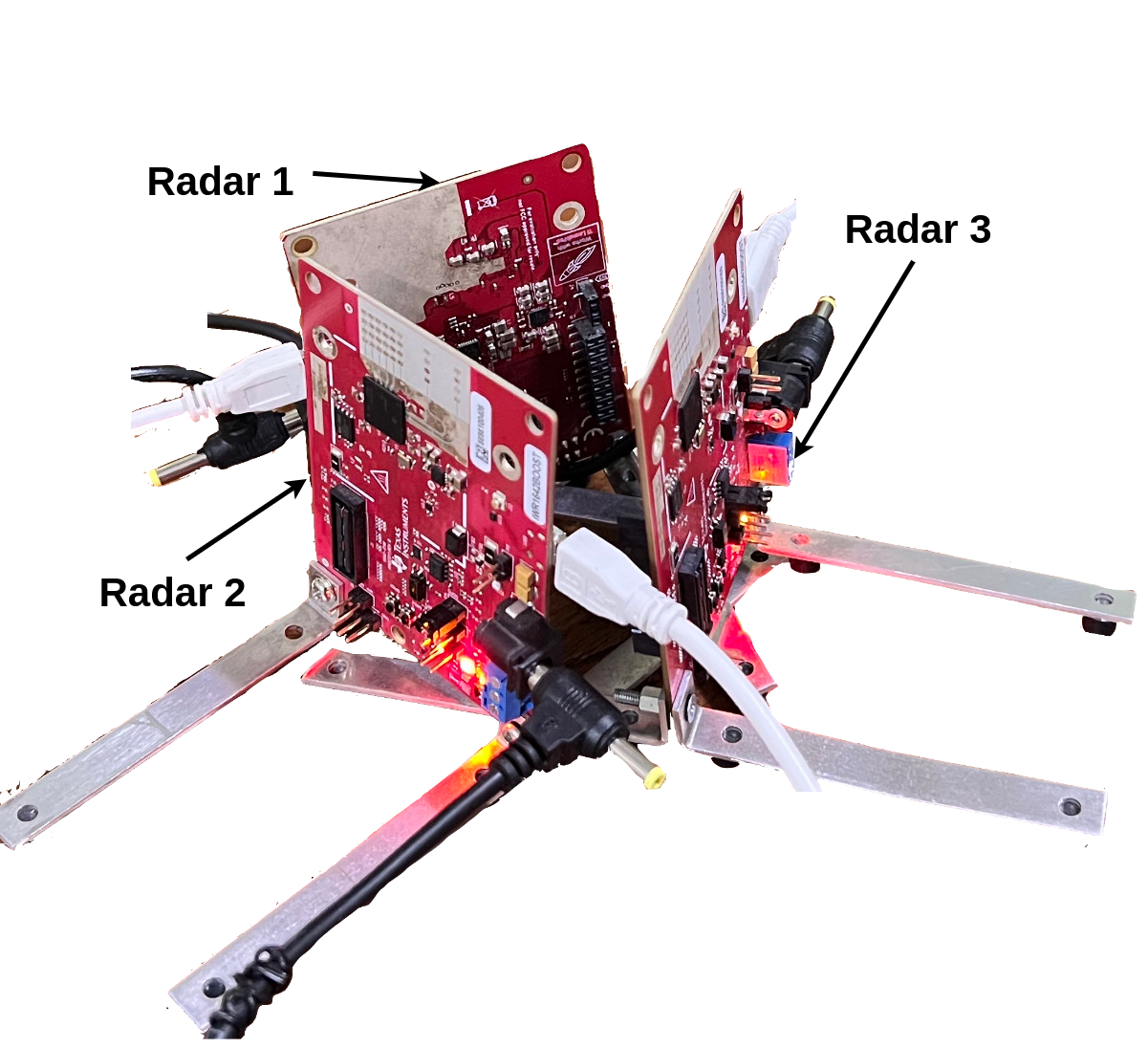}
        \subfloat[Jumping]{
            \includegraphics[width=0.3\columnwidth,angle=-90]{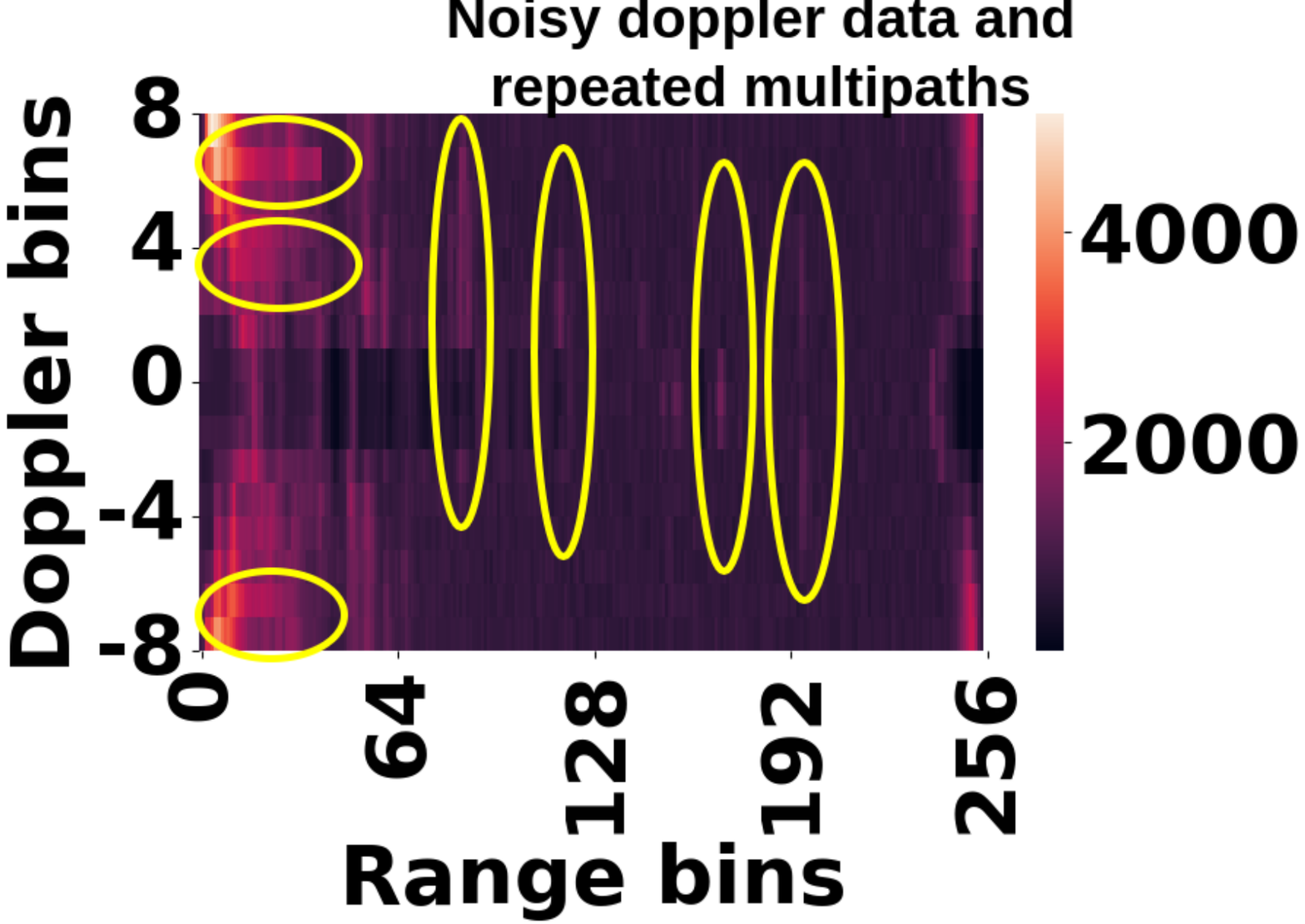}
        }
        \subfloat[Sitting]{
            \includegraphics[width=0.3\columnwidth,angle=-90]{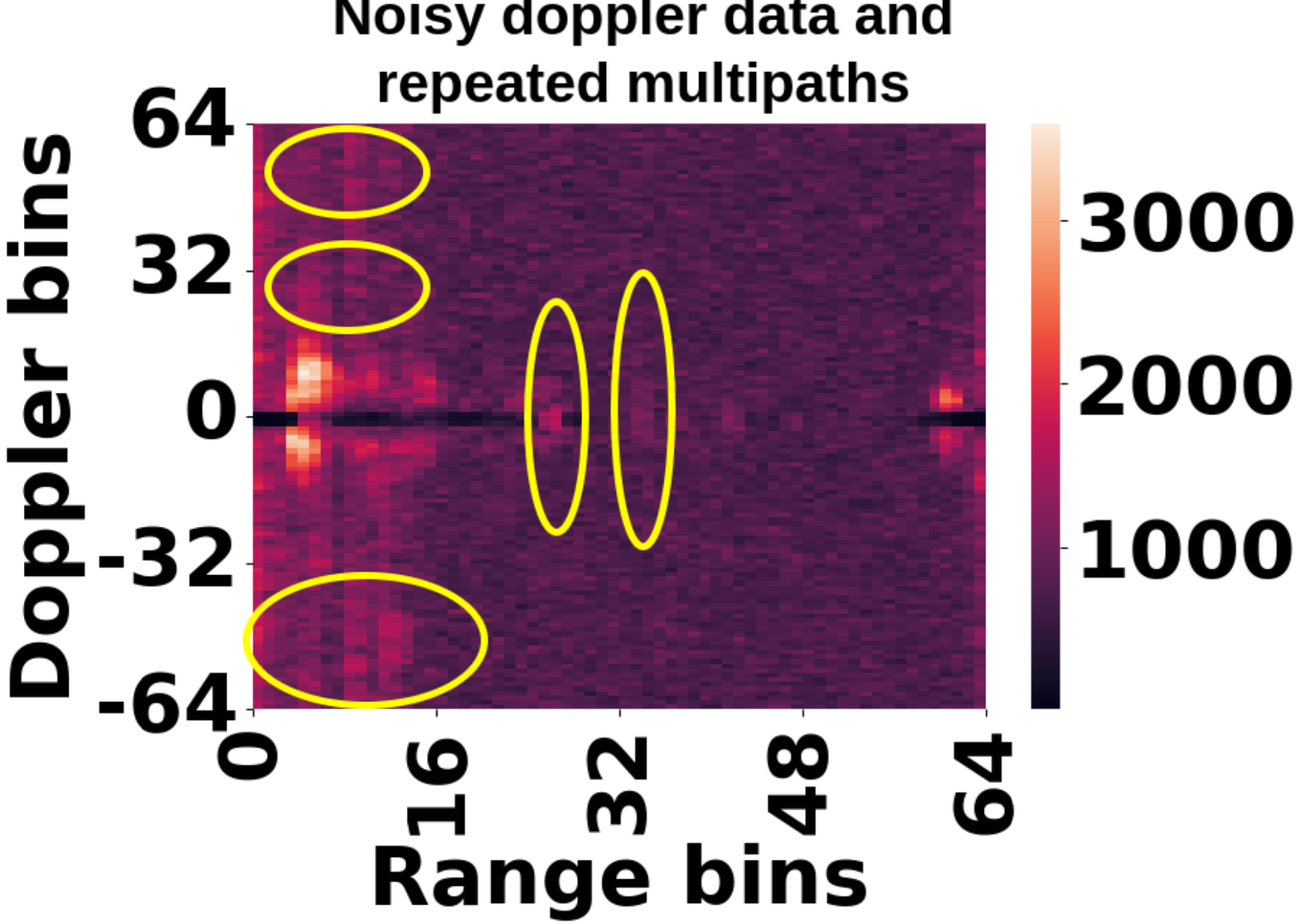}
        }
        % figures/macro_multi_radar.png
        % figures/micro_multi_radar.png
    \caption{Multi-radar challenges}
    \label{fig:multiradarexp}
    \end{minipage}
\end{figure*}

% SImilar activity related concern. time-axis details... 
% The range-doppler information effectively captures motion and thus can help extract body movement patterns. 
\subsubsection{Feasibility study for range-doppler}
\figurename~\ref{fig:dopplerpatterns} shows the standard deviation in the range-doppler heatmaps captured during the activity. Notably, standard deviation technique removes static powers (-3 to 2 doppler bins) in the heatmap; thus, we see a low-power value in these doppler bins. We observe that each activity has different signatures captured by the range-doppler heatmap. Although the plotted standard deviation looks similar for some activity pairs with similar body movements (like walking/running, jumping/lunges), there are temporal changes in the heatmaps; for example, ``running'' induces a faster change than ``walking''. Thus, combining the observations from range, doppler, and time, we can get different signatures. Notably, the macro activities have more robust patterns due to the magnitude of movement involved. Even though the micro activities have relatively weaker signatures, they can be distinctively captured with a \textit{higher doppler resolution} ($-64$ to $+64$ doppler bins, in contrast to $-8$ to $+8$ doppler bins used for macro activities).

% Therefore, from \figurename~\ref{fig:dopplerpatterns}, it is evident that range-doppler can provide sufficient information to assess an activity pattern at different scales (macro vs. micro) of activity granularity with a judicious selection of the doppler resolution (detailed in Table~\ref{tab:radar_conf}). 

\subsubsection{Impact of static clutters} 
Static clutters are any object (walls, furniture, etc.) that are stationary but can reflect the mmWave signal and therefore, generates unwanted signatures in the range-doppler data. We consider a scenario with two subjects -- \textit{Subject 1 and Subject 2}, both sitting inside the room, as shown in \figurename~\subref*{fig:static_scenario_data}. The room also contains multiple static clutters, such as wooden sheets and walls. From the corresponding range-doppler heatmap, we observe multiple peaks at the range bins corresponding to both the subjects and the static clutters. Indeed, the static clutters produce a higher magnitude along the zero doppler axis, thereby signifying zero or no movement. On the other hand, the dynamic movements of the subjects are positioned across non-zero doppler bins. A major takeaway from the range-doppler heatmap is that static clutters are easily identifiable by their zero-doppler signatures.

%*********** Fig: Range doppler in different cases
%  \begin{figure}[!ht]
%  \centering
%   \setlength\tabcolsep{4pt}
%   \def\arraystretch{0.5}
%   \def\wide{0.2}
%  \begin{tabular}{ll}

%  % &\multicolumn{1}{c}{Static clutter}& \multicolumn{1}{c}{Dynamic Clutter}&\multicolumn{1}{c}{Zombie user}& \multicolumn{1}{c}{Low DR} & \multicolumn{1}{c}{High DR}\\

%  % \parbox[b]{6mm}{\multirow{3}{*}{\rotatebox[origin=c]{90}{Scenario\hspace{20pt}}}}
%  % & 
% \raisebox{-\totalheight}{\includegraphics[width=\wide\textwidth]{figures/static_clutter_scenario.jpg}}\label{fig:static_scenario}
% % & 
% % \raisebox{-\totalheight}{\includegraphics[width=\wide\textwidth]{figures/static_clutter_scenario.jpg}}\label{fig:static_data}
% & 
% \raisebox{-\totalheight}{\includegraphics[width=\wide\textwidth]{figures/zombie_user_scenario.jpg}}\label{fig:zombie_data}
% \\[-8pt]
%  \subfloat[With Static clutter]{\raisebox{-\totalheight}{\includegraphics[width=\wide\textwidth]{figures/range_doppler_heatmaps/plot_static.jpg}}\label{fig:static_scenario}}
% % & 
% % \subfloat[Dynamic Clutter]{\raisebox{-\totalheight}{\includegraphics[width=\wide\textwidth]{figures/place_holder.png}}\label{fig:dynamic_scenario}}
% & 
% \subfloat[NLoS subject movement]{\raisebox{-\totalheight}{\includegraphics[width=\wide\textwidth]{figures/range_doppler_heatmaps/plot_zombie.jpg}}\label{fig:zombie_scenario}}
%  \end{tabular}
%  \caption{Range-doppler signatures for two different scenarios. Subject 1 and Subject 2 perform Macro (M) activities in either case.}
% \label{fig:pilot}
%  \end{figure}
%\AD{new photos to be added}

\subsubsection{The effect of \ac{NLoS} movements} 
To study the NLoS reflections, we first ask a single subject (Subject 1) to stand close to a wall and make some movements (macro-scale) as shown in \figurename~\subref*{fig:zombie_scenario}. From the corresponding range-doppler heatmap, it can be observed that the subject's movements are captured at two different instances at two different range bins. Of the two visible peaks, the more substantial peak belongs to the actual user's movement, whereas the other instance, also termed as a \textit{zombie subject}, occurs due to the multi-path reflection from the wall.

\subsubsection{Impact of radar configurations on determining users' activity}
To understand how the radar configuration affects the patterns in the activity signatures, we ask one subject to switch his activity from jumping to two micro activities, namely, sitting in a chair and phone typing, and finally, walking out of the room. The subject is asked to repeat the pattern twice to collect the corresponding range-doppler data under \textit{low and high doppler resolution}. From the std in the heatmap across the entire activity time axis \figurename~\ref{fig:conf_motivation}, it is evident that low doppler resolution is adequate for capturing macro activities like walking and jumping. Still, typing and sitting does not have any significant signatures. On changing the radar configuration to \textit{high doppler resolution}, we observe that micro activities like typing and sitting have better visibility. However, with this, the macro activities (walking, jumping) generate noisy data due to the higher  resolution. Therefore, \textit{different doppler resolutions} is crucial to capture the signatures corresponding to different activities.  

\subsubsection{Impact of multiple radars}
To have entire room coverage, we have taken three radars and kept them in a colocated position with 120\textdegree to each other as shown in \figurename~\ref{fig:multiradarexp}. We observe that incorporating multiple radars within the same room leads to complex interference patterns in the range-doppler heatmaps. The same or overlapping frequency bands lead to interference in the mmWave chirps and also cause more multipath effects. As shown in \figurename~\ref{fig:multiradarexp}, the range-doppler heatmaps are very noisy and have complex interference patterns which are not easily separable. This indicates that using multiple radars for 360\textdegree coverage makes the system complex; therefore, we need some alternate solution.

\section{Methodology}
\begin{figure}
    \centering
    \includegraphics[width=0.4\textwidth]{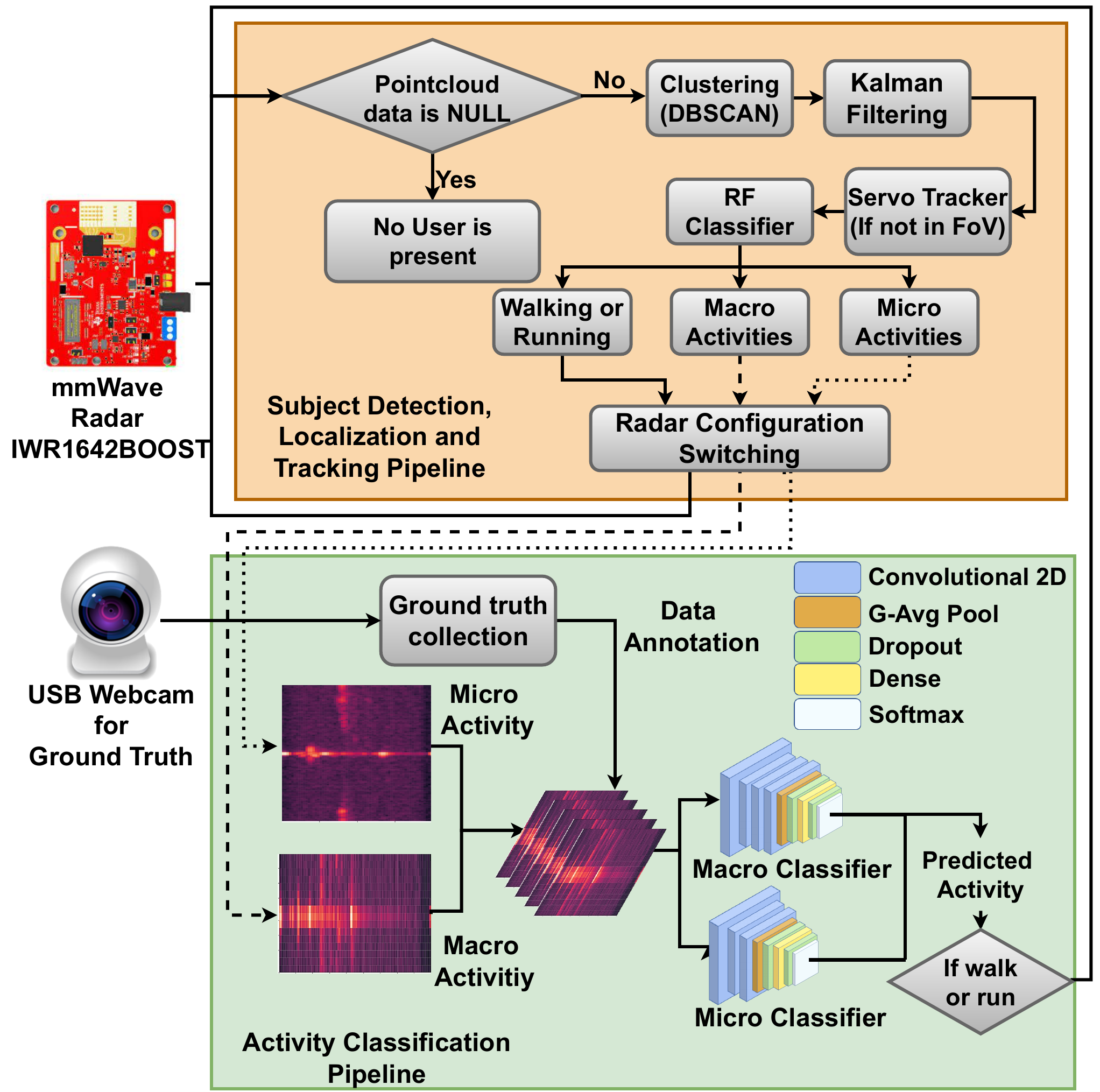}
    \caption{System architecture}
    \label{fig:system_architect}
\end{figure}

To have the end-to-end user localization and activity monitoring pipeline, we divide the problem into two sub-problems as highlighted in \figurename~\ref{fig:system_architect}: (i) subject detection followed by the localization and tracking of the subjects, and (ii) activity classification for individual subjects. We next discuss the two modules corresponding to these two sub-problems. %Algorithm~\ref{algo:selection} provides an overview of \ourmethod{} workflow. 

\subsection{Localization and Tracking}\label{sec:user_presence}
\ourmethod{} relies on the \textit{pointcloud data} to localize subjects and track their movements. Motivated by the challenges discussed in Subsection~\ref{sec:pilot_study}, we perform the following steps.
%the main challenges in this step are -- (i) isolation of the subject from the static clutters, (ii) scenarios when the subject is present inside the room but not within the field-of-view (FoV) of the radar, (iii) associating each subject with their RF reflections, and (iv) removal of zombie subjects introduced by multipath reflections. We start with separating the static clutters that generate unnecessary pointclouds from the radar. 

\subsubsection{Isolate subjects from static clutters} In practice, multiple static objects can be present within the FoV of the radar. As we are interested in identifying subjects' movement, the background, corresponding to stationary objects, needs to be removed. For this, we remove the zero-valued doppler bins for segregating the static objects (clutters). With this step, the mmWave radar can generate a pointcloud that does not contain static obstacles to isolate the subjects. Once \ourmethod{} starts receiving the pointcloud data, it tracks the subject by converting its pointcloud coordinate to a global coordinate. 

% \begin{figure}
    
% \end{figure}

\subsubsection{Global Coordinate Conversion} 
\begin{figure}[!ht]
    \centering
    \includegraphics[width=0.45\columnwidth]{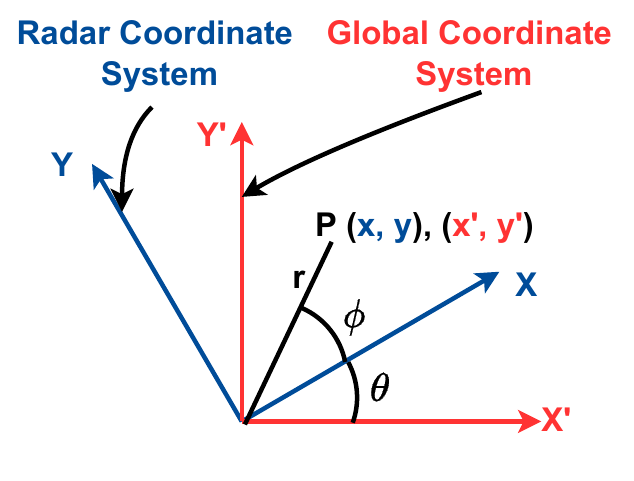}
    \caption{Transformation of the coordinate system}
    \label{fig:coordinate_transform}
\end{figure}

When the subject is present within the room but outside the radar's FoV, localization, and activity recognition of the subject is not feasible. As a solution, we mount the mmWave radar on top of the rotor axis of a servo motor. This enhances the FoV of the radar to 360\textdegree. However, rotating the sensor will directly change the reference coordinate system of the estimated pointclouds. Therefore, instead of keeping the local coordinate system w.r.t. the radar, we use a magnetometer to keep a global reference coordinate system. The magnetometer provides the reference azimuthal angle w.r.t. the earth's magnetic pole. Consider a user at $P(x, y)$ in the radar coordinate system as shown in \figurename~\ref{fig:coordinate_transform}. The radar is oriented by an angle of $\theta$ w.r.t. the magnetometer. So in the global coordinate system, the angular position of the object is at $(\theta+\phi)$, where $\phi = \tan^{-1}(\frac{y}{x})$. Equation~\ref {eq:transform_coord} illustrates the transformation of the radar coordinate system to the global coordinate system. 
\begin{equation}\label{eq:transform_coord}
    \begin{bmatrix}x'\\y' \end{bmatrix} 
    = \begin{bmatrix} r\cos(\theta+\phi)\\r\cos(\theta+\phi) \end{bmatrix}
    = \begin{bmatrix} \cos \theta & -\sin \theta \\ \sin \theta & \cos \theta \end{bmatrix} \begin{bmatrix} x\\y \end{bmatrix}
\end{equation}
%\swadhincomment{[SP: Needs some 2-3 lines explanation of the transformation above]}
With the above transformation matrix, the pointclouds are now referenced w.r.t. the magnetometer and does not suffer from any coordinate shift due to the rotation of the radar. %After applying the preprocessing steps of static clutter removal and global coordinate transformation, we next discuss the methodology adopted in \ourmethod{} to localize multiple subjects. 

\subsubsection{Tracking multiple subjects}
Based on the pointcloud data, we have information about all the subjects; however, we also get noisy pointclouds due to the movements of the subjects. To tackle this, we take the pointcloud data in a queue format and pass this information to \ac{DBSCAN}~\cite{ester1996density} for clustering. Each cluster is associated with a unique ID to associate subjects with their respective clusters. Now, to detect the presence of a new subject, \ourmethod{} compares the detected pointcloud clusters between two consecutive frames received from the radar. If the Euclidean distance between the centroid of a cluster over a new frame and that over the previous frame is less than $\epsilon$, we keep the respective cluster ID the same as before. In the case of a newly discovered cluster, we assign it a new cluster ID, indicating a new subject. Note, $\epsilon$ is a hyperparameter, and in our setup, we keep it as $10$cm, signifying the minimum range resolution for a subject. %\sandipcomment{SC: Impacts of such hyperparameters must be discussed in the evaluation.}

\subsubsection{Tracking movement of individual subjects}
After the clustering step in the pipeline, each cluster corresponds to the pointcloud information associated with each subject. However, some of these clusters may correspond to zombie subjects, as discussed earlier. We observe that \textit{pointclouds for zombie clusters have a low reflection power} and thus are generated less frequently when compared to the pointclouds caused by the actual subject. So, we first apply a mode function on the pointcloud queue for each cluster to filter out the pointclouds generated more frequently due to the actual subject's presence. The remaining noisy outliers get removed with this approach. However, due to the uncertain movements of the subjects, two subjects may impede each other while crossing. This may lead to \textit{blind spots} in the pointcloud data. We handle such blind spots as follows. 

\subsubsection{Handling blind spots during multi-user tracking}
To track each subject seamlessly, we apply a Kalman filter~\cite{pegoraro2021real} on the pointcloud queue. The Kalman filtering technique uses the prior knowledge of the state of an object and then \textit{predicts} and \textit{updates} the location and velocity of that object for the next frame. For precise tracking of individual clusters instead of a static Kalman gain, we opted for \ac{RKF} to estimate the subjects' motion states. \ac{RKF} can recursively generate the error covariance matrix and Kalman gain at each stage of the update process. With this step, we can estimate the subjects' state when the actual pointcloud data is unavailable due to occlusion by other subjects' movement or errors in the former pipeline.
 %In the prediction step, it tries to predict the subject's current state (position, velocity) given the previous state and the time that has passed since. The update step takes the predicted and actual state as its input (in the form of a Gaussian probability distribution function) and updates its Kalman gain for the next prediction step.
% It uses a state transition matrix for relating the previous state to the current state, a covariance matrix for calculating the uncertainty of the state, along with a diagonal noise matrix.
% The tracking phase is required to keep track of people as they move around in an indoor arena and to maintain precise and trustworthy measurements. 
% Since the tracking comes after the first set of preprocessing due to clustering; a precise tracking technique is needed to tackle scenarios with errors in the former pipeline. Thus, in this work, we employ a \ac{RKF} to forecast and estimate the subject's motion states. Instead of using a static Kalman gain, we opted for \ac{RKF} to recursively generate the error covariance matrix and Kalman gain at each stage of the update process. With this step, we can estimate the subjects' state apriori under scenarios when the actual point-cloud data is unavailable due to occlusion by other subjects' movement or errors in the former pipeline.

\subsubsection{Servo-based tracking}
Usually, the azimuthal FoV of the radar is $120^{o}$ which can localize and track subjects. To enhance this FoV, we rely on servo-based tracking. As soon as we have the final coordinates of the denoised pointclouds, we check if each subject is within the main lobe of the radar, i.e., $\leq \pm 15$\textdegree. Otherwise, we rotate the servo towards the subject by an angle of $\tan^{-1}(\frac{y}{x})$, to generate high-fidelity pointclouds and range-doppler heatmaps which are needed for activity classification. 
% Here, $(x,y)$ is the global coordinate of the subject generated after applying all the pre-processing steps. 
We stop the rotation when the subject is within $\pm15^{o}$  so that the doppler remains unaffected for the next activity classification task. Gradually with the rotation process, a new pointcloud queue is generated for subjects that were earlier outside the FoV.
% Once we have the final coordinates of the denoised pointclouds, we take each subject's location information and check if the subject's position is within the main lobe of the radar, i.e., $\leq \pm 15$\textdegree. Otherwise, we rotate the servo towards the subject by an angle of $\tan^{-1}(\frac{y}{x})$. Here, $(x,y)$ is the global coordinate of the subject generated after applying all the pre-processing steps. This method helps track a subject and generates a pointcloud for a new subject, which came into the FoV after the rotation.\\
%\swadhincomment{[SP: Although we talked about point-cloud estimation in preliminary, we need talk about in the intro and a bit preliminary, what we actually mean by point-cloud]}\AD{AD: Included in preliminaries}

\subsubsection{Monitoring state change of a subject} \label{ref:rf_classifier} 
Once a subject is tracked, \ourmethod{} monitors the possible state changes of that subject by utilizing the pointcloud data. Broadly, it performs a high-level classification to check whether (i) the subject is walking or running inside the room or (ii) the subject is static and performing some macro/micro-activities. For this purpose, we capture the mean, standard deviation, kurtosis, and skewness in each of the denoised pointcloud clusters for a time window of $1$ seconds. These features are fed to a Random Forest Classifier to predict the subject's activity scale. Based on the prediction, we continue the localization and tracking if the subject is walking or running. Else, \ourmethod{} enables the macro or micro activity classifier opportunistically based on the inference. Thus we name this classifier as Opportunistic Classifier.

% Let $\epsilon$ be the radius of a neighborhood for some point. 
% \begin{enumerate}
%     \item A point $\it{p}$ is a core point if the number of points within a distance of $\epsilon$ is above a certain threshold.
%     \item A point $\it{q}$ is directly reachable from $\it{p}$ if point $\it{q}$ is within distance $\epsilon$ from core point $\it{p}$.
%     \item A point $\it{q}$ is reachable from $\it{p}$ if there is a path $\it{p_1}, ..., \it{p_n}$ with $\it{p_1} = \it{p}$ and $\it{p_n} = \it{q}$, where each $\it{p_{i+1}}$ is directly reachable from $\it{p_i}$.
%     \item All points not reachable from any other point are outliers or noise points.
% \end{enumerate}

\subsection{Macro/Micro Activity Monitoring}
\label{sec:activity}We keep two different radar configurations to capture the classes of micro and macro activities. For macro activities, \ourmethod{} uses a low doppler resolution of $16$ doppler bins (captures major body movements but eliminates the details that may generate noise), while for capturing micro activities, it uses a high doppler resolution of $128$ doppler bins (captures minor body movements with finer details). Using range-doppler enables us to easily switch the radar configurations at different resolutions to recognize both macro and micro activities from different users. 

\subsubsection{Segregation of individual subject's activity signatures} As shown in~\figurename~\ref{fig:dopplerpatterns}, range-doppler is represented as a heatmap image, where the abscissa is the range, the ordinate is the doppler speed containing the power value of subjects' movement. Each subject's activity has its activity signatures in the range-doppler heatmap (See \figurename~\ref{fig:dopplerpatterns}). To classify the activity of individual subjects, we first segregate these activity signatures based on the range bins. From the pointcloud data (collected along with the range-doppler), we check if there is a non-zero doppler value in the range profile where the subject is present. If a doppler variation exists, we slice out that Range-doppler heatmap information with padding of $\pm10$ range bins. Additionally, we define another copy of the Range-doppler heatmap for each subject, replacing the remainder with the minimum heatmap value for the subject. In this way, each subject has its own activity signatures, and the remaining signatures corresponding to other subjects are suppressed. This modified range-doppler data is fed to the classification model.

\subsubsection{Differentiated frame stacking for macro/micro activities classification} 
These macro or micro activities span over a short period, affecting range-doppler values temporally. We stack $1$ sec range-doppler data to capture temporal features, thus achieving a two-dimensional (2D) multichannel array. However, for macro activities, the doppler resolution is low, resulting in a heatmap of size $16 \times 256$, while for micro activities, the doppler resolution is high, resulting in a heatmap of size $128 \times 64$. This diversity results in different \ac{FPS} for the range-doppler computation and data transfer. For low-resolution doppler, the \ac{FPS} is $5$, while for the high-resolution doppler, the \ac{FPS} is $2$. Therefore, we stack $5$ frames together in the case of the macro activity classifier, while for the micro activity classifier, we stack $2$ frames together. This enables us to have the range-doppler for a consistent time period of $1$ sec for both scenarios.

\subsubsection{Model Architecture}
\label{sec_model}
The 2D range-doppler heatmaps have different spatial patterns for each activity. So, we employ a 2D Convolutional Neural Network architecture (2D-CNN). Convolution 2D operation considers the dependency of neighboring spatial values and the temporal relationship of past $t$ ($t = $ \ac{FPS}) frames. We use four and three 2D convolutional layers with `same' padding and Relu activation for the macro and micro activity classifiers. Next, a global average pooling layer is added to extract the average spatial activation across the entire feature map. Finally, we add two successive dropout and dense layers, where the dropout rate is kept as $20\%$ and $10\%$, respectively. The last layer outputs a joint probability distribution over all possible activities with a softmax activation (detail in \tablename~\ref{tab:2d_cnn}). Although the subject's orientation may not impact the detection of macro activities, the micro activities need precise signatures. As we collect the range-doppler at a higher resolution for micro activity classification, it can sense the movements even when the signal strength is low. As a result, the proposed 2D-CNN model can capture micro activities even when the subject is not directly facing the radar. 

%Finally, we describe the entire flow of \ourmethod{} after the activity classification.
%\sandipcomment{SC: Need a subsection at the end to connect all the components and conclude the methodology, like overall how the radar timeshares across multiple users, with a configuration switching between tracking, macro and micro classification.}(\AS{I have added Sec 4.3 for this purpose})

\begin{table}[!t]
\scriptsize
\centering
\caption{2D-CNN architecture (M: macro, $\mu$: micro)}
\label{tab:2d_cnn}
\begin{tabular}{|p{4.8em}|p{2em}p{2em}p{2em}p{2em}p{2em}p{2em}p{2em}p{2em}|}
\hline
\multirow{3}{*}{\textbf{CNN Layer}} & \multicolumn{8}{c|}{Parameters} \\ 
\cline{2-9}
 & \multicolumn{2}{c|}{Kernel} & \multicolumn{2}{c|}{Stride} & \multicolumn{2}{c|}{Channel} & \multicolumn{2}{c|}{Dropout}\\ 
 \cline{2-9} 
 & \multicolumn{1}{c|}{M} & \multicolumn{1}{c|}{$\mu$} & \multicolumn{1}{c|}{M} & \multicolumn{1}{c|}{$\mu$} & \multicolumn{1}{c|}{M} & \multicolumn{1}{c|}{$\mu$} & \multicolumn{1}{c|}{M} & $\mu$ \\ \hline
Input Layer & \multicolumn{1}{c|}{-} & \multicolumn{1}{c|}{-} & \multicolumn{1}{c|}{-} & \multicolumn{1}{c|}{-} & \multicolumn{1}{c|}{5} & \multicolumn{1}{c|}{2} & \multicolumn{1}{c|}{-} & - \\ \hline
Conv1 & \multicolumn{1}{c|}{2 x 5} & \multicolumn{1}{c|}{3 x 2} & \multicolumn{1}{c|}{1 x 2} & \multicolumn{1}{c|}{2 x 1} & \multicolumn{1}{c|}{32} & \multicolumn{1}{c|}{32} & \multicolumn{1}{c|}{-} & - \\ \hline
Conv2 & \multicolumn{1}{c|}{2 x 3} & \multicolumn{1}{c|}{3 x 3} & \multicolumn{1}{c|}{1 x 2} & \multicolumn{1}{c|}{2 x 2} & \multicolumn{1}{c|}{64} & \multicolumn{1}{c|}{64} & \multicolumn{1}{c|}{-} & - \\ \hline
Conv3 & \multicolumn{1}{c|}{2 x 3} & \multicolumn{1}{c|}{3 x 3} & \multicolumn{1}{c|}{1 x 2} & \multicolumn{1}{c|}{2 x 2} & \multicolumn{1}{c|}{96} & \multicolumn{1}{c|}{96} & \multicolumn{1}{c|}{-} & - \\ \hline
Conv4 & \multicolumn{1}{c|}{2 x 3} & \multicolumn{1}{c|}{-} & \multicolumn{1}{c|}{1 x 2} & \multicolumn{1}{c|}{-} & \multicolumn{1}{c|}{96} & \multicolumn{1}{c|}{-} & \multicolumn{1}{c|}{-} & - \\ \hline
G-avg Pool & \multicolumn{1}{c|}{-} & \multicolumn{1}{c|}{-} & \multicolumn{1}{c|}{-} & \multicolumn{1}{c|}{-} & \multicolumn{1}{c|}{-} & \multicolumn{1}{c|}{-} & \multicolumn{1}{c|}{-} & -  \\ \hline
Dropout1 & \multicolumn{1}{c|}{-} & \multicolumn{1}{c|}{-} & \multicolumn{1}{c|}{-} & \multicolumn{1}{c|}{-} & \multicolumn{1}{c|}{-} & \multicolumn{1}{c|}{-} & \multicolumn{1}{p{2em}|}{20\%} & 20\% \\ \hline
Dense1 & \multicolumn{1}{c|}{-} & \multicolumn{1}{c|}{-} & \multicolumn{1}{c|}{-} & \multicolumn{1}{c|}{-} & \multicolumn{1}{c|}{32} & \multicolumn{1}{c|}{32} & \multicolumn{1}{c|}{-} & - \\ \hline
Dropout2 & \multicolumn{1}{c|}{-} & \multicolumn{1}{c|}{-} & \multicolumn{1}{c|}{-} & \multicolumn{1}{c|}{-} & \multicolumn{1}{c|}{-} & \multicolumn{1}{c|}{-} & \multicolumn{1}{c|}{10\%} & 10\% \\ \hline
Softmax & \multicolumn{1}{c|}{-} & \multicolumn{1}{c|}{-} & \multicolumn{1}{c|}{-} & \multicolumn{1}{c|}{-} & \multicolumn{1}{c|}{6} & \multicolumn{1}{c|}{6} & \multicolumn{1}{c|}{-} & - \\ \hline
\end{tabular}
\end{table}
% \subsection{Opportunistic Sensing Design}
% \begin{table}
% \scriptsize
% \centering
% \caption{Opportunistic Configuration Switching}
% \label{tab:conf_switch}
% \begin{tabular}{|l|l|l|l|}
% \hline
% \textbf{Configuration} & \textbf{Conf1} & \textbf{Conf2} & \textbf{Conf3} \\ \hline
% \textbf{Static Clutter Removal} & Yes & No & No \\ \hline
% \textbf{Pointcloud} & Yes & Yes & Yes \\ \hline
% \textbf{FPS} & High & Low & Low \\ \hline
% \textbf{Range-Doppler} & No & Yes & Yes \\ \hline
% \textbf{Doppler resolution} & No & Low & High \\ \hline
% \textbf{Primary Task} & \begin{tabular}[c]{@{}l@{}}Localization \& \\ Tracking\end{tabular} & Macro Activity & Micro Activity \\ \hline
% \end{tabular}
% \end{table}

\subsubsection{Opportunistic Configuration Switching}
For each macro and micro activity, \ourmethod{} switches the configuration accordingly (as derived from the step mentioned in Section~\ref{ref:rf_classifier}). Once the activity classification is performed, it checks whether the subjects are still in their activity state. If any subject starts walking or running, the micro and macro classifiers can detect that and switch the configuration back to capture the pointcloud data to reinitiate the \textit{Localization and Tracking} Pipeline. The clustering and denoising filters get restarted to track the subjects' movement.

\section{Implementation}~\label{sec:impl}
\begin{figure*}[t]
    \centering
            \subfloat[]{
	    \includegraphics[height=3.03cm,width=0.22\textwidth]{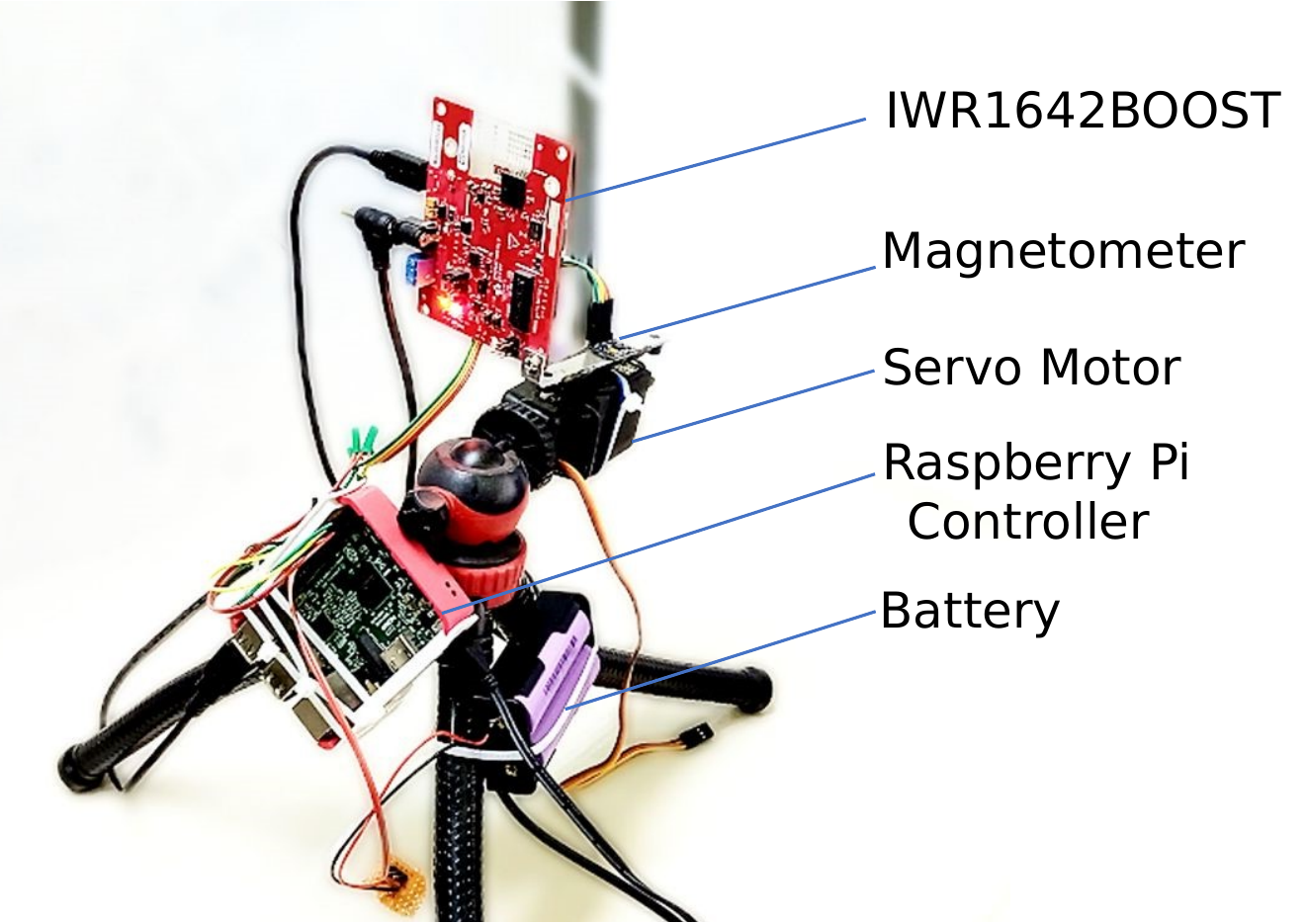}
    \label{fig:implementation_setup}}\hfil
	\subfloat[]{
	    \label{fig:r1setup}
	    \includegraphics[height=3.03cm,width=0.22\textwidth]{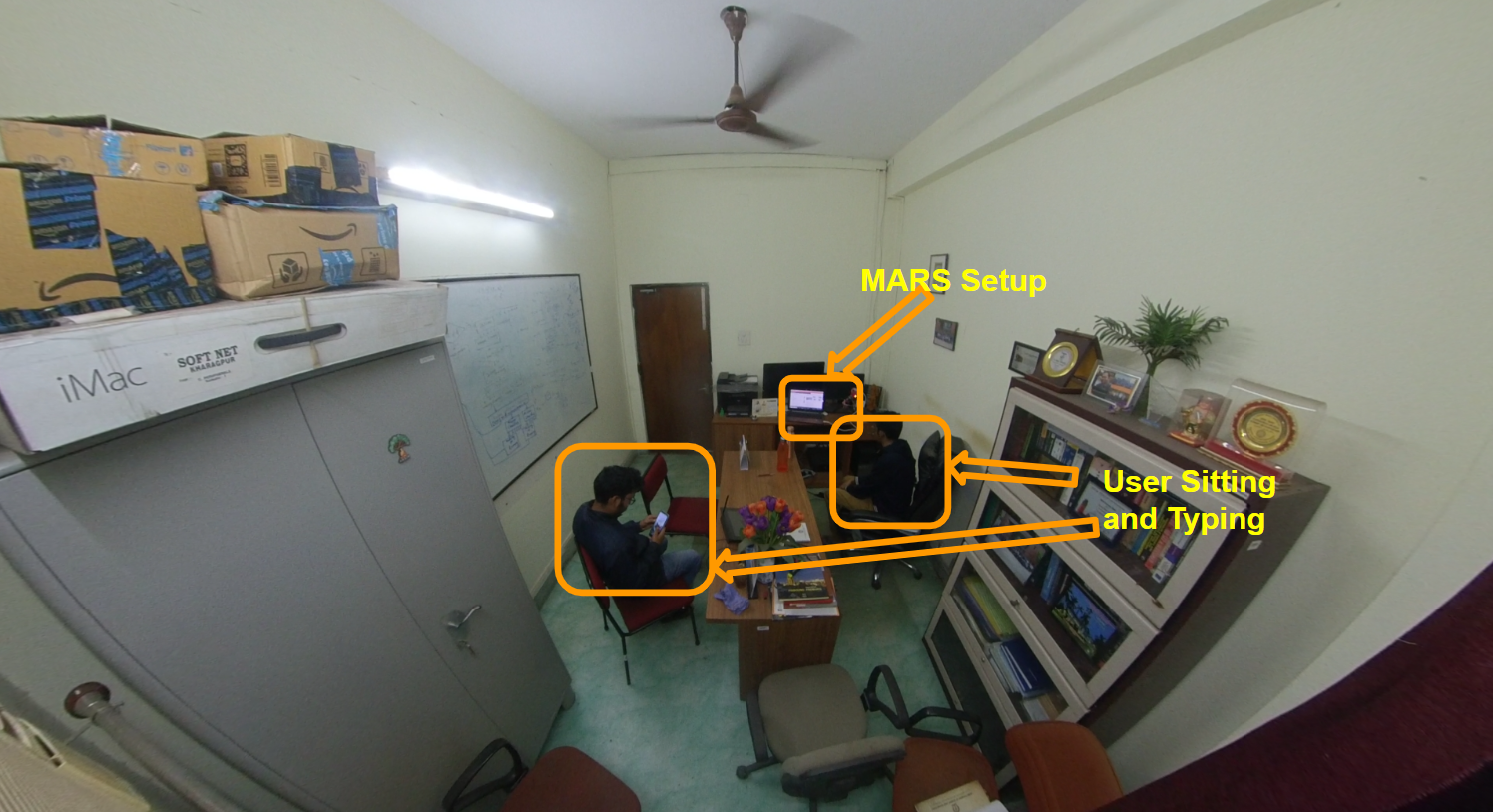}
	}\hfil
        \subfloat[]{
	    \label{fig:r2setup}
	    \includegraphics[height=3.03cm,width=0.22\textwidth]{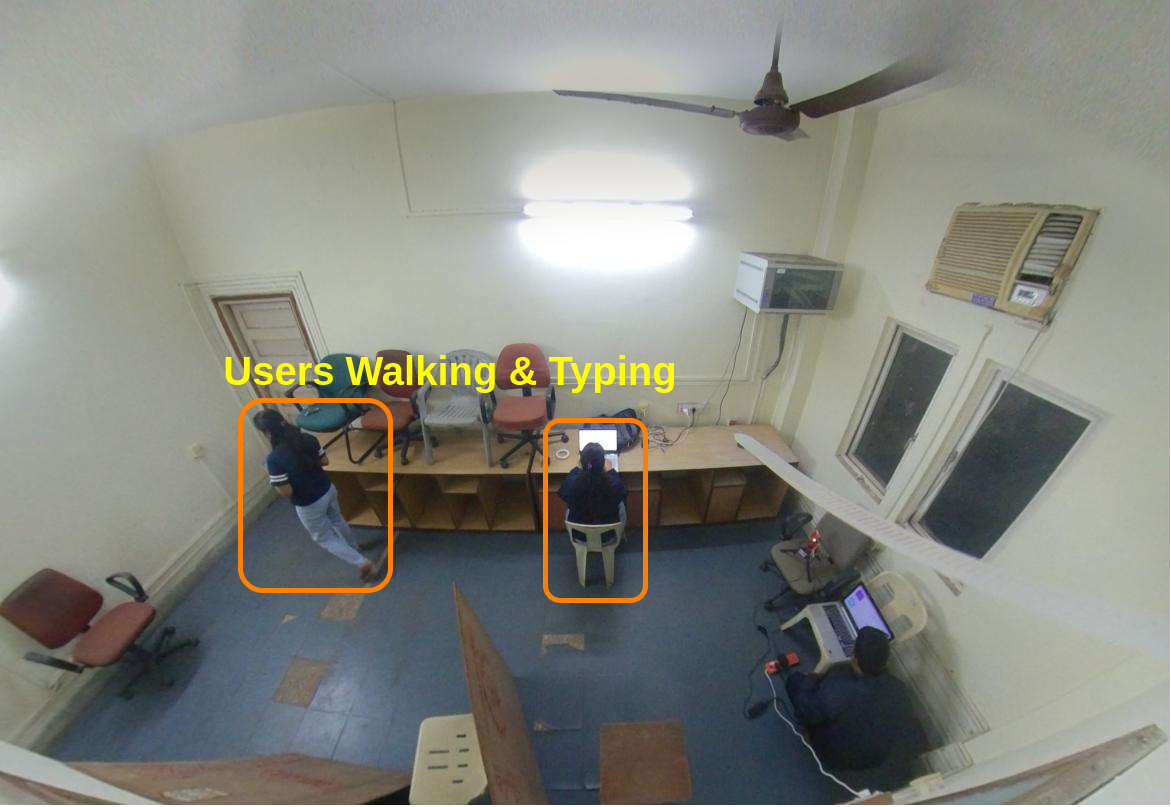}
	}\hfil
        \subfloat[]{
	    \label{fig:r3setup}
	    \includegraphics[height=3.03cm,width=0.22\textwidth]{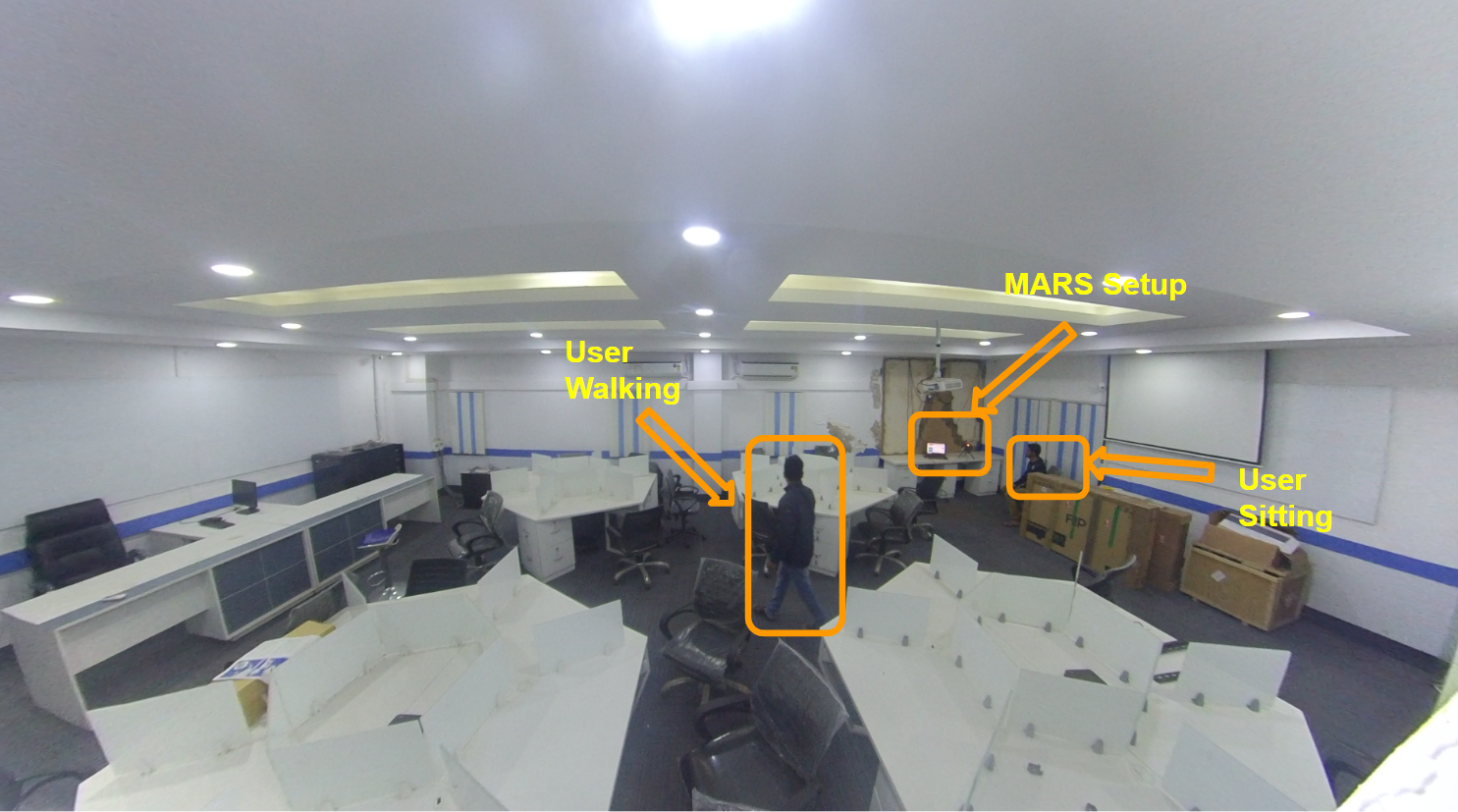}
	}\hfil
    \caption{(a) \ourmethod{} hardware setup; and data collection in different rooms: (b) R1 (c), R2 (d), R3}
    \label{fig:roomSetup}
\end{figure*}

As shown in \figurename~\ref{fig:implementation_setup}, \ourmethod{} is developed on top of a \ac{COTS} millimeter wave radar, \texttt{IWR1642BOOST}~\cite{iwr1642boost}. The system is tested in three different rooms (see \figurename~\ref{fig:r1setup}, \ref{fig:r2setup}, \ref{fig:r3setup})  -- (i) R1, an office cabin of size $4 \times 3$ m$^2$, (ii) R2, a classroom of size $8 \times 5$ m$^2$, and (iii) R3, a laboratory of size $12 \times 6.5$ m$^2$. The ground truth activity of each subject is manually annotated with the help of the video captured using a USB camera. Overall, \ourmethod{} consists of: the front-end radar and the backend processing unit. The radar senses data and generates 2D pointclouds and range-doppler heatmaps. These data entries are transferred via a USB cable with a baud rate of $921600$ to the backend \texttt{Raspberry Pi-4 Model B}~\cite{raspberrypiRaspberryModel} with 1.5GHz Broadcom BCM2711 64bit CPU and 8 GB RAM. We have used \texttt{Python 3.9.6}, \texttt{TensorFlow v2.10.0}, and \texttt{Scikit-learn v1.1.2} for implementing the macro and micro activity classifiers and the opportunistic Random Forest classifier. The models are trained on an iMac-M1 (with 16 GB primary memory running macOS v12.6 with base-kernel version: 21.6.0) and then deployed on the Raspberry Pi-4 for live inference. The training takes $10$ minutes for the opportunistic classifier and $20$ \& $25$ minutes for the case of macro and micro classifiers, respectively, with a model size of $7.8$ MB, $460$ KB, and $334$ KB, respectively, for the three cases.    %\sandipcomment{Give some stat about the training -- training time, model size, etc.}

\subsection{Hardware setup}
\subsubsection{Radar Configuration}
The IWR1642BOOST radar is configured to use two transmitter and four receiver antennas with frequencies of $77$-$81$ GHz (bandwidth $4$ GHz). For the three different use cases, i.e., (i) localization and tracking, (ii) macro activity classification, and (iii) micro activity classification, we have used three different radar configurations (\tablename~\ref{tab:radar_conf}). For the localization and tracking, we set the frame periodicity as $33.33$ millisecond to have $30$ \ac{FPS} to fill the localization queue fast so that clustering and Kalman filter-based tracking can be performed with minimal error. This configuration provides a range resolution of $4.36$ cm, with a maximum unambiguous range of $9.02$ m. It can measure a maximum radial velocity of $1$ m/s, with a doppler resolution of $0.13$m/s. The sensor is set to transmit $32$ chirps per frame. We use the same radar configuration for the macro activity classification, except we reduce the \ac{FPS} to $5$ to allow the flow of larger range-doppler heatmaps (matrix of size $16 \times 256$) via USB. The doppler resolution is kept at $0.01$ m/s for the micro-scale activity classification. The size of the range-doppler heatmap is $128 \times 64$, which supports a frame rate of $2$ FPS. 

\begin{table}
\scriptsize
\centering
\caption{Radar configuration}
\label{tab:radar_conf}
\begin{tabular}{|c|ccc|}
\hline
\textbf{Parameters} & \multicolumn{1}{c|}{\textbf{Localization}} & \multicolumn{1}{c|}{\textbf{Macro}} & \textbf{Micro} \\ \hline
\textbf{Start Frequency} & \multicolumn{3}{c|}{77 GHz} \\ \hline
\textbf{End Frequency} & \multicolumn{3}{c|}{81 GHz} \\ \hline
\textbf{Range Resolution (cm)} & \multicolumn{2}{c|}{4.36} & 12.5 \\ \hline
\textbf{Maximum Range(m)} & \multicolumn{2}{c|}{9.02} & 6.4 \\ \hline
\textbf{Maximum Radial Velocity (m/s)} & \multicolumn{2}{c|}{1} & 0.64 \\ \hline
\textbf{Velocity Resolution (m/s)} & \multicolumn{2}{c|}{0.13} & 0.01 \\ \hline
\textbf{Azimuthal Resolution (Degree)} & \multicolumn{3}{c|}{14.5\textdegree} \\ \hline
\textbf{Frames per Second} & \multicolumn{1}{c|}{30} & \multicolumn{1}{c|}{5} & 2 \\ \hline
\textbf{Chirps Per Frame} & \multicolumn{2}{c|}{32} & 64 \\ \hline
\textbf{ADC Samples per Chirp} & \multicolumn{3}{c|}{256} \\ \hline
\end{tabular}
\end{table}
% \begin{figure}
%     \centering
%     \includegraphics[width=0.55\columnwidth]{figures/device.pdf}
%     \caption{\ourmethod{} hardware setup}
%     \label{fig:implementation_setup}
% \end{figure}

\subsubsection{Localization and Tracking Setup}
To enhance the radar field-of-view to $360$\textdegree, we have mounted the radar on top of the rotor axis of a \texttt{TowerPro MG995 Servo Motor}~\cite{towerproMG995x2013}, powered using a 1200mAh Li-ion Rechargeable Battery. This enables the localization and tracking of subjects for the entire indoor space. \texttt{GY-273 Compass Magnetometer Sensor}~\cite{amazonRobodoSEN40} is used to transform the pointcloud coordinates to a global coordinate system, i.e., w.r.t. the earth's magnetic pole. 
% \subsubsection{Activity Classifier Setup}.
% We have used \texttt{Python 3.9.6}, \texttt{TensorFlow v2.10.0}, and \texttt{Scikit-learn v1.1.2} for implementing the 2D-CNN-based macro and micro activity classifier and the Opportunistic Random Forest classifier. The models are trained on an iMac (with 16 GB primary memory running macOS v12.6 with base-kernel version: 21.6.0). The generated knowledge file is then hosted on a Raspberry Pi board with 1.5GHz Broadcom BCM2711 64bit CPU and 8 GB RAM.
\subsection{Data Collection Setup}\label{sec:data_collection}
Data collection is carried out for $7$ subjects ($3$ female and $4$ male), with ages ranging from $23$ to $35$, for a total duration of $44$ hours across $19$ different activity classes (details in \sectionname~\ref{sec:pilot_study}) involving both macro and micro activities. We have experimented over different controlled, semi-controlled, and in-the-wild setups, as we explained later for individual evaluations. To generate the ground truth for localization and tracking pipeline of \ourmethod{}, we manually marked the positions of users' movement in the room's floor map. We asked the users to move in the marked path. We have evaluated the MAE in the marked coordinates and the denoised pointcloud coordinates. Further, we have used \textit{mmWave-Demo-Visualizer}~\cite{tiMmWaveDemo} tool, and implemented a patch to extract raw data, containing pointcloud and range-doppler heatmap under different radar configurations. Annotating the video footage captured via another USB camera installed in the room was done with the help of two volunteers.
\subsection{Baseline}
We compare \ourmethod{} activity classifier with three baselines, (i) \textbf{Pointcloud-based: RadHAR}~\cite{singh2019radhar}, which is based on voxelized 3D pointclouds for classifying six macro activities. For developing the baseline, we have collected the 3D pointclouds using a TI IWR1443ISK~\cite{iwr1443boost}, and we train the classifier (as provided in~\cite{singh2019radhar}). (ii) \textbf{Range-Doppler: Vid2Doppler}~\cite{ahuja2021vid2doppler}, which used range-doppler data to classify 12 different activities. With our collected datasets, we transfer-learn the model weights using the open-sourced Vid2Doppler classifier model. (iii) \textbf{VGG-16 network}~\cite{simonyan2014very}  which is pre-trained on the
ImageNet~\cite{deng2009imagenet} dataset, we apply transfer learning to learn new model weights w.r.t. our collected range-doppler matrix. This transfer learning approach helps in reducing the feature extraction part, as all the trained convolutional layers in VGG-16 are used as feature extractors and do not require retraining. The base VGG-16 model has been enhanced with 2D-Global Average Pooling and successive Dropout and Dense layers as done in the 2D-CNN Architecture (see \S\sectionname~\ref{sec_model}). The models are trained with a train-test split of $70\%$-$30\%$ and a validation split of $20\%$ from the training set.
\begin{figure*}
    \centering
        \subfloat[]{
	    \label{fig:diff_baseline_response_on_activity_switching}
	    \includegraphics[width=0.24\textwidth]{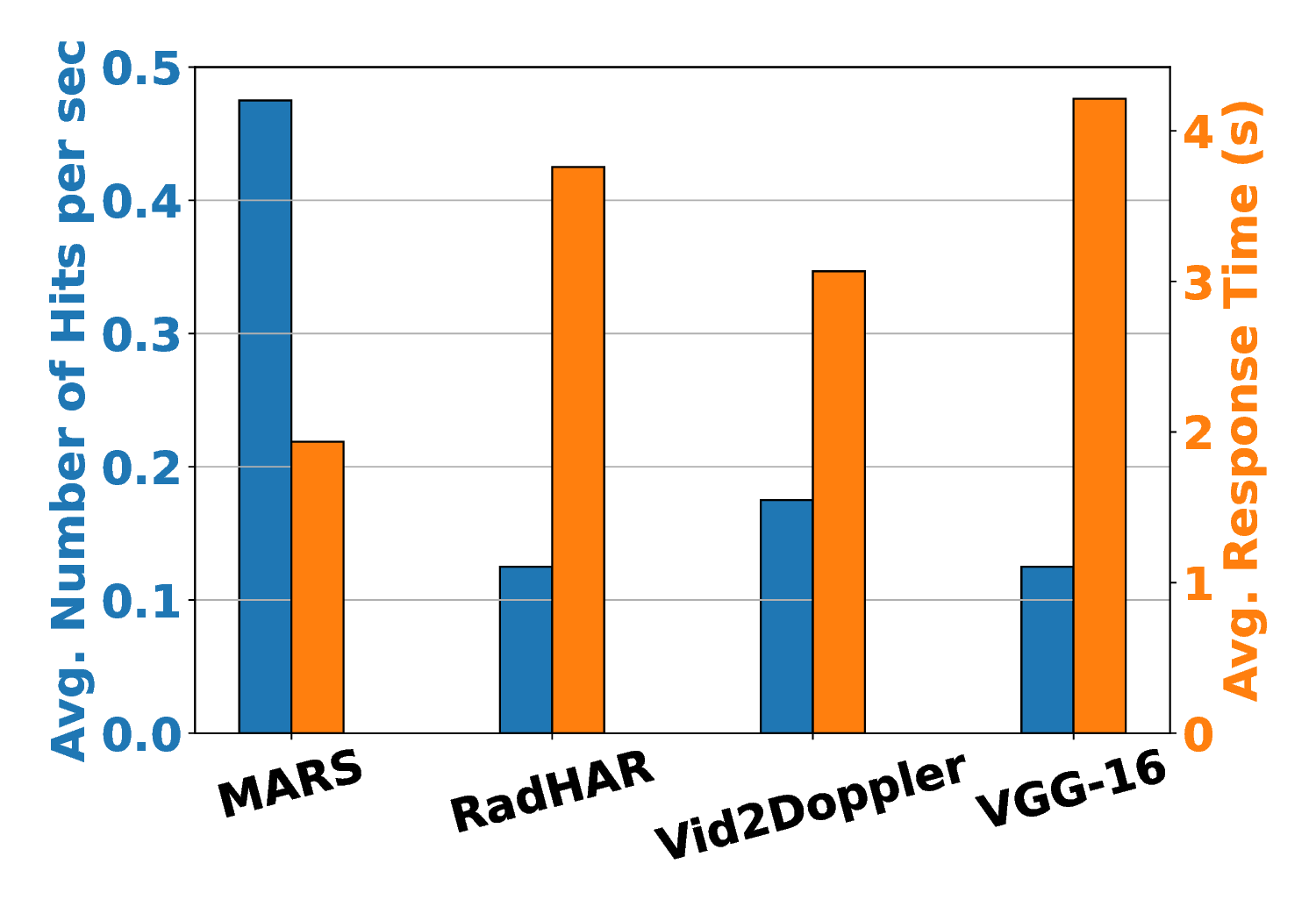}
	}
        \subfloat[]{
	    \label{fig:diff_baseline_response_on_activity_space}
	    \includegraphics[width=0.24\textwidth]{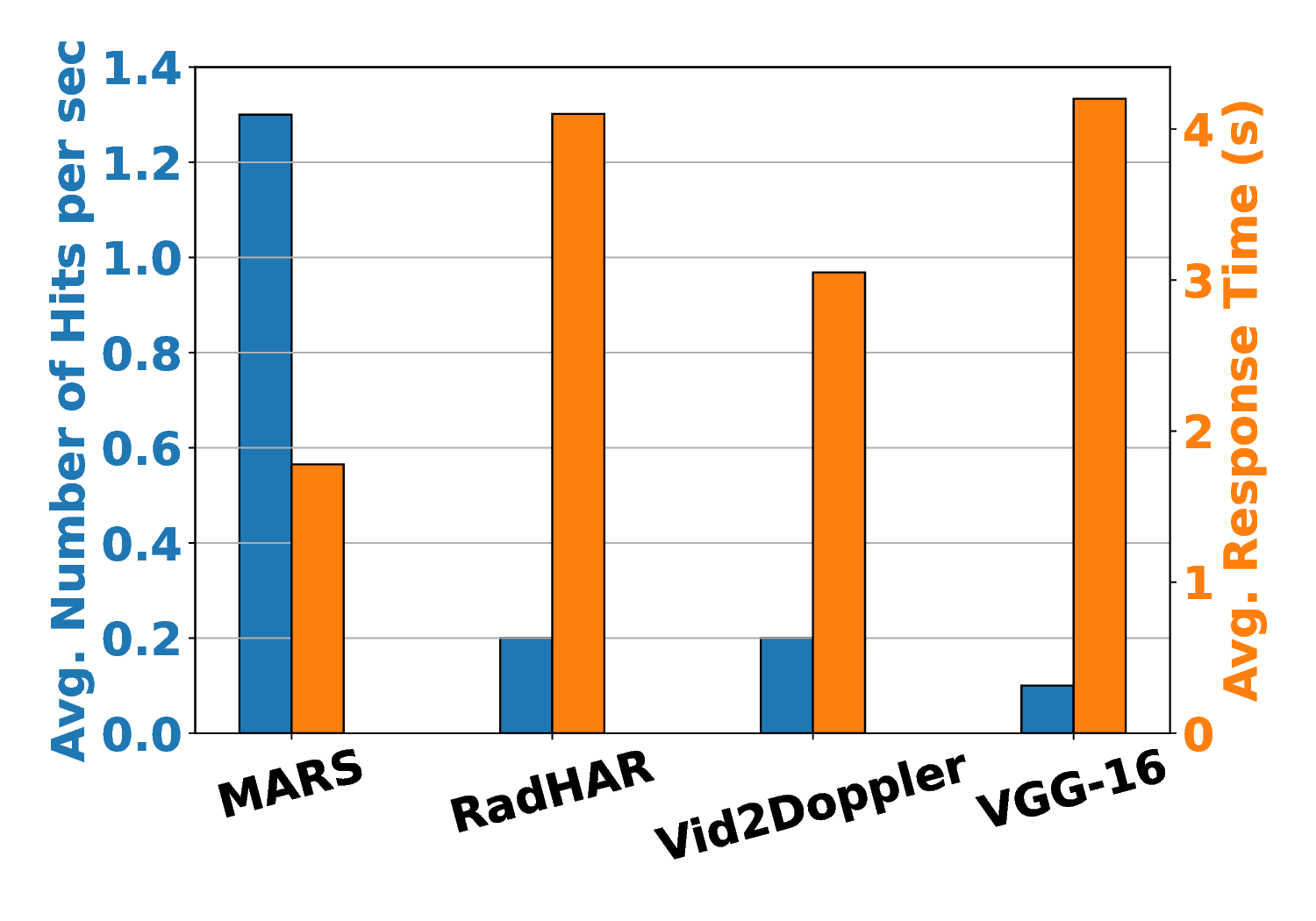}
	}
     \subfloat[]{
	    \label{fig:diff_baseline_response_on_activity_switching_space}
	    \includegraphics[width=0.24\textwidth]{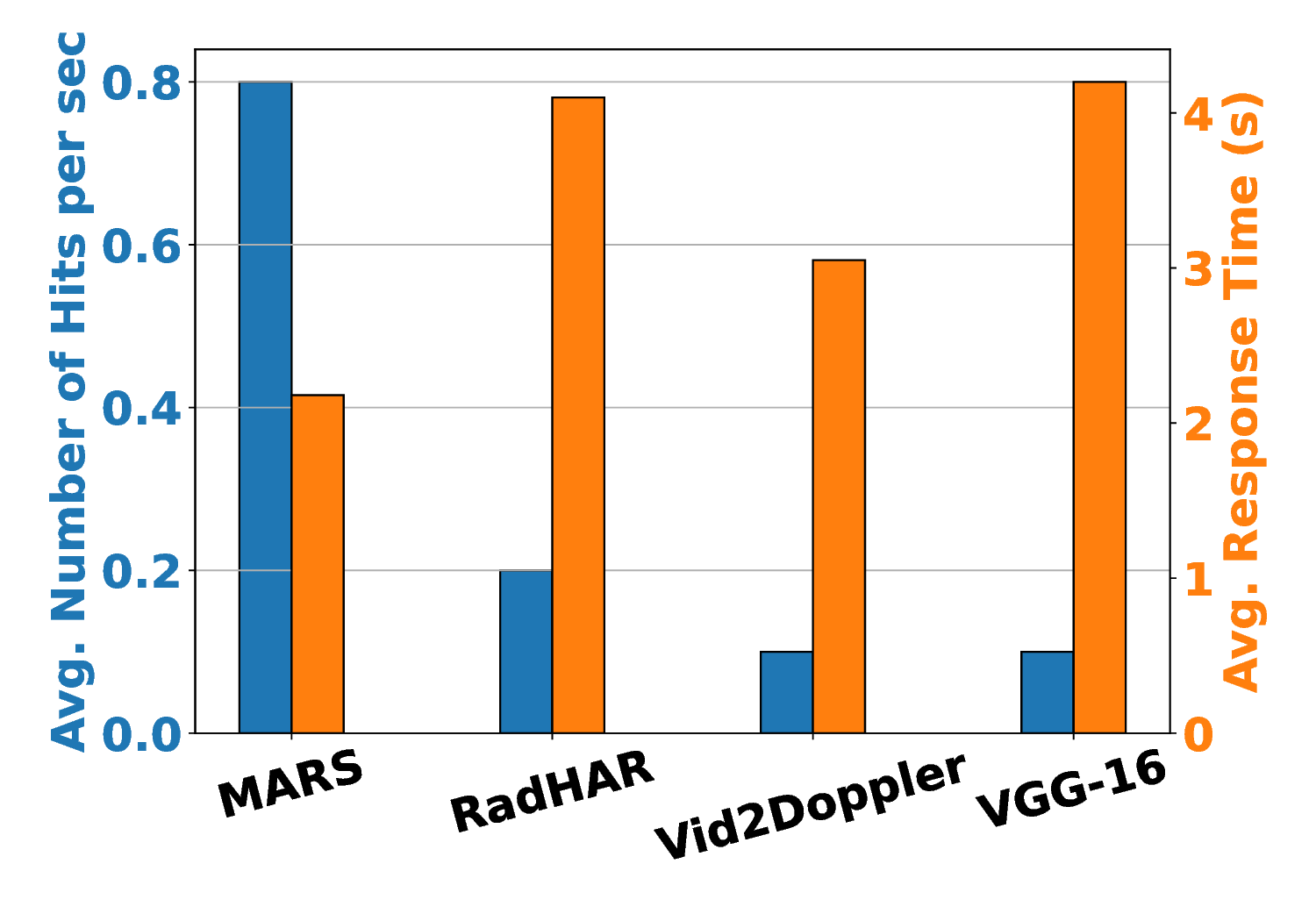}
	}
    \subfloat[]{
	    \label{fig:baseline_activity}
	    \includegraphics[width=0.24\textwidth]{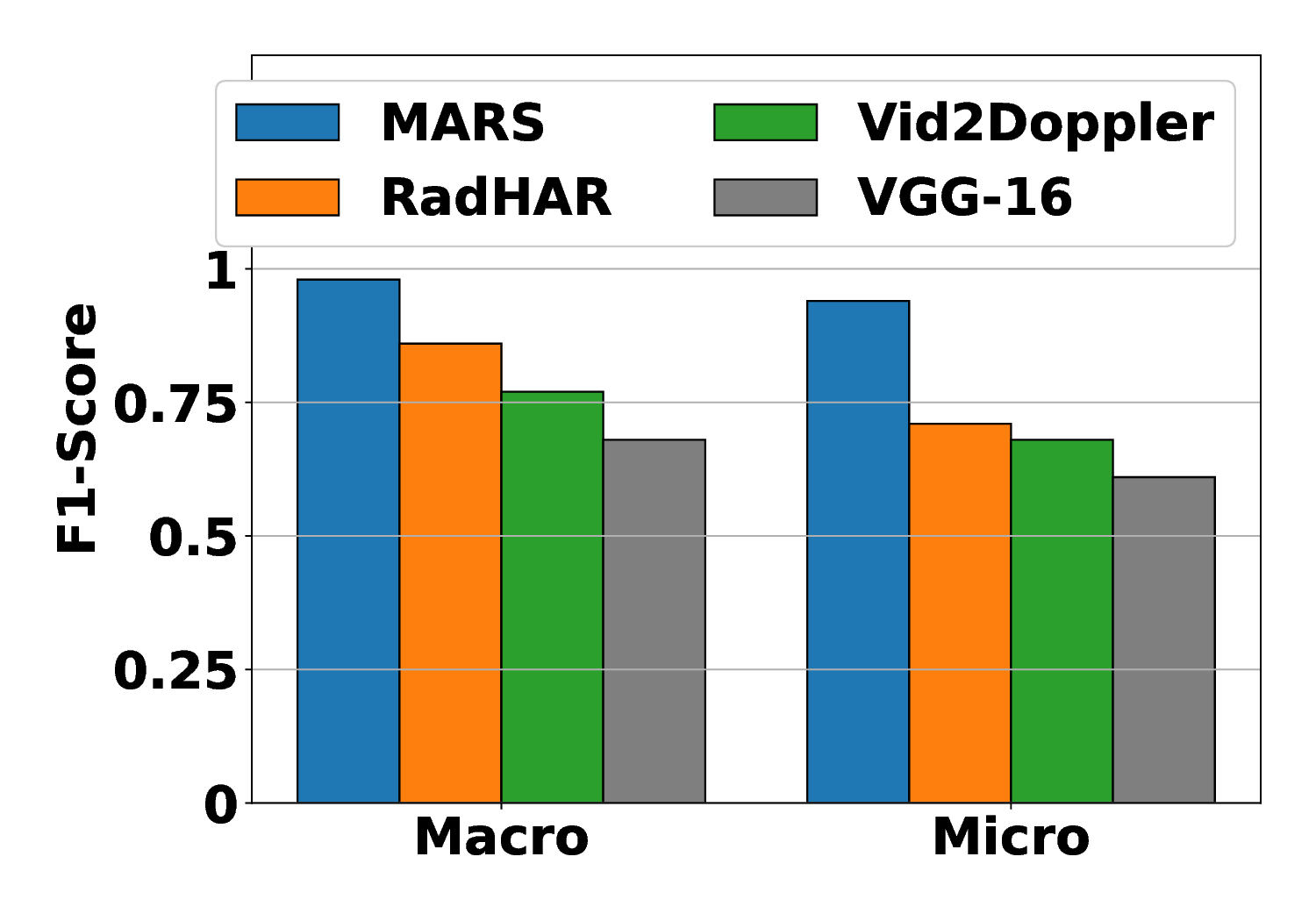}
	}\hfil
    \caption{Average number of hits and average response time while predicting (a) different activities over time (single-user), (b) multi-user, (c) different activities over time for multi-user, (d) overall weighted F1-Score of \ourmethod{}}
    \label{fig:res_time}
\end{figure*}
\section{Evaluation}
%We first evaluate the overall performance of \ourmethod{} on continuous activity monitoring of multiple subjects over time.
\subsection{Overall Performance}
We consider three different scenarios to evaluate the overall performance of \ourmethod{} in comparison to the baselines -- (i) single subject, multiple activities over time (\textit{Temporal activity diversity}), (ii) multiple subjects, individual subject performs a single activity over time but different subjects may perform different activities (\textit{Spatial activity diversity}), (iii) multiple subjects, each performing different activities over time (\textit{Spatio-temporal activity diversity}). We performed these experiments in a room $R2$; later, we discuss the impact of the room size of \ourmethod{} performance with a \textit{leave one out} train-test method. To evaluate the efficacy, we use two different metrics. We measure the \textit{average number of hits per second}, indicating the rate at which the individual methods are able to correctly report the activity being performed. As the activities are performed in a temporal sequence, this metric indicates the efficacy for real-time prediction of the activities. We also compute the \textit{average response time}, the amount of time a method needs to produce the first correct output after a subject starts performing an activity. 

\subsubsection{Impact of different activities over time}
Here, we asked the subjects to choose four activities (two macro and two micro) in a logical sequence and perform each for at least $10$ sec within a room. For example, a subject may first do some exercise through jumping (macro), then sits (micro), then take their phone and type a message (micro), and finally walk to leave the room (macro). As shown in \figurename~\subref*{fig:diff_baseline_response_on_activity_switching}, \ourmethod{} takes the least response time in inferring the activities with the highest number of hits per second in comparison to the baselines. We observe that the response time for the first activity takes $\approx 2$ sec, which involves the bootstrap time to denoise and cluster the data for localizing the subject. When a configuration switch is necessary (macro to micro or vice versa), the average response time is $\approx 3.14$ seconds. Without a configuration switch, the average response time is $\approx 1.08$ seconds. In comparison, the baselines perform worst in the response time due to more extended frame stacking ($2$ sec and $3$ sec, respectively, for RadHAR and Vid2Doppler) and longer classifier inference time ($\approx 4$ sec for VGG-16). Longer response times of the baselines directly impact lowering the number of hits in the activity time window, as shown in \figurename~\subref*{fig:diff_baseline_response_on_activity_switching} w.r.t. \ourmethod, which has a low response time due to smaller frame stacking ($1$ sec) and reduced inference time ($\approx 0.08$ sec) with a light-weight model architecture. 

\subsubsection{Impact of multi-user activities}
In the second scenario, we pick four subjects and ask three of them to choose one activity from the set of macro activities and the remaining one to choose one from the set of micro activities. After determining the subjects' location and states, \ourmethod{} configures the low doppler resolution and classifies the macro activities simultaneously with a response time of $\approx 3$s at the beginning. However, using $1$ sec of frame-stacked data, it can gradually infer the three macro activities simultaneously with a response time of $1.04$s. For the subject performing the micro activity, it switches the configuration to high doppler resolution and classifies the same with a response time of $2.08$ sec, resulting in an average response time of $1.9$ sec and on an average of $13$ hits in the entire activity time window of $10$ sec (see \figurename~\subref*{fig:diff_baseline_response_on_activity_space}). RadHAR and Vid2Doppler show poor performance as they are built focused on macro activities only and are trained only for single-user activity classification. Interestingly, we observe that the average number of hits per second for \ourmethod{} is more in this case (spatial diversity) compared to the previous one (temporal diversity), as the radar needs less configuration switching.

\subsubsection{Impact of different activity over time for multi-user} In the final scenario, we ask four subjects to simultaneously perform four different activities of their choices (with at least one micro activity and one macro activity) in sequence within a room, where they switch the activity in approximately every $10$ sec. As shown in \figurename~\subref*{fig:diff_baseline_response_on_activity_switching_space}, the average response time of \ourmethod{} in this scenario is $\approx 2$ sec with $8$ average number of hits, in a time window of $10$s. Thus the overall performance of \ourmethod{} demonstrates its potential to be adopted as a real-time system for multi-subject scenarios.  

% \begin{figure}[t]
%     \centering
%     \includegraphics[width=0.28\textwidth]{results/rf_classification.pdf}
%     \caption{Confusion matrix for the Opportunistic Classifier}
%     \label{fig:rf_cfm}
% \end{figure}

% \begin{figure}[t]
%     \centering
%     \subfloat[]{
% 	    \label{fig:baseline_activity}
% 	    \includegraphics[width=0.16\textwidth]{results/baseline_comp.eps}
% 	}\hfil%
%     \subfloat[]{%
% 		\label{fig:macro_cfm}%
% 		\includegraphics[width=0.16\textwidth]{results/macro_classification_1.pdf}}
% 		\hfil%
% 	\subfloat[]{%
% 		\label{fig:micro_cfm}%
% 		\includegraphics[width=0.16\textwidth]{results/micro_classification_1.pdf}}
%         \hfil
%         \subfloat[]{
% 	    \label{fig:diff_activity_macro_baseline}
% 	    \includegraphics[width=0.23\textwidth]{results/baseline_comp_diff_activity_macro.eps}
% 	}\hfil
%         \subfloat[]{
% 	    \label{fig:diff_activity_micro_baseline}
% 	    \includegraphics[width=0.23\textwidth]{results/baseline_comp_diff_activity_micro.eps}
% 	}
%     \caption{(a) Weighted-F1 Score of \ourmethod{} w.r.t. baselines; Confusion matrix for (b) macro, (c) micro classifiers.}
%     \label{fig:loc_track}
% \end{figure}

\begin{figure}
    \centering
    \includegraphics[width=0.40\textwidth]{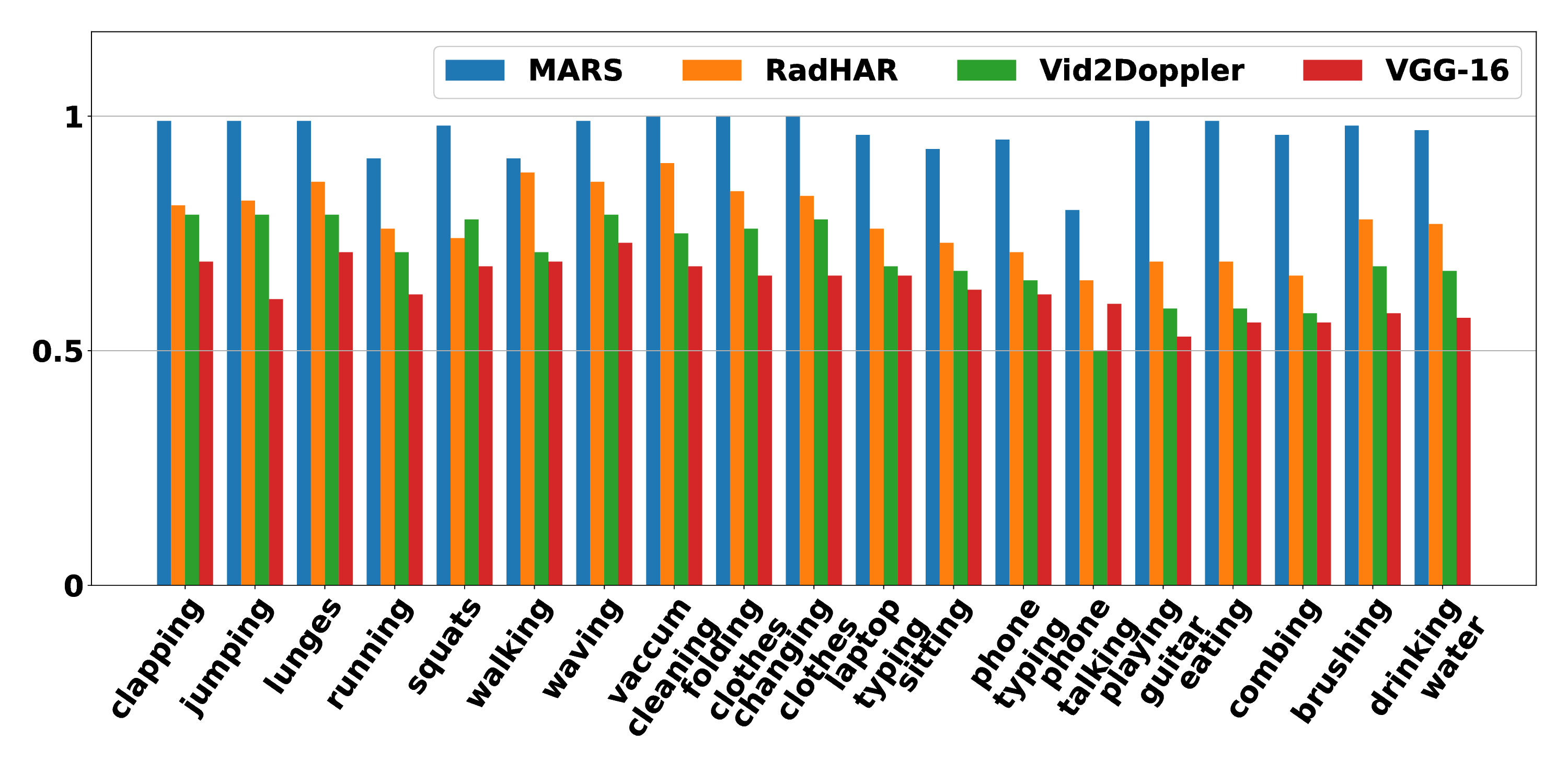}
    \caption{F1-Score across different activities}
    \label{fig:f1_macro_micro}
\end{figure}

Next, we evaluate individual components of \ourmethod{}, beginning with the Opportunistic Random Forest Classifier.   

\subsection{Performance of Opportunistic Classifier}
With the dataset collected across different scenarios, under the localization configuration (as mentioned in Table~\ref{tab:radar_conf}), we first perform a train-test split of $70\%:30\%$. The Opportunistic Classifier (as discussed in Sec.~\ref{ref:rf_classifier}) is trained with the $70\%$ training dataset with a validation split of $20\%$ from the training set. According to our observations, the pointcloud dataset can accurately classify macro and micro activity sets with $90\%$, and $99\%$ accuracy, respectively. However, a slight overlap exists (of $10\%$) between the macro activity class and the \textit{walking} or \textit{running} class, as under both the scenarios, there exists a significant variation in the pointcloud data. Interestingly, this variation is similar during the activity initiation period, and gradually, the variation becomes more separable with time. 

\begin{figure}[t]
    \centering
	\subfloat[]{
	    \label{fig:diff_user_number_activity}
	    \includegraphics[width=0.4\linewidth]{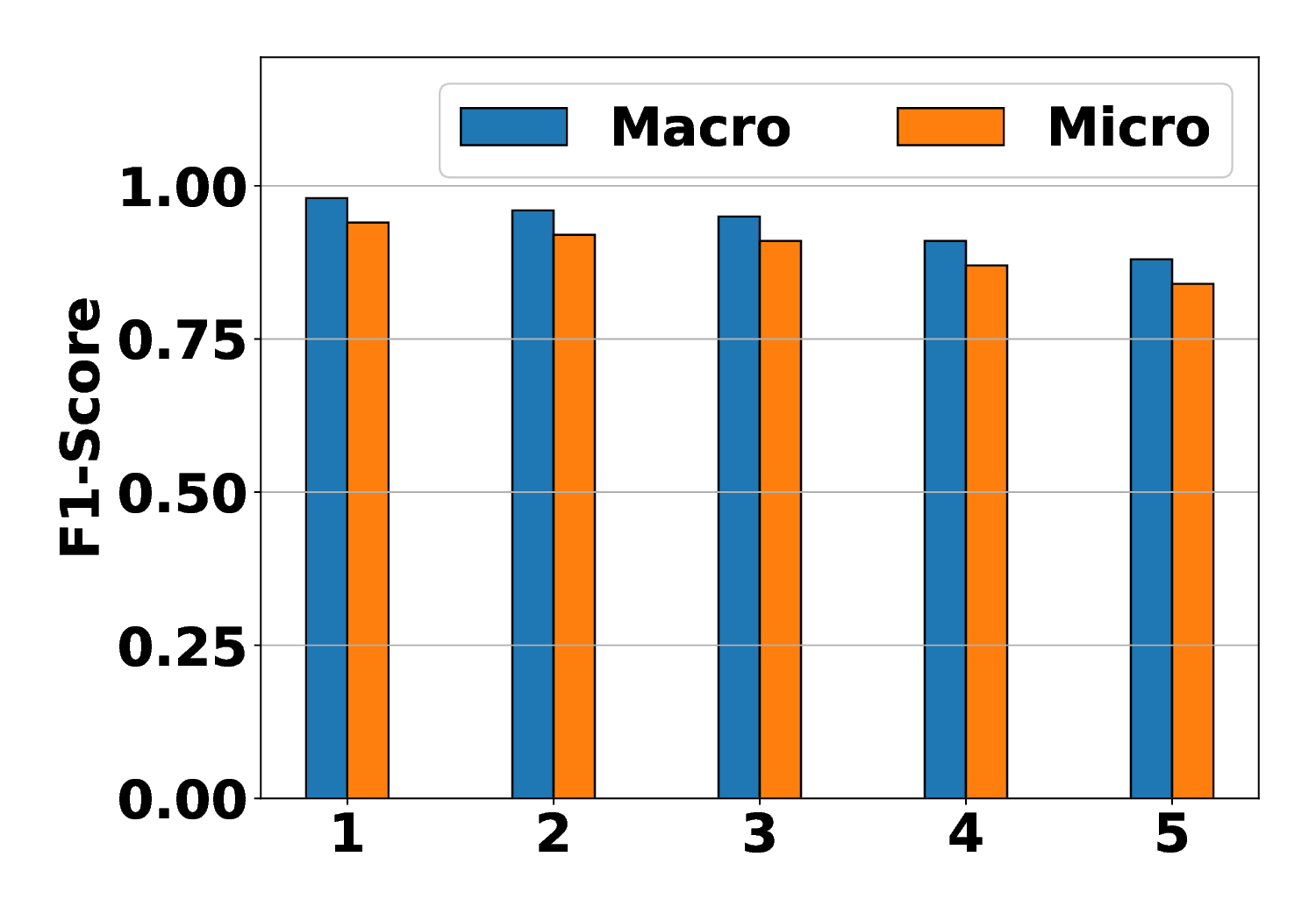}
	}\hfil
        \subfloat[]{
	    \label{fig:room_f1}
	    \includegraphics[width=0.4\linewidth]{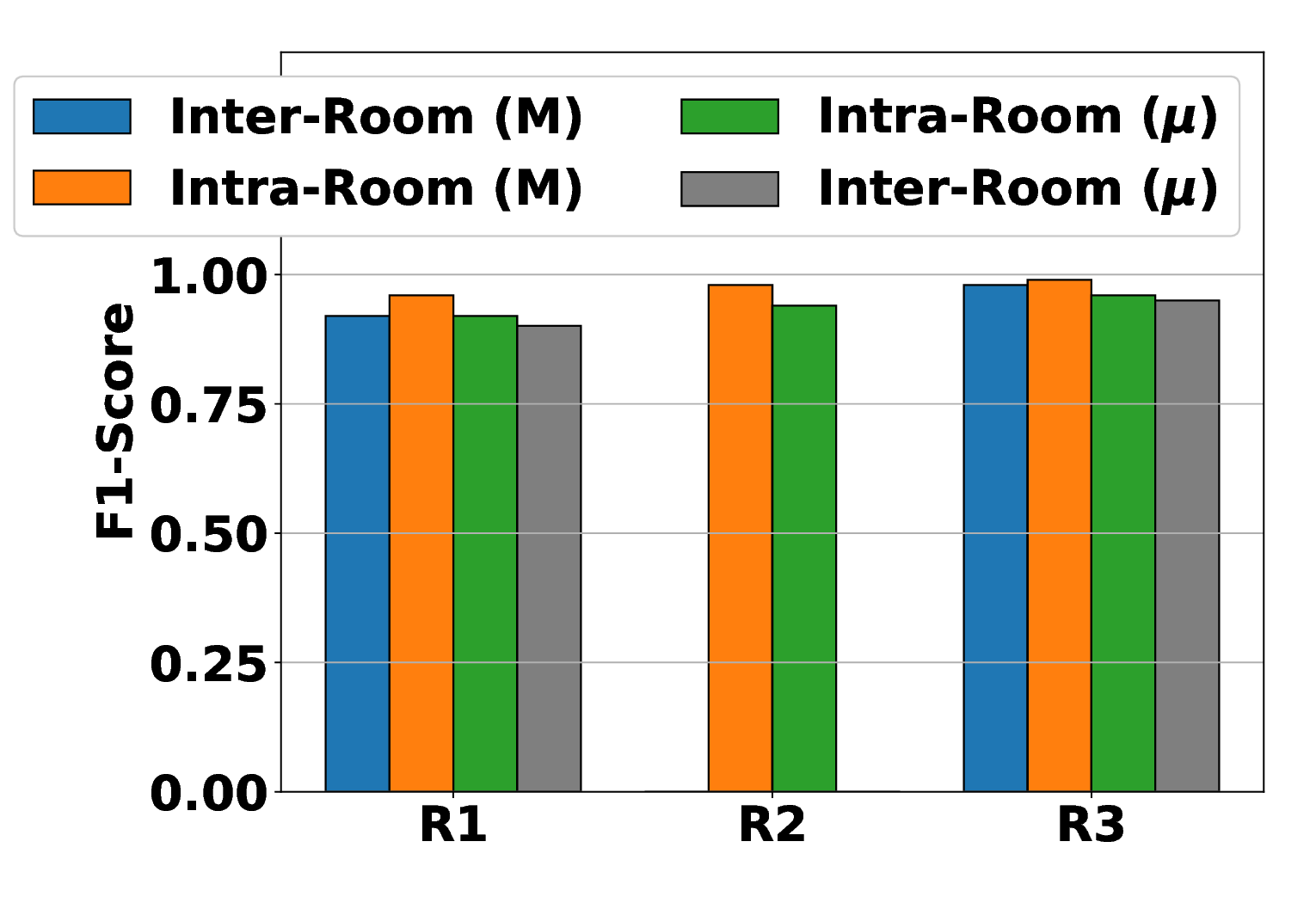}
	}\hfil
        \subfloat[]{
	    \label{fig:blockage_f1}
	    \includegraphics[width=0.4\linewidth]{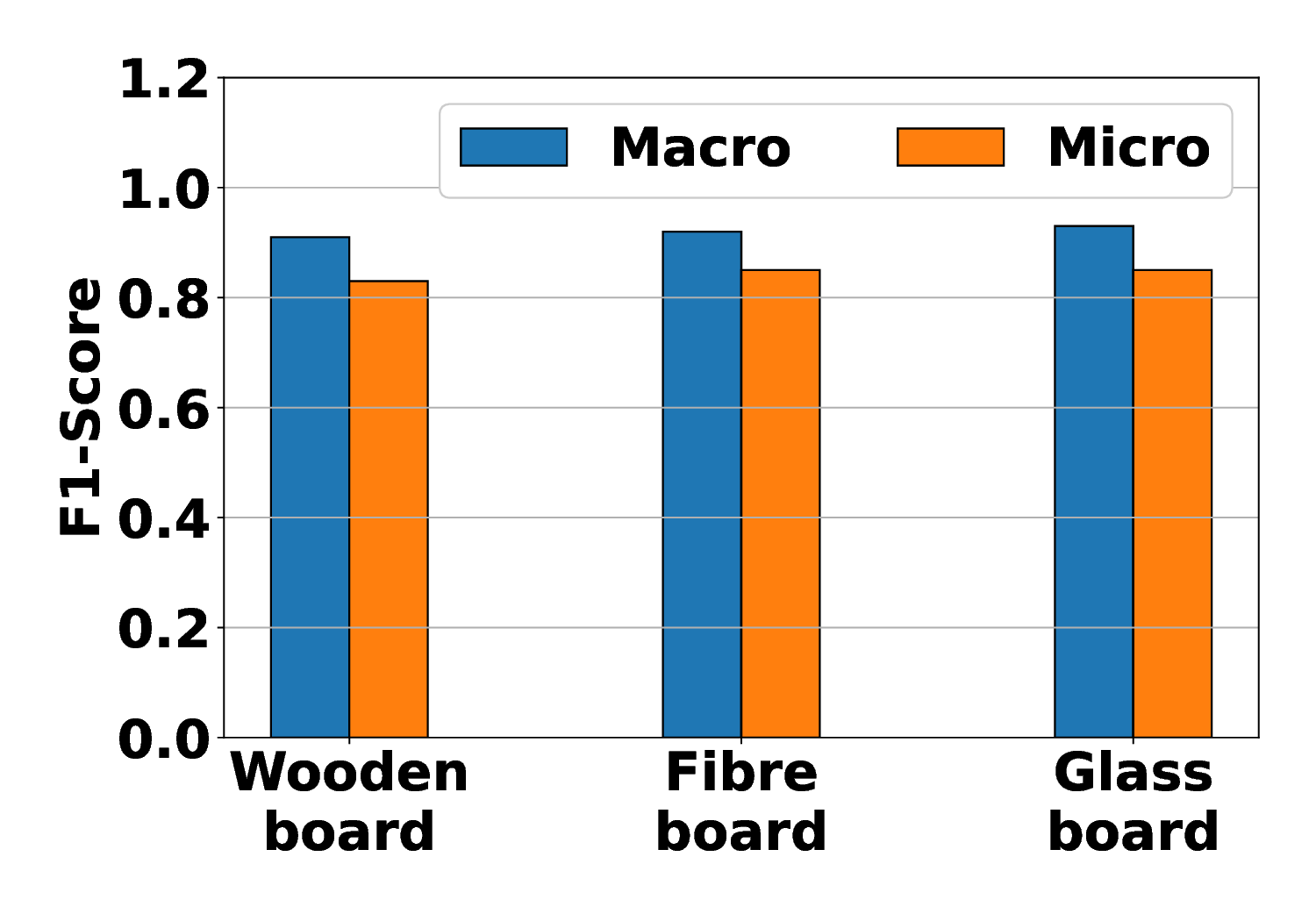}
	}\hfil
        
    \caption{Impact on (a) $\#$ of subjects, (b) different rooms (M: macro, $\mu$: micro), (c) different blockages}
    \label{fig:loc_track}
\end{figure}

\begin{figure*}[t]%
	\centering
	\subfloat[Macro classifier]{%
		\includegraphics[width=0.23\textwidth]{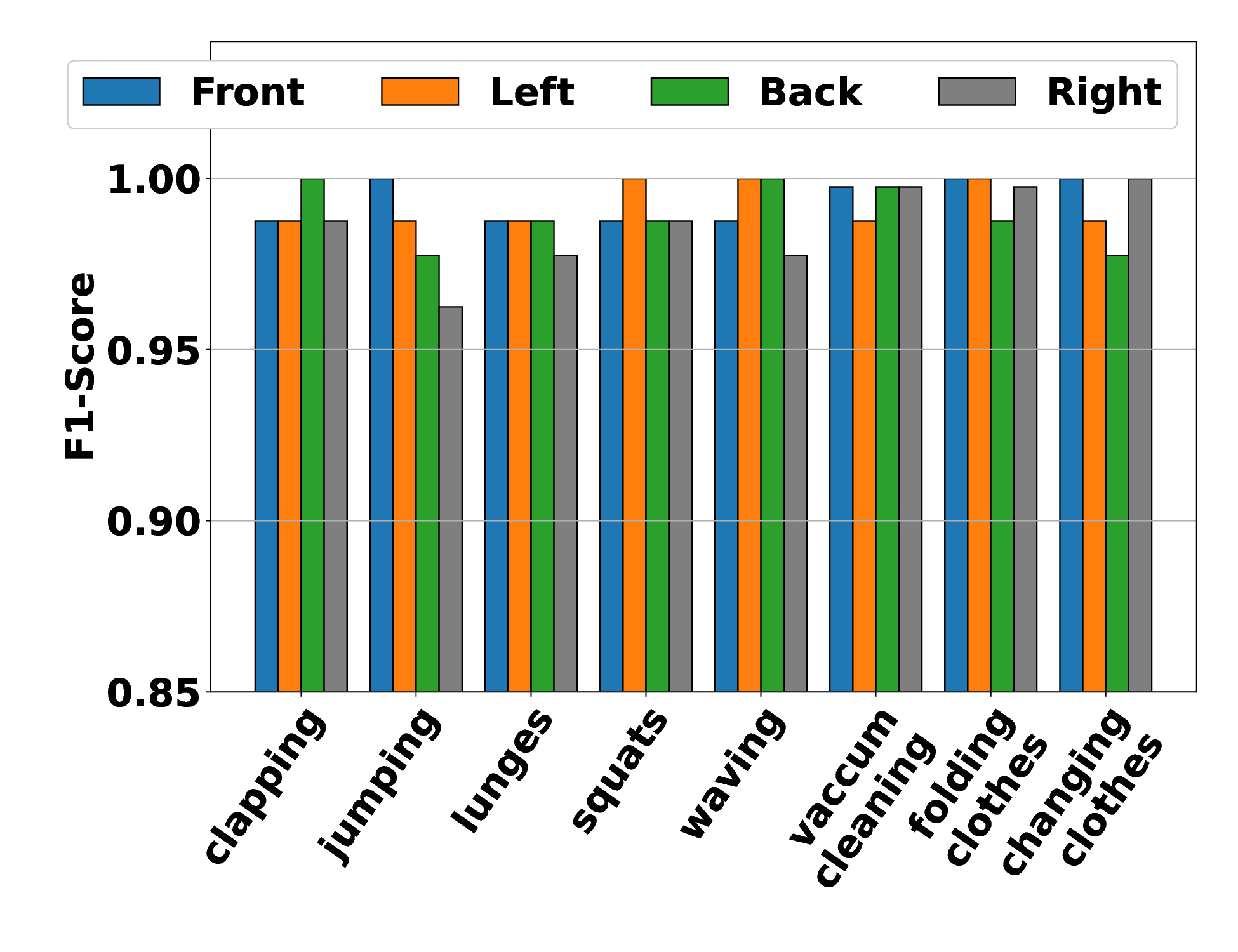}\label{fig:diff_orient_macro}}
		\hfil%
	\subfloat[Micro Classifier]{%
		\includegraphics[width=0.23\textwidth]{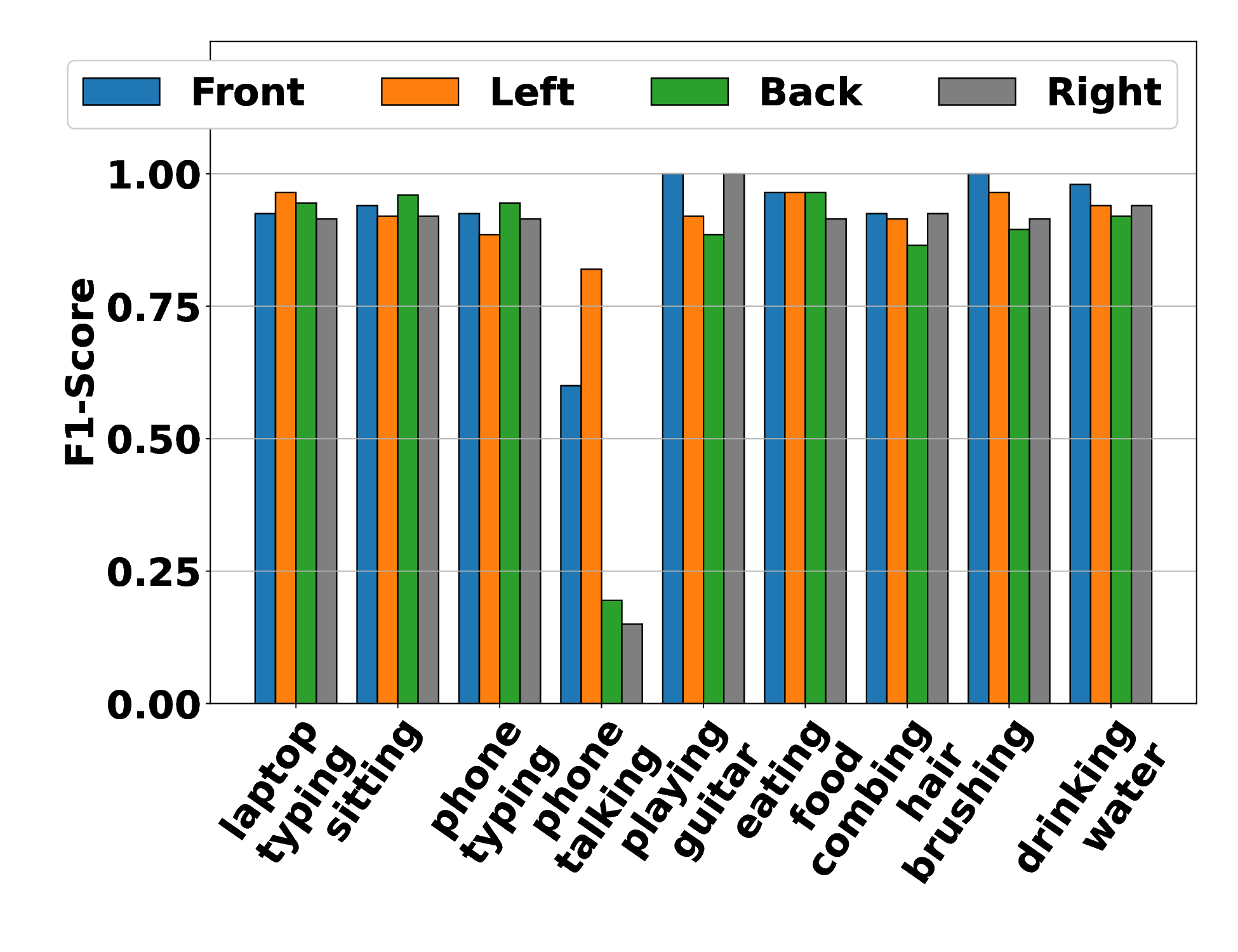}\label{fig:diff_orient_micro}}
		\hfil%
	\subfloat[Macro classifier (wrt distance)]{%
		\includegraphics[width=0.23\textwidth]{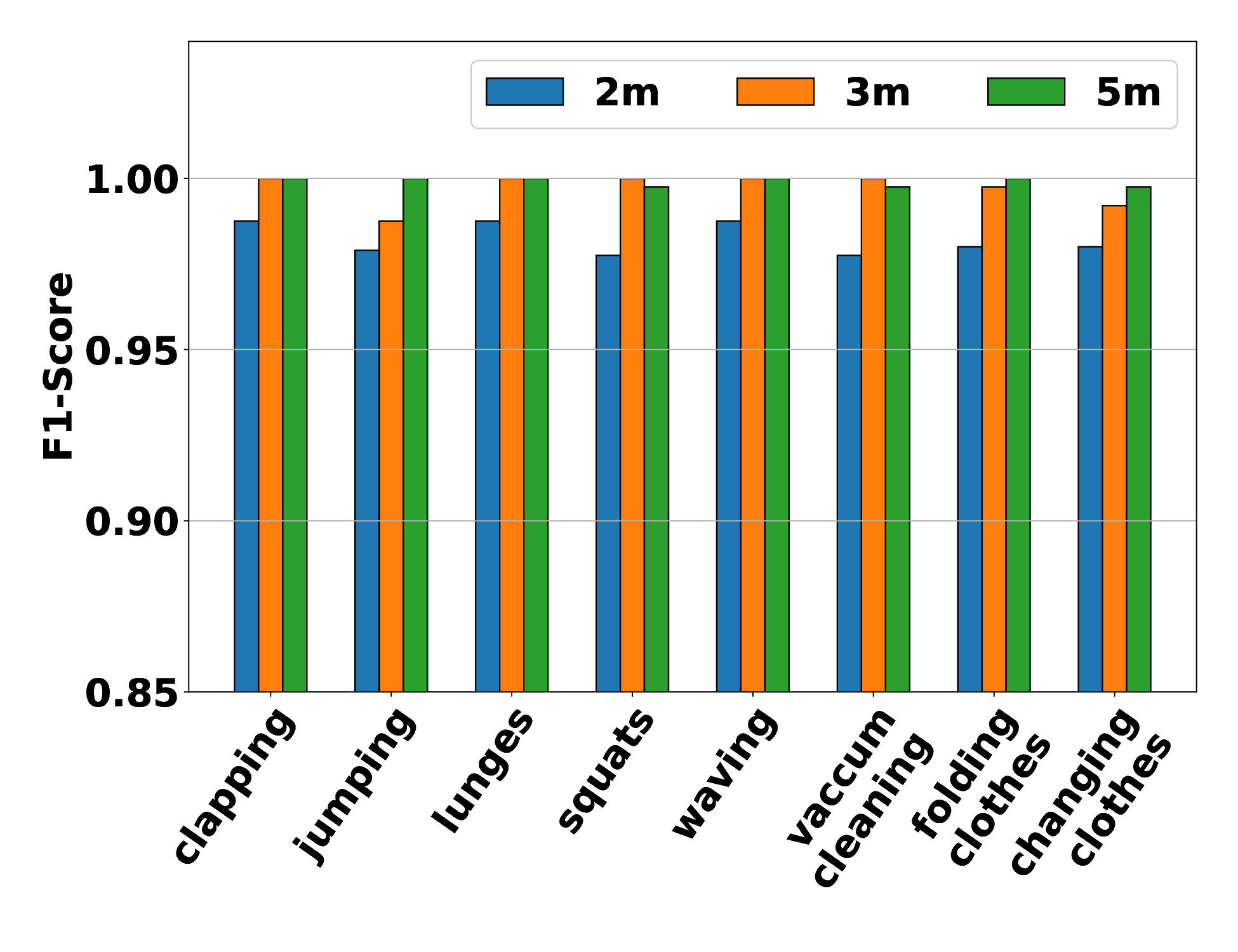}\label{fig:diff_dist_macro}}\hfil%
        \subfloat[Micro Classifier (wrt distance)]{%
		\includegraphics[width=0.23\textwidth]{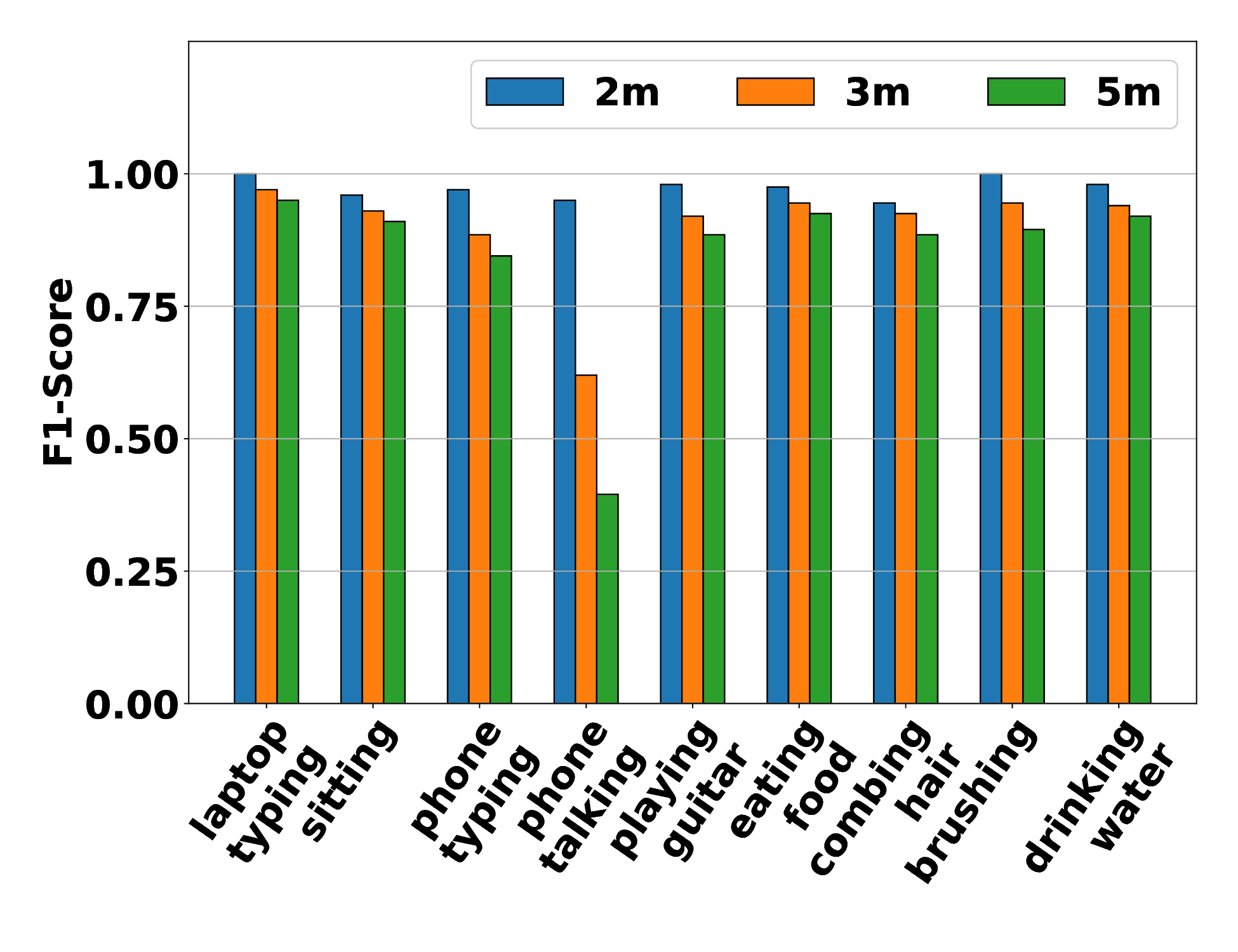}\label{fig:diff_dist_micro}}
	\caption{Weighted F1-Score at different orientations}
	\label{fig:micro_benchmark}
\end{figure*}

\begin{figure}[t]%
	\centering
	\subfloat[]{
    \includegraphics[width=0.23\textwidth]{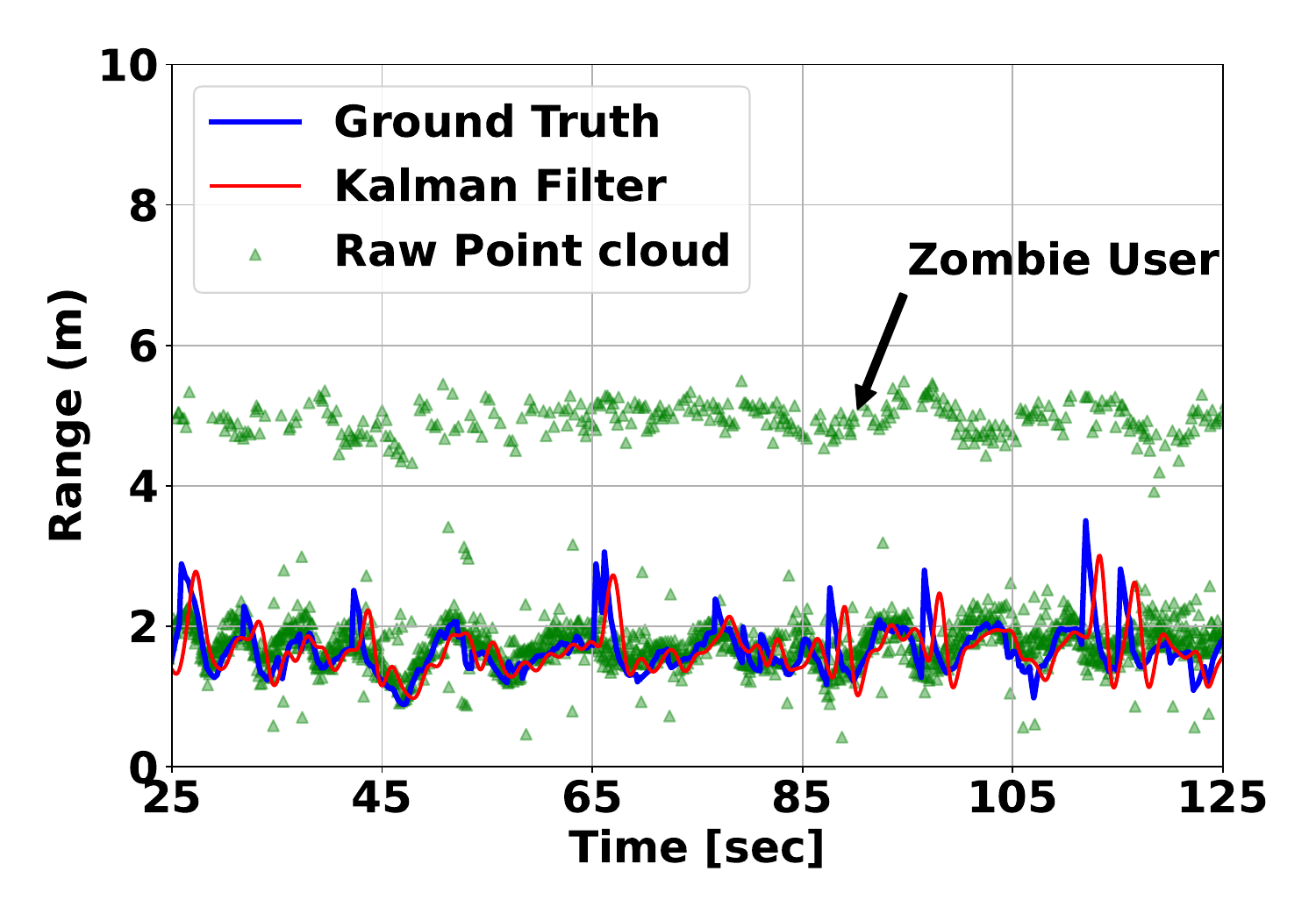}\label{fig:track_u_cls}}
    \subfloat[]{
    \includegraphics[width=0.23\textwidth]{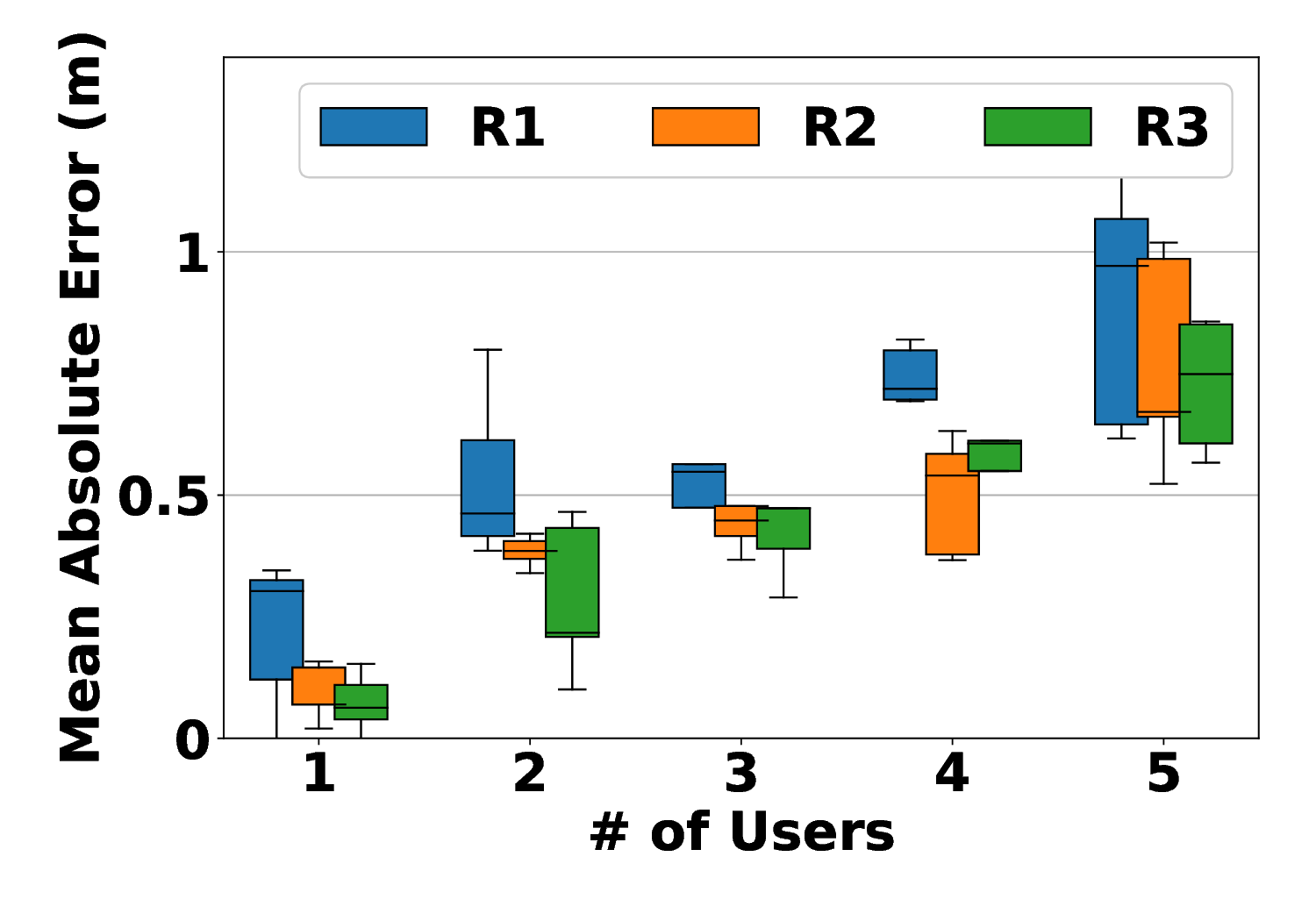}\label{fig:box_mae}
    }
	\caption{(a) Qualitative analysis of Kalman tracking pipeline, (b) MAE with different walking subjects}
	\label{fig:cfm}
\end{figure}

\subsection{Performance of Activity Classifiers} 
We next evaluate the performance of the macro and the micro activity classifiers w.r.t. the baseline in terms of the weighted F1-Score. As shown in \figurename~\subref*{fig:baseline_activity}, the weighted F1-Score for \ourmethod{} is $98\%$ in the case of macro activities and $94\%$ in the case of micro activities. The lower weighted F1-Score for RadHAR is primarily because it relies on the pointcloud dataset for the voxel formation and generates sparse pointclouds in case of micro activities. For Vid2Doppler, the poor accuracy is primarily because it takes only 32 doppler bins which are unsuitable for micro activity monitoring, and the model feature extraction part is pre-trained on macro activity datasets. As the body movements in the case of macro activities are significant, thus the classifier can segregate individual classes with an excellent weighted F1-Score (close to $\approx 0.98$). In the case of the micro activities, the body movements are less significant, but with the proposed classification pipeline with a higher doppler resolution, we can achieve a weighted F1-Score of $0.94$. Among the micro activities, laptop typing, eating food, and playing guitar involve higher body movements, and thus for these particular activities, we observe higher F1-Score (see \figurename~\ref{fig:f1_macro_micro}). Activities such as sitting, typing, and talking on a phone are carried out while subjects sit on a chair. Thus, the doppler shift for these activities is very low, and minimal variation exists compared to other micro activities. When the subject talks on a phone, the overlap with the \textit{sitting} class is more significant ($\approx10\%$). In~\figurename~\ref{fig:f1_macro_micro} we show activity wise F1-Score of \ourmethod{} w.r.t. the baselines. In \cite{singh2019radhar} authors have considered only five activities, and for Vid2Doppler~\cite{ahuja2021vid2doppler} 12 activities are considered, while in our case with a total of 19 activities, the classification F1-Score of the baselines significantly drops in comparison to \ourmethod{}. We look deeper at the activity classification pipeline of \ourmethod{} with the micro benchmarks detailed below.

\subsection{Results: Micro Benchmarks}
\subsubsection{Impact of number of subjects} In \figurename~\subref*{fig:diff_user_number_activity}, we show the variation in the weighted F1-Score for both the macro and micro activity classifiers with the different number of subjects present inside the room. With an increase in the number of subjects up to $5$, we observe a direct impact on the weighted F1-Score for both macro and micro classifiers, but by only $10$\%. 

\subsubsection{Impact on different rooms} We studied \ourmethod{} extensively in the three rooms R1, R2, R3 as discussed in Sec.~\ref{sec:impl}. We consider two cases -- (i) \textit{Inter-room Training}, where we train \ourmethod{} over the data collected at R2 and test over R1 and R3, and (ii) \textit{Intra-room Training}, where we train and test the models using the data collected over the same room. As shown in the \figurename~\subref*{fig:room_f1}, the weighted-F1 Score for all the rooms is $> 90\%$ for both the classifiers. Interestingly, for R1, the multipath-reflection effect is more significant due to the smaller room size, so the F1-Score is lower; however, the impact of intra-room and inter-room is not very significant as \ourmethod{} learns the features related to the doppler patterns of the subject's activity rather than room-specific features. Inter-room training and testing (leave one out) also indicate that the model does not \textbf{overfit} on datasets.

\subsubsection{Impact of blockages} We have tested \ourmethod{} with different blockages such as (i) Wooden board, (ii) Fibreboard, (iii) Glass board, etc. However, mmWave at higher frequency shows higher penetration loss, although it can penetrate materials like plywood, glass, and fiber. Thus for macro activities, at least, we can achieve weighted-F1 Score $>90\%$ (see \figurename~\subref*{fig:blockage_f1}). However, the F1-Score for micro activities is lower, as small phase variations are attenuated more quickly than macro phase variations.
\subsubsection{Impact of subject orientation} We test \ourmethod{} under different body orientations of the subject, i.e., (i) front, (ii) left, (iii) right, (iv) back, w.r.t. the setup. As observed in \figurename~\subref*{fig:diff_orient_macro} under different body orientations, the macro activity classifier can recognize the activity class with a weighted F1-Score of $\approx 0.98$. But in the case of the micro activity classifier (shown in \figurename~\subref*{fig:diff_orient_micro}, the weighted F1-Score is lower, especially for the case of phone talking. During talking on the phone, the subject's body orientation, such as back or right (while holding the phone in left hand) concerning the radar, significantly impacts the activity classification due to the occlusion of small-scale body movements. Nevertheless, it is comforting to see that for other micro-activities, the weighted F1-score is always $>0.80$, even when the subject is at a complete opposite orientation from the radar.      
\subsubsection{Impact of distance} \figurename~\subref*{fig:diff_dist_macro} shows the variation in the weighted F1-Score for the macro activity classifier under different distances from the subject. The classification is reported for up to a distance of $5$m. The average weighted F1-Score is $\approx 0.98$ as observed. However, \figurename~\subref*{fig:diff_dist_micro} indicates that the F1-Score for micro activity classifier sometime drops (up to $10\%$) with the increase in the distance. Due to signal attenuation, the doppler shift for micro activities sometimes becomes undetectable with increasing distance. Interestingly, the macro activities provide better results with higher distance (>3m, as shown in \figurename~\ref{fig:diff_dist_macro}), since the spread of macro body movements can be easily captured within radar's conical FoV with the increased distance. 
\subsection{Localization and Tracking Performance}
\figurename~\subref*{fig:track_u_cls} shows a snapshot of the subject tracking pipeline with respect to the ground truth distance. In this selected experiment, we observe the raw pointcloud data for the subject cluster contains the signature of a zombie subject arising due to multi-path reflections (present at a distance of $\approx 5$m). With the proposed localization approach, this zombie subject's pointcloud gets suppressed. In \figurename~\subref*{fig:box_mae}, we show the mean absolute error (MAE) in subject localization in three different rooms R1, R2, R3, w.r.t. the ground truth under different numbers of subjects present inside the room, who are walking simultaneously within the FoV of the radar at an average speed of $0.7$-$1.1$ m/s. Although the MAE in the localization is $<60$ cm with three simultaneous subjects, the MAE gradually increases with increasing the number of subjects walking simultaneously. Since R1 is smaller, the MAE is higher because the pointcloud data is noisier with more multi-path reflections. However, it is comforting to see that even with the five users walking randomly at the normal to moderately high speed, the MAE is within $\approx1m$.

\subsection{Resource and Energy Benchmarks}
\begin{figure}[t]%
	\centering
	\subfloat[CPU]{%
		\includegraphics[width=0.16\textwidth]{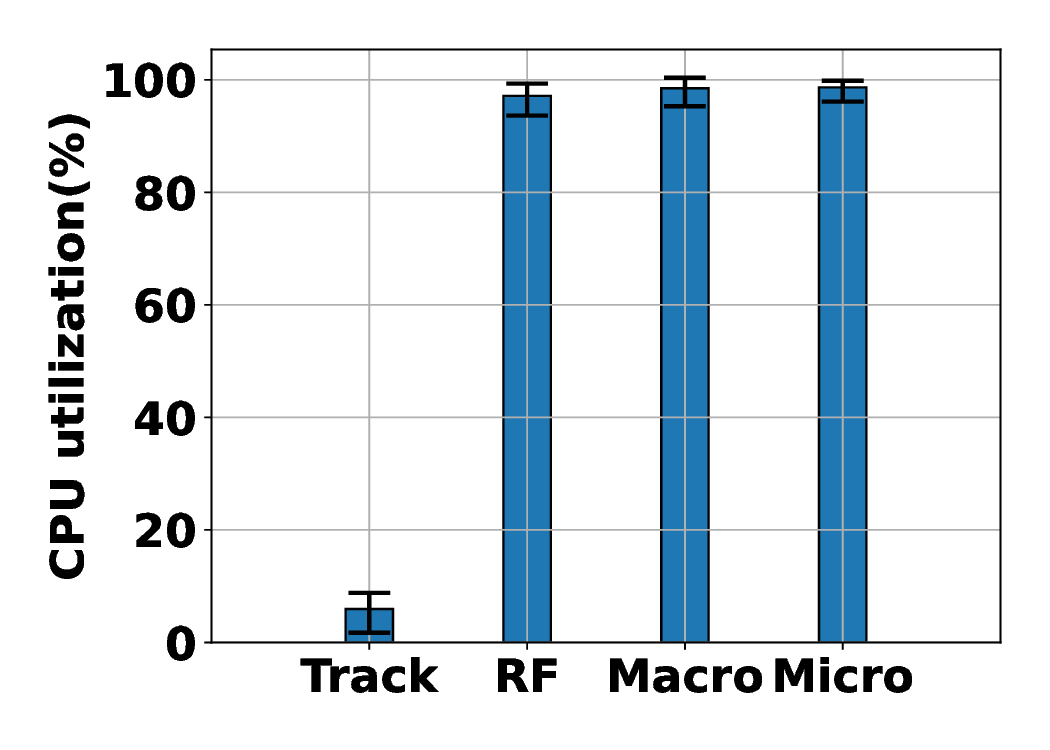}\label{fig:cpu}}
	\subfloat[Memory]{%
		\includegraphics[width=0.16\textwidth]{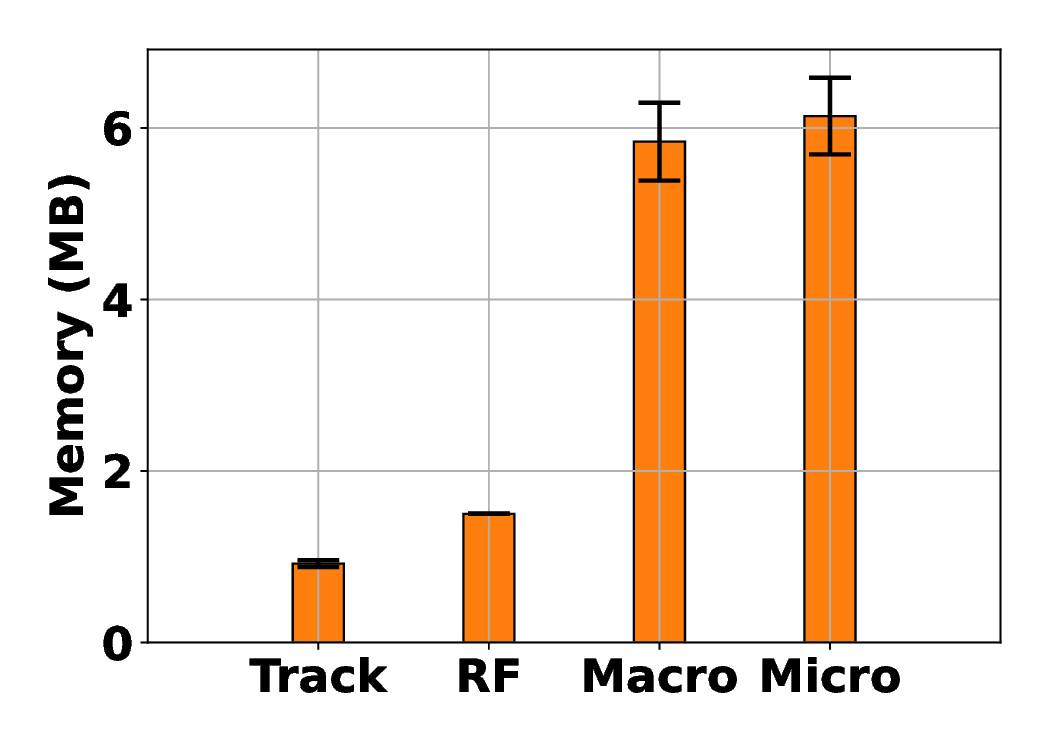}\label{fig:mem}}
	\subfloat[Energy]{%
		\includegraphics[width=0.16\textwidth]{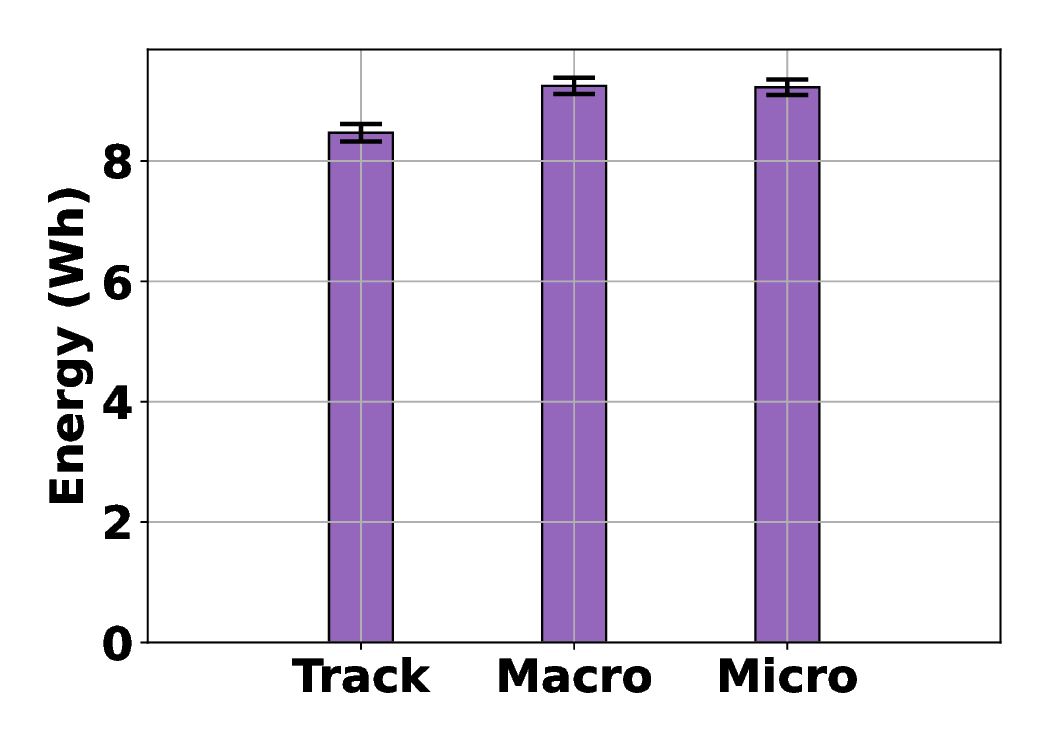}\label{fig:power}}
	% \subfloat[]{%
	% 	\includegraphics[height=2.5cm, width=0.23\textwidth]{results/temp.eps}\label{fig:temp}}\hfil%
	\caption{Resource consumption of \ourmethod{}}
	\label{fig:usage}
\end{figure}
We measure the resource and energy consumption of the back-end processing unit, i.e., RPi-4, under different scenarios. As observed in \figurename~\subref*{fig:cpu},~\subref*{fig:mem}, the CPU and the memory utilization in case of the localization and the tracking pipeline is low when the subject is not present inside the room. As the subject enters the room, the Opportunistic classifier gets initiated. As a result of feature computation and pipeline initiation, we observe significantly higher CPU and memory utilization. After that, when the activity classification pipeline gets initiated, we observe that memory utilization increases significantly due to large-scale feature computation and loading of the trained model in the memory. Finally, using a Monsoon Power Monitor~\cite{msoonHighVoltage}, we measure the overall energy consumption of the RPi under the three scenarios -- (i) localization and tracking, (ii) macro activity classification, (iii) micro activity classification. As observed in \figurename~\subref*{fig:power}, the energy consumption is comparatively higher in the case of macro and the micro activity classification because of the higher CPU and RAM utilization. 
% However, it is comforting to observe that the CPU temperature is within safe limits across all the steps of the \ourmethod{} pipeline (See \figurename~\subref*{fig:temp}).
\section{Related Work}
Some studies have used wearables for active sensing techniques, mostly to detect human activity~\cite{sandhu2021solar,lawal2019deep,casale2011human,fujimoto2013wearable}. Even though such methods are useful, they are not seamless and pervasive enough. While vision-based activity assessments such as~\cite{han2005human,harville2004fast,cheng2005multiperspective,caba2015activitynet, mukherjee2020human} rely primarily on video data, these methods have operational limitations due to LoS issues, occlusions, and privacy concerns. Our next discussion explores alternative passive sensing methods, mainly those based on acoustics and radio frequency.

\begin{table}[t]
\scriptsize
\caption{Comparison of the state-of-the-art systems}
\label{tab:comparison}
\resizebox{\columnwidth}{!}{%
\begin{tabular}{|l|c|c|c|c|c|}
\hline
\multicolumn{1}{|c|}{Title} &  \begin{tabular}[c]{@{}c@{}}Macro\\ Activities\end{tabular} & \begin{tabular}[c]{@{}c@{}}Micro \\ Activities\end{tabular} & \begin{tabular}[c]{@{}c@{}}Continuous \\ Monitoring\end{tabular} & \begin{tabular}[c]{@{}c@{}}Real-time\\ Inference\end{tabular} & \begin{tabular}[c]{@{}c@{}}Multi-Person\\ Monitoring\end{tabular} \\ \hline
IMU2Doppler~\cite{bhalla2021imu2doppler}                                                   & \textbf{$\checkmark$}                                                 & \textbf{$\times$}                                                  & \textbf{$\checkmark$}                                                       & \textbf{$\times$}                                                    & \textbf{$\times$}                                                        \\ \hline
Mobi2Sense~\cite{zhang2022mobi2sense}             & \textbf{$\checkmark$}                                                 & \textbf{$\checkmark$}                                                  & \textbf{$\times$}                                                       & \textbf{$\checkmark$}                                                    & \textbf{$\checkmark$}                                                        \\ \hline
RF-Action~\cite{li2019making}                      & \textbf{$\checkmark$}                                                 & \textbf{$\checkmark$}                                                  & \textbf{$\times$}                                                       & \textbf{$\checkmark$}                                                    & \textbf{$\checkmark$}                                                        \\ \hline
RF-Net~\cite{ding2020rf}                         & \textbf{$\checkmark$}                                                 & \textbf{$\times$}                                                  & \textbf{$\times$}                                                       & \textbf{$\times$}                                                    & \textbf{$\times$}                                                        \\ \hline
RF-Pose~\cite{zhao2018through}                        & \textbf{$\checkmark$}                                                 & \textbf{$\times$}                                                  & \textbf{$\checkmark$}                                                       & \textbf{$\times$}                                                    & \textbf{$\checkmark$}                                                        \\ \hline
Vibrosight~\cite{zhang2018vibrosight}                 & \textbf{$\checkmark$}                                                 & \textbf{$\times$}                                                  & \textbf{$\times$}                                                       & \textbf{$\times$}                                                    & \textbf{$\times$}                                                        \\ \hline
Vid2Doppler~\cite{ahuja2021vid2doppler}                                    & \textbf{$\checkmark$}                                                 & \textbf{$\times$}                                                  & \textbf{$\checkmark$}                                                       & \textbf{$\checkmark$}                                                    & \textbf{$\times$}                                                        \\ \hline
RadHAR~\cite{singh2019radhar}                         & \textbf{$\checkmark$}                                                 & \textbf{$\times$}                                                  & \textbf{$\checkmark$}                                                       & \textbf{$\times$}                                                    & \textbf{$\times$}                                                        \\ \hline
m-activity~\cite{wang2021m}                     & \textbf{$\checkmark$}                                                 & \textbf{$\times$}                                                  & \textbf{$\times$}                                                       & \textbf{$\times$}                                                    & \textbf{$\times$}                                                        \\ \hline
RF-Diary~\cite{fan2020home}                   & \textbf{$\checkmark$}                                                 & \textbf{$\times$}                                                  & \textbf{$\checkmark$}                                                       & \textbf{$\times$}                                                    & \textbf{$\times$}                                                        \\ \hline
Jiang et. al.~\cite{jiang2018towards}                                      & \textbf{$\checkmark$}                                                 & \textbf{$\times$}                                                  & \textbf{$\times$}                                                       & \textbf{$\times$}                                                    & \textbf{$\times$}                                                        \\ \hline
Cominelli et. al.~\cite{cominelli2023exposing}                  & \textbf{$\checkmark$}                                                 & \textbf{$\times$}                                                  & \textbf{$\times$}                                                       & \textbf{$\times$}                                                    & \textbf{$\times$}                                                        \\ \hline
\rowcolor[HTML]{DFE1E2} \ourmethod
            & \textbf{$\checkmark$}                                                 & \textbf{$\checkmark$}                                                  & \textbf{$\checkmark$}                                                       & \textbf{$\checkmark$}                                                    & \textbf{$\checkmark$}                                                        \\ \hline
\end{tabular}%
}
\end{table}
%\subsection{Acoustic-based approaches}

\noindent\textbf{Acoustic-based:} Passive acoustic sensing~\cite{wang2018c, li2022lasense, wang2023df} involves the creation of audio chirps that are reflected away from nearby surfaces and detected by a \textit{microphone}. In this direction, both macro~\cite{blumrosen2014noncontact}, as well as micro activities~\cite{li2022lasense}, were studied. Acoustic-based sensing involves several frequencies, including ultra-sounds~\cite{jiang2018towards}. Due to the reliance on audio frequency, acoustic-based approaches are susceptible to environmental noise, interference, and microphone orientation. In addition, multiple individuals affect the acoustic signature in an unpredictable manner~\cite{bai2020acoustic, wang2018c}. %So, the existing works~\cite{wang2018c} focus on a single individual's activity. %Finally, it might raise privacy concerns whenever microphones are involved, as potential eavesdropping is difficult to prevent.
%CSI information proved to be an essential direction for exploring passive monitoring of human activities as several works~\cite{arshad2017wi, fang2016bodyscan, ali2015keystroke} leveraged it to analyze human activities. It also supports real-time~\cite{wang2015understanding} estimation. UWB offers a better estimation accuracy~\cite{monica2016comparison} as compared to WiFi.   Here, we avoid studies involving gesture recognition~\cite{yu2020mmwave} due to their application-specific nature.
% \subsection{light-based approaches}
% \NEWTEXT{Some researchers have explored lights for studying activities. By capturing reflected lights from surfaces, various movements, and involved activities can be classified. In this direction, the use of lasers~\cite{zhang2018vibrosight} as well as visible light~\cite{jiang2018towards}, can be seen. However, with the use of lasers on surfaces, safety is a critical concern. Similarly, visible lights are often affected by ambient illuminance. After the signatures have been collected, they are processed for classification into different types of activities. 
% }

% \usepackage{graphicx}
%\subsection{RF-based approaches}

\noindent\textbf{RF-based:} Radio-frequency (RF) in the form of Wi-Fi~\cite{ding2020rf, wang2015understanding}, RFIDs~\cite{kellogg2014bringing}, UWB radars~\cite{zhang2022mobi2sense, bouchard2020activity} has been studied for capturing human dynamics. 
Many have explored Wi-Fi Channel State Information (CSI)~\cite{cominelli2023exposing, wang2015understanding, tan2019multitrack,cominelli2023exposing, soltanaghaei2020robust}; as the phase and amplitude of radio waves are impacted by the movements of objects in their path~\cite{chen2018wifi}. Both macro and micro-level~\cite{ali2015keystroke} activities have been studied in this direction. Alternatively, UWB is suitable for penetrating walls~\cite{zheng2021siwa} and capturing movements~\cite{chen2021movi} including small movements~\cite{zhang2022mobi2sense}. In some cases, RFIDs are used to capture human activity~\cite{huang2019id, PradhanRIO}. %Information such as phase and RSSI extracted from RFID signals has been leveraged to identify movement patterns. 

Regarding Wi-Fi, CSI extraction from signals is usually a complex process. Other environmental dynamics, such as door movement, furniture movement, and electromagnetic interference, can also affect the signal. %\textcolor{red}{Need to add a few more lines on wifi sensing with FMCW systems}. \NEWTEXT{In this regard, FMCW radio frequencies are more } 
Some works~\cite{fan2020home, zhao2018through} rely on FMCW techniques with specialized hardware aiming for a relatively higher depth resolution~\cite{fan2020home}. However, this specialized hardware is usually expensive compared to COTS hardware~\cite{chen2021rf}. A COTS FMCW mmWave sensor such as IWR1642~\cite{barrett2017using} demonstrates a better range resolution as compared to the specialized device used in \cite{fan2020home} ($\sim$3.75cm vs. $\sim$8cm)~\cite{rao2017introduction, fan2020home}. Compared to Wi-Fi CSI, UWB has a higher achievable resolution~\cite{chen2021rf}; however, it has a well-known spectrum coexistence issue. RFIDs also have a limited range of around 5 meters. Instead, mmWave-based sensors can detect small movements at a finer level. Due to its shorter wavelength, mmWave can create stronger reflections even from smaller objects~\cite{jiang2018towards}. The works~\cite{sen2023mmassist,sen2023mmdrive,gu2019mmsense, singh2019radhar, wang2021m, bhalla2021imu2doppler, gong2021mmpoint, xie2023mm3dface, shuai2021millieye, lu2020see} employing this modality usually rely on emitted mmWave signals in the form of chirps and exploit the received signal reflected by the surroundings to capture activity signatures. COTS mmWave radars often use FMCW chirps for this purpose. Features, such as pointclouds~\cite{singh2019radhar, wang2021m}, range-doppler~\cite{sen2023mmassist,sen2023mmdrive,yu2020mmwave, bhalla2021imu2doppler, ahuja2021vid2doppler}, etc., have been proven to be effective in movement detection. %In fact, mmWave sensing with Doppler sensing can demonstrate higher fidelity commensurate with the trade-off of invasiveness~\cite{ahuja2021vid2doppler} compared to the other relevant modalities. %However, mmWave does suffer relatively from attenuation~\cite{norouzian2019rain} compared to UWB and WiFi frequencies. Lastly, the literature also highlights the effectiveness of mmWave for tracking and localization~\cite{kong2022m3track} even in real-time. %Multiple individuals can be tracked with~\cite{soltanaghaei2021millimetro} or without~\cite{meng2020gait} the involvement of additional modalities.

In contrast to previous works, this work uses mmWave sensing to continuously track activities since it is minimally intrusive on privacy and captures micro-movements. The single modality is sufficient for continuous activity monitoring of multiple individuals. Our approach also detects the most number of activities (both macro and micro simultaneously) in the mmWave domain with a dynamic environment when multiple users are present. Table~\ref{tab:comparison} highlights the advantages of \ourmethod compared to the state-of-the-art contributions in the relevant domain. 
\section{Conclusion}
We need simple yet effective ways for humans to interact with our smart spaces. Existing ideas, however, use techniques that are both invasive and difficult to integrate. The key insight prompted us to design and develop \ourmethod, a lightweight yet highly effective mmWave-based continuous activity monitoring system. Through experiments, \ourmethod proves its effectiveness of single subject tracking with a mean absolute error of just 45cm despite supporting global coordinates. After that, it demonstrates field-deployable accuracy of 98\% and 94\%, respectively, for multiple macro and micro-scale activities. Based on the results, we are confident that \ourmethod will seamlessly adapt to human activities in all situations encountered in real-world scenarios. 

\ourmethod{} can be upgraded to a higher resolution by using devices like DCA1000EVM~\cite{dca1000evm}; however, it will lead to higher latency because of the increased volume of data. The existing form of \ourmethod is incapable of capturing micro activities beyond five meters due to signal attenuation with an increasing range, which is a fundamental challenge in mmWave. Instead of limiting our evaluation to a simplistic functional accuracy, we evaluated the performance of \ourmethod based on different counts, orientations, distances, and even energy consumption footprints, comparing it to the state-of-the-art baselines which demonstrate its superiority.

%Secondly, since the hardware transfers the captured data over USB, it suffers from a limited baud rate and, therefore, a low frame capture rate.\ourmethod provides a robust, scalable, and, most importantly, adaptive solution. Instead of limiting our evaluation to a simplistic functional accuracy, we take into account different counts, orientations, distances, and even energy consumption footprints.
%\input{tex/Sec_8_Limitations_and_discussions}

\balance
% \newpage
\bibliographystyle{ACM-Reference-Format}
\bibliography{refs.bib}
\begin{acronym}
	\acro{mmWave}{Millimeter Wave}
	\acro{RF}{Radio Frequency}
	\acro{4G}{4$^\text{th}$ Generation}
	\acro{5G}{5$^\text{th}$ Generation}
	\acro{COTS}{Commercial off-the-shelf}
	\acro{FMCW}{Frequency Modulated Continuous Wave}
	\acro{ADC}{Analog to Digital Conversion}
	\acro{DBSCAN}{Density-Based Spatial Clustering of Applications with Noise}
	\acro{EKF}{Extended Kalman Filter}
	\acro{RKF}{Recursive Kalman Filter}
	\acro{LoS}{Line of Sight}
	\acro{NLoS}{Non Line of Sight}
	\acro{FPS}{Frames Per Second}
\end{acronym}
\end{document}